\documentclass[aps,prl,reprint,amsmath,amssymb,superscriptaddress,longbibliography]{revtex4-1}

\usepackage[utf8]{inputenc}
\usepackage{graphicx}% Include figure files
\usepackage{dcolumn}% Align table columns on decimal point
\usepackage{bm}% bold math
\usepackage{amsmath}
\usepackage{color}
\usepackage{xspace}
\usepackage{ulem}
\usepackage{xfrac}
\usepackage{color,soul}
\usepackage{graphicx}
\usepackage[colorlinks=true,urlcolor=blue,citecolor=blue,linkcolor=blue,bookmarks=false,pdfstartview={FitH}]{hyperref}

\begin{document}
	
	\title{Multiband Quantum Criticality of Polar Metals}

	\author{Pavel A. Volkov}
	\email{pv184@physics.rutgers.edu}
	\affiliation{Department of Physics and Astronomy, Rutgers University, Piscataway, NJ 08854, USA}
	\affiliation{Center for Materials Theory, Rutgers University, Piscataway, New Jersey, 08854, USA}
	\author{Premala Chandra}
	\affiliation{Department of Physics and Astronomy, Rutgers University, Piscataway, NJ 08854, USA}
	\affiliation{Center for Materials Theory, Rutgers University, Piscataway, New Jersey, 08854, USA}
	
	\date{\today}
	
	\begin{abstract}
		Motivated by recent experimental realizations of polar metals with broken inversion symmetry, we explore the emergence of strong correlations driven by criticality when the polar transition temperature is tuned to zero. Overcoming previously discussed challenges, we demonstrate a robust mechanism for coupling between the critical mode and electrons in multiband metals.  We identify and characterize several novel interacting phases, including non-Fermi liquids, when band crossings are close to the Fermi level and present their experimental signatures for three generic types of band crossings.
	\end{abstract}
	
	\maketitle
	
	Metals close to quantum critical points (QCPs) are strongly
	correlated systems that often exhibit non-Fermi liquid behavior \cite{sachdev.1999} and novel orderings including 
	unconventional superconductivity \cite{scalapino2012}. 
	Studies of metals near
	spin density wave \cite{abanov2003,lohneysen.2007}, 
	ferromagnetic \cite{brando.2016} and nematic QCPs \cite{shibauchi.2014} 
	indicate that the behaviors of quantum critical metals depend crucially
	on the nature of the QCPs involved. Recent discoveries \cite{kolod.2010,shi2013,rischau.2017,jiang2017,yu2018,cao2018,fei2018} and predictions \cite{shirodkar2014,fei2016,ding2017} of a number of {\sl polar metals} \cite{benedek2016} that undergo an inversion symmetry breaking transition, structurally similar to a ferroelectric one \cite{anderson.1965} (whose QCP properties are also actively studied \cite{khmelnitskii1973,roussev.2003,Rowley2014,chandra2017,chandra2018,narayan2019})
	suggest a novel avenue of metallic quantum criticality to be explored.

\begin{figure}[h]
		\includegraphics[width=0.5\textwidth]{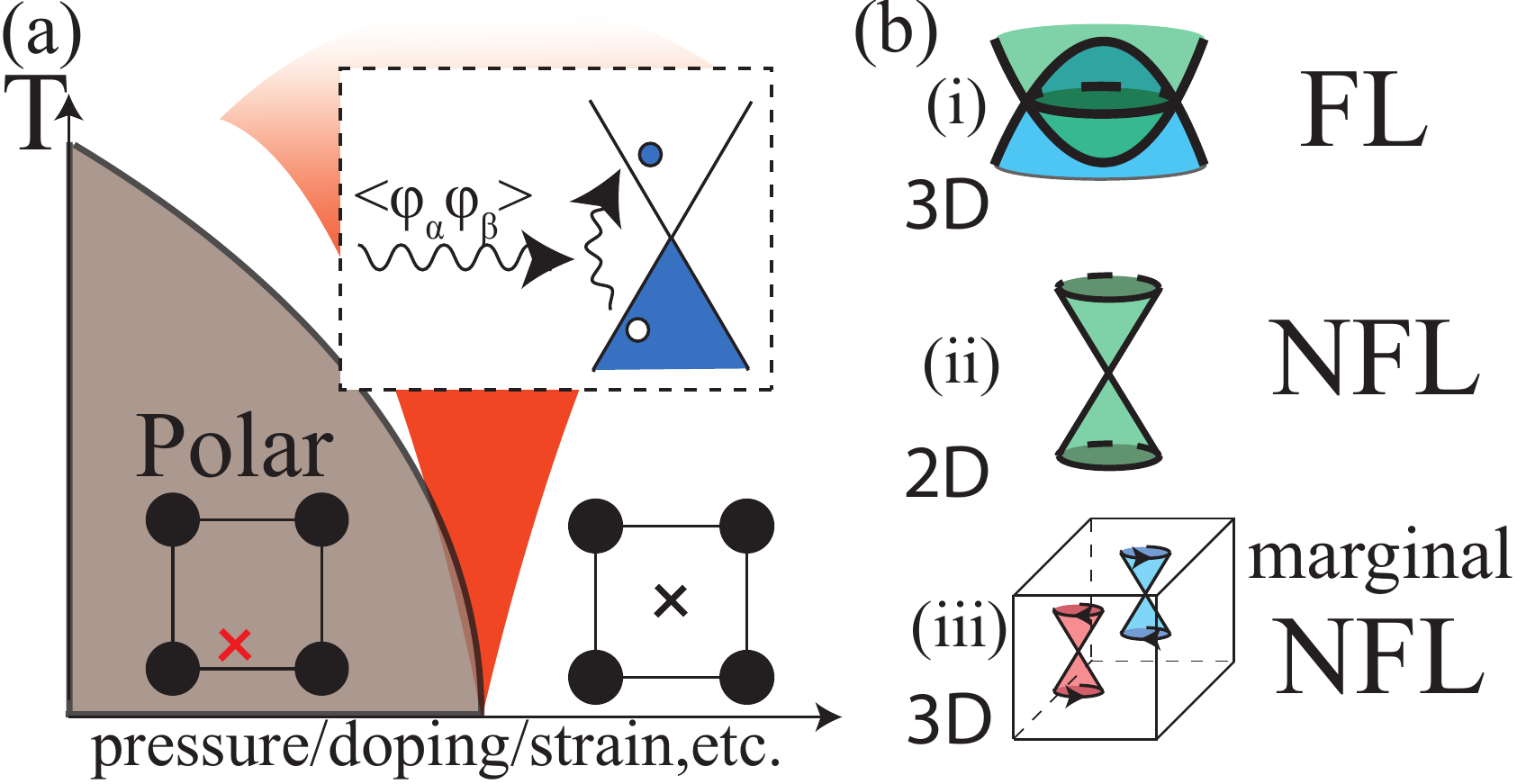}
		\caption{(a) Schematic phase diagram of a polar metal with a critical region around the QCP. Inset illustrates that the critical fluctuations couple to an interband excitation. (b) Summary of the QCP behaviors near typical band crossing: (i) a 3D nodal line, (ii) a 2D nodal point, and
			(iii) 3D Weyl points.%, where our results for (i) and (ii) do {\sl not} rely on the presence of spin-orbit coupling.  
			(N)FL is (non-)Fermi liquid, and in all cases the polar mode is strongly renormalized. Coulomb interactions introduce anisotropy for (i) and (ii), and
			gap the longitudinal mode for (iii).}
		\label{fig:cartoon0}
	\end{figure}

	Here we perform a systematic study of quantum critical polar metals 
	and possible strong correlations therein.  We show the critical polar
	mode to be strongly coupled to {\sl interband} particle-hole 
	excitations (Fig. \ref{fig:cartoon0}).
	Since unconventional metallic quantum criticality  
	occurs when critical bosons are coupled to gapless excitations, we study
	quantum critical polar metals with Fermi energies pinned to electronic
	band crossings; we present evidence of strong renormalization of the
	polar phonon spectra and non-Fermi liquid behavior of the charge carriers
	(Fig. \ref{fig:cartoon0}).
	
	Experimentally, intrinsic \cite{yoshida.2005,shi2013,lei.2018}
	and engineered \cite{kolod.2010,kim2016polar,rischau.2017,nukala2017,cao2018} 
	polar metals exist;
	the former include several layered transition metal 
	dichalchogenides \cite{jiang2017,fei2018,sharma2019}
	with more predicted \cite{shirodkar2014,fei2016,ding2017}.  
	Furthermore the search for Weyl semimetals has led to more
	polar (semi-)metals \cite{hasan2017}.
	Chemical tuning of polar transition temperatures has been 
	demonstrated  \cite{rischau.2017,sakai2016critical,barraza2018,narayan_2019}.
	Here we make predictions for experimental signatures of the novel
	phases that we identify in  quantum
	critical polar metals.
	
	An important challenge for the realization of correlated polar metals is that the critical boson (a polar optical phonon \cite{cochran.1960}) 
	of a polar metallic QCP
	does not couple easily to the electronic
	degrees of freedom; proposed couplings that involve order parameter 
	gradients \cite{Gorkov2016,ruhman.2016,wolfle.2018,wolfle.2018com,wolfle.2018repl} and/or nonlinearities \cite{marel2019} are 
	usually irrelevant in the scaling sense at a QCP \cite{sachdev.1999}. Additionally, Coulomb interactions play a special role here, leading to a splitting between
	longitudinal and transverse modes when the screening is weak 
	\cite{cochran.1960,mooradian.1966,larkin1969}, although this effect may be smaller for certain ferroelectric systems \cite{garrity2014,tagantsev1988}.

	Here we show that a Yukawa coupling of the order parameter ($\varphi$) to carriers, $H_Y = \lambda \int d{\bf r} \varphi({\bf r}) c^\dagger({\bf r})c({\bf r})$, known to induce strong correlations for other types of QCPs \cite{abanov2003,lohneysen.2007,brando.2016,shibauchi.2014}, can be generically realized in multiband systems even without 
	spin-orbit coupling (SOC) 
	(that has been previously considered \cite{kozii.2015,wu2017,yanase2018,yanase2019}), 
	leading to the most pronounced interaction effects 
	at band crossings.
	Using symmetry-based classification of such crossings, we analyze possible strongly coupled metallic behaviors near
	polar QCPs, including long-range Coulomb effects.

	{\sl Yukawa Coupling to the Polar Critical Mode.-}  
	We look for fermionic bilinears $\hat{O}^i({\bf k})$ such that 
	$\varphi^i \int d{\bf k}\hat{O}^i({\bf k})$ respects inversion symmetry, first assuming time-reversal symmetry $\mathcal{T}$. 
	Since the order parameter breaks inversion 
	symmetry, $\mathcal{P}^{-1}\varphi^i\mathcal{P}=-\varphi^i$, we thus seek a
	fermionic bilinear $\hat{O}^i({\bf k})$ that breaks inversion but not 
	time-reversal symmetry.
	
	For a single conduction band without SOC, the only possible form of $\hat{O}({\bf k})$ is $\hat{c}^\dagger_{\bf k} f_0({\bf k}) \hat{c}_{\bf k}$. 
	Since both $\mathcal{P}$ and $\mathcal{T}$ require $f_0$ to be even, it is 
	not possible for $\hat{O}({\bf k})$ to break only inversion symmetry. 
	However with SOC present, bilinears of the form 
	$\hat{c}^\dagger_{\bf k} f_i({\bf k}) s_i \hat{c}_{\bf k}$, 
	where $s_i$ are the Pauli matrices in spin space, are allowed:  an odd 
	in ${\bf k}$ choice for $f_i({\bf k})$ results in a bilinear that is 
	odd under $\mathcal{P}$ only. We thus conclude that SOC is necessary for
	a Yukawa polar coupling in a single-band model.
	
	By contrast, in a multiband system a Yukawa coupling can exist without SOC. In a two-band model ignoring spin, $\mathcal{T}$ is complex conjugation and $\mathcal{P}$
	acts in band space: 
	$\mathcal{P}\sim\sigma_0$ for bands with the same parity or 
	$\mathcal{P}\sim\sigma_3$ (up to a unitary transformation) in the opposite case.
	Writing a generic fermionic bilinear as $\hat{c}^\dagger_{\bf k} [f_0({\bf k})+\sum_{i=1}^3 f_i({\bf k}) \sigma_i ]\hat{c}_{\bf k}$, we find that the 
	terms breaking inversion, but not time-reversal, symmetries are even in 
	${\bf k}$ $f_1({\bf k})$ for $\mathcal{P}\sim\sigma_3$ or 
	odd in ${\bf k}$ $f_2({\bf k})$ for $\mathcal{P}\sim\sigma_0$. 
	We can thus have the following Yukawa couplings to the polar mode at ${\bf q} \approx 0$
	
	\begin{equation}
	\begin{array}{lr}
	H_{coupl}^{(a)} =
	\sum_{i,{\bf q},{\bf k}}
	f_a^i({\bf k})
	\varphi^i_{\bf q}
	c^\dagger_{{\bf k}+{\bf q}/2} \sigma_1 c_{{\bf k}-{\bf q}/2},
	&
	\mathcal{P}\sim\sigma_3
	\\
	H_{coupl}^{(b)} =
	\sum_{i,{\bf q},{\bf k}}
	f_b^i({\bf k})
	\varphi^i_{\bf q}
	c^\dagger_{{\bf k}+{\bf q}/2} \sigma_2 c_{{\bf k}-{\bf q}/2},
	&
	\mathcal{P}\sim\sigma_0,
	\end{array}
	\label{eq:Yukawa}
	\end{equation}
	
	\noindent where $f_{a(b)}^i({\bf k})$ is even(odd) in ${\bf k}$, and 
	the order parameter couples to an {\it interband} bilinear (Fig. \ref{fig:cartoon0}(a), inset).
	
\begin{figure}[h]
		\includegraphics[width=0.45\textwidth]{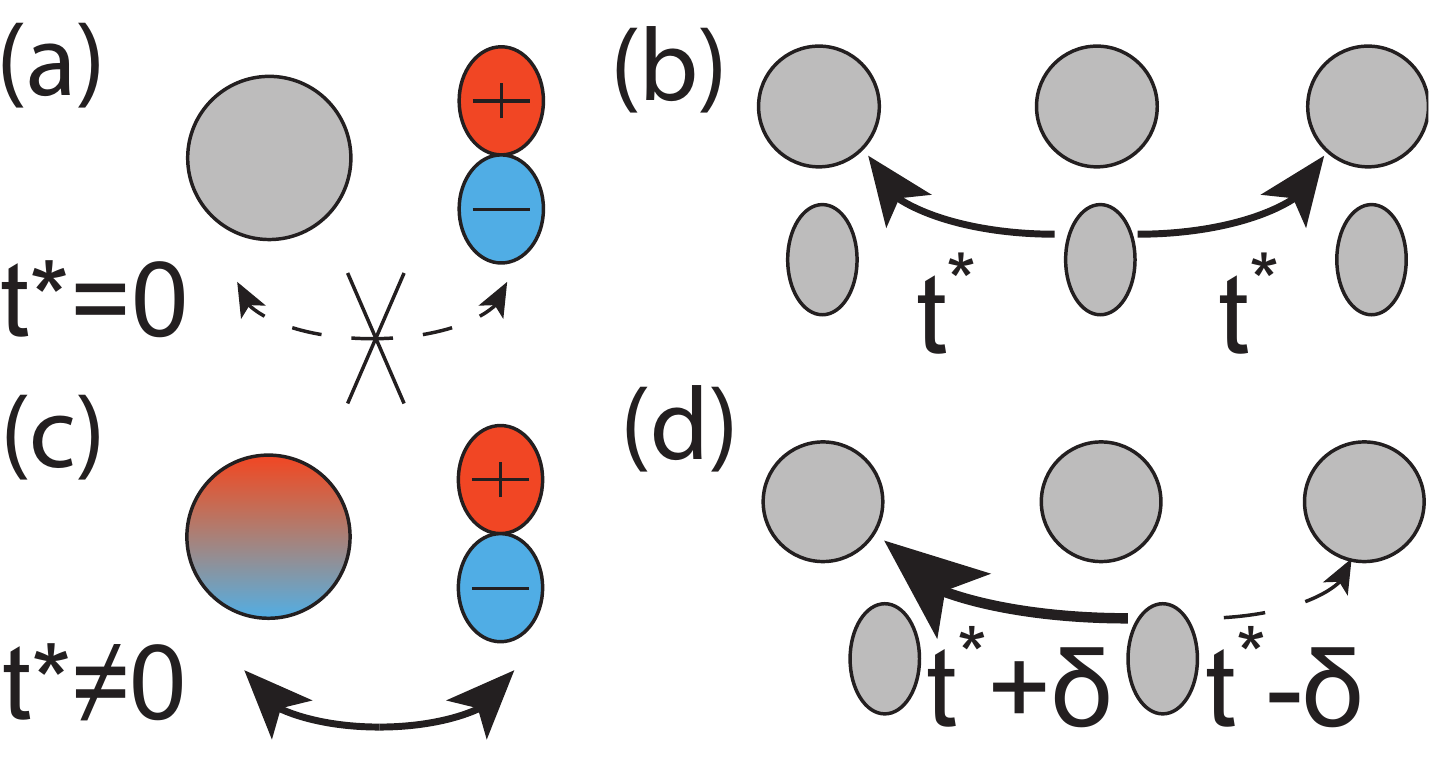}
		\caption{ Illustration of the coupling to the polar order parameter for two orbitals having (a,c) opposite and (b,d) same parity under inversion. (a) and (b) show the symmetric phase $\varphi^i=0$, while (c) and (d) show the state for $\varphi^i\neq0$. In both cases, interorbital hopping changes, as reflected in \eqref{eq:Yukawa}.
		}
		\label{fig:cartoon}
	\end{figure}

	If we assume the bands to originate from two distinct orbitals,
	the physical mechanism of this Yukawa polar coupling can be 
	illustrated (Fig. \ref{fig:cartoon}). 
	If the orbitals have different parity (e.g. s and p) 
	(Fig. \ref{fig:cartoon} (a)), they  are mixed 
	linearly by an inversion-breaking perturbation 
	(similar to the Stark effect). This mixing is reflected 
	in a nonzero constant hybridization between the resulting bands, 
	forbidden in the symmetric phase (Fig. \ref{fig:cartoon} (a)). 
	Due to the necessity of ${\bf k}$-dependence, the same-parity
	case (Fig. \ref{fig:cartoon} (b)) cannot be viewed as local.
	We exemplify it by a nearest-neighbor hopping between the orbitals 
	(Fig. \ref{fig:cartoon} (b)); absence of inversion symmetry 
	yields distinct left and right interorbital hoppings from a given site,
	similar to the dimerization occurring in the SSH model \cite{su1979}. Similar effects have been considered in studies of SrTiO$_3$ interfaces \cite{joshua2012,diez2015}.

	{\sl Band Crossings and Low-Energy Theory.-} To drive unconventional metallic behavior already at weak coupling, the interband particle-hole excitations coupled to the critical mode with \eqref{eq:Yukawa} need to be gapless. This is possible close to momenta where the two bands cross; a low-energy theory can then be constructed around these band crossings if they occur close to the Fermi energy. This situation can be realized due to filling considerations (as e.g. in graphene) or by carrier doping. Here we study the polar QCP when the Fermi energy is at the 
	band crossing. Neglecting SOC, $\mathcal{PT}$ symmetry leads to protected 
	line crossings in 3D systems \cite{kimrappe.2015,fang.2015,bzdusek.2017} 
	and point nodes in 2D \cite{bzdusek.2017}. For completeness, 
	we also study a polar QCP in a $\mathcal{T}$-breaking driven Weyl semimetal
	\cite{burkov.2011,halasz.2012,armitage.2018}.

	Having identified the three generic types of band crossings, we next 
	turn to their emergent metallic behaviors at the polar QCPs. 
	We study the limit $T = 0^+$ 
	at the QCP itself, 
	analyzing each case both with and without long-range Coulomb interactions. The latter situation is relevant when (i) inversion symmetry-breaking in the insulating system does not produce a macroscopic dipole moment (e.g., when the transition is to a structure with a nonpolar point group that does not allow for a macroscopic dipole moment \cite{cryst} or in the case of elemental materials) or if (ii) there exist additional Fermi pockets that lead to strong screening. In the absence of (ii), screening of the Coulomb interaction will depend on the type of the band crossing.

	{\sl 3D Nodal Lines.-} We consider a minimal Hamiltonian with a 
	circular nodal line in the $k_x,k_y$ plane
	\begin{equation}
	\begin{gathered}
	H_{line} =
	\sum_{\bf k}
	c^\dagger_{\bf k}
	\left[
	\frac{k_x^2+k_y^2-k_F^2}{2m}\sigma_3+\gamma k_z\sigma_2
	\right]
	c_{\bf k},
	\\
	H_{coupl} =
	\lambda \sum_{{\bf q},{\bf k}}
	{\varphi}_{\bf q}
	c^\dagger_{{\bf k}+{\bf q}/2} \sigma_1 c_{{\bf k}-{\bf q}/2},
	\end{gathered}
	\label{eq:ham:nline}
	\end{equation}
	which corresponds to taking $f_i=const$ in \eqref{eq:Yukawa} (as is discussed below, this does not affect the qualitative results). Only a single order parameter component can
	couple to fermions near a single nodal line. 
	In principle isotropy is restored when two additional nodal lines that couple 
	to the other order parameter components are present e.g. three $p_{x,y,z}$-like 
	bands crossing an $s$-like one; then there are three nodal lines, 
	each coupled to the corresponding component of the order 
	parameter $\varphi_{x,y,z}$. However, as we will show now, 
	it is sufficient to consider \eqref{eq:ham:nline} with straightforward 
	generalizations.
	
	We begin by considering the lowest-order bosonic self-energy, following the Hertz 
	approach \cite{hertz1976}. Our calculations 
	\footnote{See Supplemental Material at [URL will be inserted by	publisher] for the details of calculations, which includes Refs. \cite{fitzpatrick.2013,metlitski.2010,goldefeld1992,Thakur2018,Zhou2018}} yield
	\begin{equation}
	\begin{gathered}
	\Pi(i\omega,{\bf q}) - \Pi(0,0)=
	-\frac{\lambda^2 m \rho(\omega,q_r,q_z)}{4\pi \gamma}
	E\left(\frac{(v_F q_r)^2}{\rho(\omega,q_r,q_z)^2}\right)
	\end{gathered}
	\label{eq:nline:rpa}
	\end{equation}
	where $\rho(\omega,q_r,q_z) = \sqrt{\omega^2+(v_F q_r)^2+(\gamma q_z)^2}$ and $E(x)$ is an elliptic integral of the second kind. 
	Gapless fermions lead to the damping of the critical mode:  $\Pi(i\omega,{\bf q})\sim |\omega|$ at low ${\bf q}$ ($v_Fq,\gamma q\ll\omega$) similar to the situation at spin-density wave QCPs \cite{abanov2003}. However, unlike that case, \eqref{eq:nline:rpa} 
	is strongly momentum-dependent, leading to an unchanged dynamical critical exponent, i.e. $z=1$. Once the bosonic self-energy is taken into account, a scaling analysis with the scheme of Ref. \onlinecite{huh.2016} yields the scaling dimensions of Yukawa and quartic interboson couplings to be $(2-d)/2$ and $1-d$, respectively, making both irrelevant in the $d=3$ nodal line case. An explicit calculation \cite{Note1} of the fermionic self-energy and vertex corrections 
	with a screened bosonic propagator also indicates the absence of infrared divergences, showing that the scaling limit of \eqref{eq:ham:nline} is captured by 
	Hertz-Millis theory. We note, however, that a recent RG analysis using a 
	different momentum shell scheme \cite{uchoa2019} suggests nontrivial 
	corrections to the Hertz-Millis fixed point away from the large-$N_f$ limit ($N_f$ being the number of fermionic flavors, for additional discussion see \cite{Note1}).
	
	We now address the role of a momentum-dependent fermion-boson coupling within Hertz-Millis theory: since at the QCP the only relevant term is the bosonic self-energy, a momentum-dependent 
	coupling would just result in replacing $\lambda^2$ in \eqref{eq:nline:rpa} with its Fermi surface average. Next we address the role of Coulomb interaction. 
	Assuming that the Fermi liquid state is robust at the QCP,
	we know from previous work \cite{huh.2016,rhim2016} that screening by 
	nodal line electrons results in a renormalized Coulomb 
	interaction $\propto q^{-1}$, while further effects of the Coulomb interaction 
	are irrelevant in the RG sense and can be considered perturbatively. 
	The coupling between renormalized Coulomb interaction and the polar phonon
	affects the critical propagator
	\[
	D_{Coul}^{-1}(i\omega,{\bf q}) = -\Pi(i\omega,{\bf q}) + V^{NL}_{Coul}({\bf q}) Q_0^2 q^2 \cos^2 \eta,
	\]
	where $\eta$ is the angle between ${\bf q}$ and the polarization direction of $\varphi$ and $Q_0$ - the effective charge of the polar mode. Since $V^{NL}_{Coul}({\bf q})\sim q^{-1}$ \cite{huh.2016,rhim2016}, the Coulomb interaction does not change the scaling properties of the critical mode but rather introduces 
	an anisotropy.

	{\sl 2D Dirac point.-} We consider a Hamiltonian $H_{point} = \sum_{\bf k} c^\dagger_{\bf k}v_F(k_x \sigma_x +k_y \sigma_y)c_{\bf k}$ and coupling $H_{coupl}=\lambda \sum_{{\bf q},{\bf k}}
	{\varphi}_{\bf q} c^\dagger_{{\bf k}+{\bf q}/2} \sigma_3 c_{{\bf k}-{\bf q}/2}$ (we take $\mathcal{P}\sim\sigma_1$ here). Again we find the order parameter to be 
	scalar, similar to what occurs at a CDW transition in 
	graphene \cite{alicea2006,Fuchs2007}, that breaks the inversion (but not translational)  symmetry and is associated with charge imbalance between two sublattices.

	In the absence of Coulomb interactions, this model is equivalent to the 
	Gross-Neveu-Yukawa (GNY) model \cite{gross.1974,zinn.1991} 
	whose critical properties have been studied 
	extensively \cite{mihaila.2017,lang.2018}. Its critical point is known
	to have emergent Lorentz invariance and $z=1$; it follows then that the 
	critical phonon velocity ($c_s$) is renormalized such that $c_s/v_F\to1$. 
	The anomalous dimensions for both the bosons and the fermions further demonstrates the non-Fermi liquid behavior at the QCP.
	
	We now include Coulomb interactions, first considering their effect on the 
	critical boson whose propagator is now 
	\begin{equation}
	D(i\omega, {\bf q}) = \frac{1}{(\omega^2+c^2 q^2)+2\pi Q_0^2 q_x^2/q},
	\label{eq:2dNP:Dcoul}
	\end{equation}
	while the Coulomb interaction takes the form:
	\begin{equation}
	V_{Coul}(i\omega, {\bf q}) = \frac{2\pi (\omega^2+c^2 q^2)}{q(\omega^2+c^2 q^2)+2\pi Q_0^2 q_x^2}.
	\label{eq:2dNP:coul}
	\end{equation}
	This renormalization changes the bare scaling dimension of $q_x$: $[q_x] = [\omega,q_y]^{3/2}$. Consequently the renormalized Coulomb interaction becomes 
	irrelevant for the fermions ($[e]=-1/4$), while the Yukawa coupling 
	remains relevant ($[\lambda]=1/4$). 
	In order to determine the critical properties at this fixed point we perform one-loop momentum-shell RG 
	calculations in 2D; details are provided in \cite{Note1}, where we introduce an additional 
	parameter, the number of nodal points $N_f$. 
	We note that the fermionic and bosonic renormalizations are determined by different dimensionless 
	couplings,  $\beta_\varphi = \frac{\lambda^2 Z_\varphi^2 a_\psi}{2\pi^2v_{Fy}^3 Z_\psi \Lambda}$ and $ 
	\beta_\psi = \frac{\lambda^2  Z_\varphi}{2\pi^2v_{Fy}^2 \sqrt{2 \pi Q_0^2 \Lambda}}$, respectively, 
	where $Z_{\psi, \varphi}$ are the quasiparticle residues and $\Lambda = \Lambda_0 e^{-l}$ is the running RG scale. We find the resulting solution of the RG equations to exhibit a runaway flow 
	for $\beta_\varphi\sim e^{2l/5}$, while $\beta_\psi\sim 2/l$ goes to zero. Most importantly, we find that the 
	Fermi velocities along the two directions are renormalized differently: 
	$\frac{d v_{Fy}}{d l}<0;\; \frac{d v_{Fx}}{d l}>0$, with $v_{Fy}$ eventually flowing to zero as $e^{l/5}$ \cite{Note1}. 
	Enhanced anisotropy is also present in the bosonic behavior, with $c_x^2 \sim e^{l};\;c_y^2\sim const$. Taking 
	$l\sim \log(k^{-1},\omega^{-1}))$, we use the asymptotic solutions to obtain the following
	forms of critical propagators (where $\alpha,\beta,\delta,\gamma,\rho$ are constants)
	\begin{equation}
	\begin{gathered}
	D(i\omega, {\bf q}) \sim \frac{1}{\alpha \omega+c_y^2 q_y^2+2\pi Q_0^2 q_x^2/q + \beta q_x} 
	\\
	G(i\varepsilon, {\bf k}) \sim \frac{1}{\delta i \varepsilon^{0.8}+ \gamma\sigma_x k_x^{0.8}+ \rho \sigma_y k_y^{1.2}}. 
	\end{gathered}
	\label{eq:2dNP:GFscaling}
	\end{equation}

	{\sl Weyl Points in 3D QC Polar Metals.-} Neglecting  possible anisotropies, we find
	the Weyl Hamiltonian and the Yukawa coupling to take the form
	\begin{equation}
	\begin{gathered}
H_{W} =  \sum_{\bf k} c^\dagger_{\bf k} v_F (\vec{k}\cdot\vec{\sigma}) c_{\bf k},
	\\
	H_{coupl} =
	\lambda \sum_{{\bf q},{\bf k}}
	\vec{\varphi}_{\bf q}
	c^\dagger_{{\bf k}+{\bf q}/2} \vec{\sigma} c_{{\bf k}-{\bf q}/2}.
	\end{gathered}
	\label{eq:ham:weyl}
	\end{equation}
	Since the interaction is marginal, we use perturbative RG to probe the system's behavior. 
	Importantly, the bosonic self-energy evaluated on the momentum shell is
	\begin{equation}
	\delta \Pi_{\alpha \beta} =  - \frac{\lambda^2}{\pi^2 v_F^3} \frac{\omega^2 +v_F^2(q^2\delta_{\alpha\beta} - q_\alpha q_\beta)}{12} dl
	\label{eq:weylrpa}
	\end{equation}
	and we see that the longitudinal mode is unrenormalized while the transverse one harderns. 
	The full RG equations are presented in the Supplemental Material \cite{Note1}; the RG flows to 
	weak coupling with the large-$l$ asymptotic $\alpha \to 2/(N_f l \log l)$, where $\alpha\equiv  \frac{\lambda^2 Z_\varphi^2}{12 \pi^2 Z_\psi v_F^3}$ and $N_f$ is the number of Weyl points in the system (cf. we neglect inter-point
	coupling since it requires finite momentum transfer). We find $c_L/c_T\to0$ due to the hardening of the transverse phonon velocity. The quasiparticle residues for bosons  and fermions both vanish 
	($Z_\psi\propto \frac{const}{l}; Z_\varphi\sim \frac{const}{\log l}$) which, in conjunction with $l=\log(k^{-1},\omega^{-1})$, suggests logarithmic non-Fermi liquid corrections. However the bosonic quasiparticle 
	residue vanishes only as $1/\log{l}$, so the bosons receive only loglog corrections and are 
	relatively well-defined at the QCP.

	The Coulomb interaction screened by the polar phonon becomes itself irrelevant, but results in the renormalized propagator for the longitudinal mode acquiring a gap.
	The RG equations are then obtained by disregarding the longitudinal mode's contribution to the 
	fermionic self-energy and vertex renormalization. 
	We note that the longitudinal mode nonetheless receives corrections due to Yukawa coupling; 
	this reflects itself in the behavior of the dielectric constant (see below). 
	The solutions of the RG equations \cite{Note1} are qualitatively similar to those without Coulomb interactions.
	
	Finally we note that for a 3D Dirac point \cite{kozii2019} (that requires additional symmetries to be realized), the RG equations are found to flow to strong coupling, and the critical mode to soften. 
	Since a 3D Dirac point can be thought of as a stable merger of two Weyl points, we attribute this result
	to inter-Weyl cone scattering that we have not considered due to finite momentum separation between cones $Q_w$.
	We thus expect that in the polar phase, where the Dirac point splits into two Weyl ones, the flow to strong 
	coupling will be cut off at a scale set by $v_F Q_W$. 
	
	{\sl Experimental Signatures.-} We next discuss simple experimental signatures of the 
	quantum critical polar metallic phases we have identified. The QC (bosonic) specific heat $C_{bos}$ can be estimated
	with Hertz-Millis theory \cite{coleman2001,Note1} leading to $C_{bos}\sim T^{d/z}$.  In the cases we study, we observe that $z=1$ except for the nodal point cases with Coulomb interactions: in 2D one momenta
	scales as $\sqrt{\omega}$ at the QCP which suggests, using \eqref{eq:2dNP:GFscaling}, that $C_{bos}\sim T^{1.5}$;
	in 3D a logarithmic correction is present.
	We also obtain T-dependent resistivity estimates with a scattering rate calculation using an RPA-screened bosonic propagator \cite{schofield1999,Note1}. These estimates are summarized in Table \ref{tab}.
	
	\begin{table}
		\begin{tabular}{lccc}
			Band crossing Type \quad\quad & $\lim_{T\rightarrow 0^+}\rho(T)$ \quad\quad& $\lim_{T\rightarrow 0^+} C_{bos}(T)$\\
			\hline
			\quad \quad \\
			3D NL &$T^3$&$ T^3$\\
			3D NP  &$T^{-1}$&$ T^3$\\
			2D NP &$\rho_0$&$T^{2-\Delta}$\\
		\end{tabular}
		\caption{Summary of Hertz-Millis estimates for the temperature dependencies of resistivity $\rho(T)$ and the bosonic contribution to the specific 
			heat $C_{bos}(T)$ close to a Polar QCP; here NL and NP are nodal lines and points respectively, $\rho_0$ is a constant and $\Delta = 0.5$ when Coulomb interactions are included and $0$ otherwise.}
		\label{tab}
	\end{table}

	In all three cases we have studied, the critical polar mode is strongly affected by interactions close to the QCP. For the nodal points the characteristic boson velocity renormalizes to a value of order $v_F$, suggesting stiffening of the transverse mode. A more complete picture can be given for the Hertz-Millis theory of the nodal line case using \eqref{eq:nline:rpa} continued to real frequencies. For $q_r=0$, the transverse has an unusual dispersion $\omega^2 = (cq_z)^2 + \omega_\lambda(\sqrt{4 (\gamma^2-c^2) q_z^2+\omega_\lambda^2} -\omega_\lambda)/2$ while for the  $q_z=0$ case one has
	\begin{equation}
	\omega \approx 
	\begin{cases}
	- i \omega_\lambda & (q\ll\omega_\lambda/v_F),\\
	\sqrt{2\omega_\lambda v q/\pi- \frac{i \omega_\lambda^2}{\pi}}& (\omega_\lambda/v_F \ll q\ll(v_F/c)^2 \omega_\lambda/v_F )\\
	cq & ((v_F/c)^2 \omega_\lambda/v_F\ll q)
	\end{cases}
	\end{equation}
	where $\omega_\lambda=\frac{\lambda^2 m}{8 \gamma}$. Observation of such dispersion renormalization and smearing of the spectral weight may be accessible by inelastic neutron scattering measurements.
	
	Additionally, for the case of 3D nodal semimetals, the bulk $\omega-$ and ${\bf q-}$dependent contribution of the polar mode to the dielectric constant $\varepsilon({\bf q},\omega)$ may be obtained from optical conductivity \cite{armitage.2018} or EELS experiments \cite{eels2017}:
	\begin{equation}
	\begin{gathered}
	\lim_{v_Fq,\gamma q\ll\omega}
	\quad \varepsilon^{Pol}_{NL}({\bf q}, \omega) = \frac{4i\pi Q_0^2 \cos^2 \eta}{\omega_\lambda\omega} 
	\\
	\lim_{{\bf q}\rightarrow 0}
	\quad \varepsilon^{Pol}_{WP}({\bf q}, \omega) = -\frac{4 \pi Q_0^2}{\omega^2(\log \omega^{-1})^{N_f/\kappa_0(N_f)}},
	%\quad \quad (\bf q\to 0),
	\end{gathered}
	\end{equation}
	where $N_f/\kappa_0(N_f)$ depends on the number $N_f$ of Weyl points but is of order $1$ \cite{Note1}.  We also note that the presence of a Yukawa-coupled band crossing actually {\sl promotes} polar ordering, since the static part $\Pi(0,0)$ of the bosonic self-energy is positive. This effect is maximal when 
	the Fermi level is at the band crossing and diminishes as it moves 
	away. Of particular 
	interest is the case when symmetries allow nodal surfaces; here the relevant
	bosonic
	self-energies would be logarithmically divergent \cite{volkov2018} leading 
	to possible realizations of an electronically driven polar order that was
	proposed some time ago \cite{bersuker1966} already at weak coupling.

	{\sl Conclusion.-} In this Letter we have shown that nodal multiband metals provide promising platforms for strongly correlated metallic behaviors near polar QCPs. We have demonstrated a generic mechanism for Yukawa-like coupling to the critical mode without spin-orbit coupling. Identifying band crossings to be most affected by the polar QCP, we have studied critical behavior for three distinct cases (2D Dirac and 3D Weyl points, and 3D nodal lines) with and without Coulomb interactions. In our study we find the critical polar mode to be strongly renormalized for all band crossing types, and we have demonstrated the emergence of non-Fermi liquid behavior for the two nodal point cases. Finally we have analyzed thermodynamic, transport, and dielectric properties and the critical mode dispersion for the quantum critical polar metallic phases we have identified. In view of the recent discovery of a number of polar metals with a multiband electronic structure such as LiOsO$_3$\cite{shi2013,yu2018}, MoTe$_2$ \cite{jiang2017} and WTe$_2$ \cite{fei2018} and predictions of many more \cite{shirodkar2014,fei2016,ding2017,narayan_2019}, we hope that our study will provide guidance for the search of exotic metallicity in future experiments on polar metals.
	
	%define D_0 somewhere
	%(3) Finish resistivity supplement
	
	\section{Acknowledgements} We thank P. Coleman for detailed discussions, particularly in the early development of
	this project.  We are grateful to P.W. Anderson, E. Christou, S. Fang, G. Jose, D. Khomskii, E. K\"onig, Y. Komijani, D. Maslov
	and B. Uchoa for their helpful comments.  P. A. V. acknowledges a Postdoctoral Fellowship from the Rutgers University Center for Materials Theory, and this work was also supported by grant DE-SC0020353 (P.C.) funded by the U.S. Department of Energy, Office of Science.

\newpage

\clearpage
\onecolumngrid
\appendix
% \setstretch{1.5}
\renewcommand{\thefigure}{S\arabic{figure}}
\addtocounter{equation}{-10}
\addtocounter{figure}{-2}
\addtocounter{table}{-1}
\renewcommand{\theequation}{S\arabic{equation}}
\renewcommand{\thetable}{S\arabic{table}}

%%%%%%%%%%%%%%%%%%%%%%%%%%%%%%%%%
%%%%%%%%%%%%%%%%%%%%%%%%%%%%%%%%%

\begin{widetext}
	\setcounter{page}{1}
	\renewcommand{\theequation}{S\arabic{equation}}
	\setcounter{equation}{0}
	\renewcommand{\thefigure}{S\arabic{figure}}
	\setcounter{figure}{0}
	
	%%%%%%%%%%%%%%%%%%
	
	\begin{center}
		\textbf{\Large
			Supplemental Material: Multiband Quantum Criticality of Polar Metals}
	\end{center}

	This supplementary material contains the details of diagrammatic and RG calculations leading to the results presented in the main text. While relying on well-established methods, we present here the derivations in an extended form to facilitate comprehension for the reader.
	
		\tableofcontents

\section{Polar QCP in a nodal line system}

We consider the simplest case a system with nodal line at the Fermi surface in Eq. (2) of the main text. In what follows we will linearize the dispersion in the radial direction such that $\frac{k_x^2+k_y^2-k_F^2}{2m} \approx v_F (k_r-k_F)$, where $v_F=k_F/m$ and ${\bf k}_r = (k_x,k_y)$. First we compute the polarization operator that enters the RPA-like correction to the boson propagator (Fig. \ref{fig:nlinerpa}) $\langle T_\tau  \varphi \varphi\rangle$:
\begin{gather*}
\Pi(i \omega_n,{\bf q})
=
-\lambda^2T\sum_{\varepsilon_n}
\int\frac{d {\bf k}}{(2\pi)^3}
{\rm Tr} \left[\sigma_1G_0(i(\varepsilon_n+\omega_n/2), {\bf k}+{\bf q}/2)
\sigma_1G_0(i(\varepsilon_n-\omega_n/2), {\bf k}-{\bf q}/2)\right]
=
\\
=
-\lambda^2T\sum_{\varepsilon_n}
\int\frac{d {\bf k}}{(2\pi)^d}
\frac{
	{\rm Tr} \left[
	\sigma_1
	[i\varepsilon_+ + \xi_+ \sigma_3+ \delta_+ \sigma_2]
	\sigma_1
	[i\varepsilon_- + \xi_- \sigma_3+ \delta_- \sigma_2]
	\right]
}
{
	(\varepsilon_+^2 + \xi_+^2 + \delta_+^2)
	(\varepsilon_-^2 + \xi_-^2 + \delta_-^2)
}=
\\
=
2\lambda^2
T\sum_{\varepsilon_n}
\int\frac{m d\xi d\theta d\delta }{\gamma(2\pi)^3}
\frac{
	\varepsilon_+\varepsilon_- + \xi_+\xi_- + \delta_+ \delta_-
}
{
	(\varepsilon_+^2 + \xi_+^2 + \delta_+^2)
	(\varepsilon_-^2 + \xi_-^2 + \delta_-^2)
}
\end{gather*}
where$\varepsilon_\pm=\varepsilon_n\pm\omega_n/2;\;\xi_\pm=v_F(k_r-k_F)\pm {\bf v}_F\cdot {\bf q}_r/2 = \xi\pm v_F q_r/2\cos(\theta);\;\delta_\pm=\gamma(k_z\pm q_z/2);\;\delta = \gamma k_z$ and ${\bf q}_r = [q_x,q_y]$.

\begin{figure}[h!]
	\includegraphics[width=0.5\linewidth]{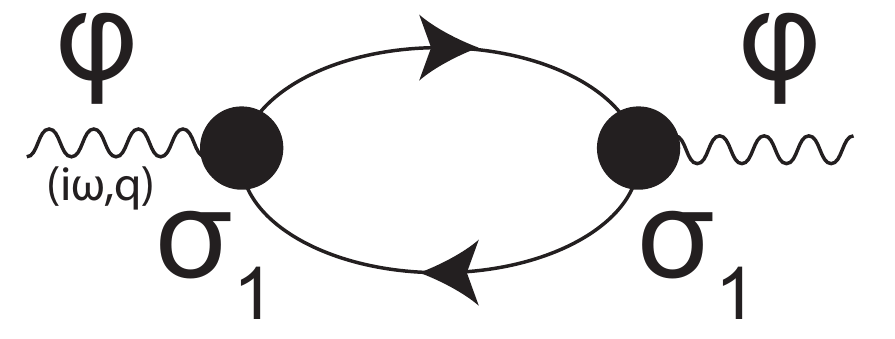}
	\caption{Feynman diagram for the lowest-order bosonic self-energy.}
	\label{fig:nlinerpa}
\end{figure}

For $T=0$ one can bring the integral to the following form:
\begin{gather*}
\int d \theta d\varepsilon d\xi d\delta \frac{\varepsilon^2+\xi^2+\delta^2 - (\omega/2)^2-(v_F q_r\cos(\theta)/2)^2-(\gamma q_z/2)^2}{
	[(\varepsilon+\omega/2)^2+(\xi+v_F q_r\cos(\theta)/2)^2+(\delta+\gamma q_z/2)^2]}
\\
\cdot \frac{1}{
	[(\varepsilon-\omega/2)^2+(\xi-v_F q_r\cos(\theta)/2)^2+(\delta-\gamma q_z/2)^2]}
=
\\
=
\int d \theta
\int r^2 dr d\cos \eta d \varphi [\equiv d\varepsilon d\xi d\delta]
\frac{r^2+r_0^2}
{(r^2+r_0^2)^2-4 r^2 r_0^2 \cos^2\eta}
\end{gather*}
It follows then that the result of integration over $\varepsilon,\xi,\delta$ can depend only on the absolute value of the vector $\vec{r}_0\equiv(\omega/2,v_F q_r\cos(\theta)/2,\gamma q_z/2)$. Thus
\[
\Pi(i \omega_n,{\bf q}) = 2\lambda^2 \int_0^{2\pi} d\theta f(\sqrt{\omega^2+(v_F q_r\cos(\theta))^2+(\gamma q_z)^2})
\]
so we can get the result by considering the case ${\bf q}=0$. Integration over $\varepsilon$ yields:
\begin{equation}
\begin{gathered}
f(\omega_n) =
\int_0^\Lambda\frac{2m r^2 dr }{\gamma(2\pi)^2}
\frac{1}{(4r^2+\omega^2)},
\\
f(0) = \int_0^\Lambda\frac{m dr }{2\gamma(2\pi)^2} = \frac{m}{8\pi^2\gamma}\Lambda,
\\
f(\omega_n) - f(0) = \frac{m}{8\pi^2\gamma}
\int_0^\infty dr \frac{-\omega_n^2 \tanh \frac{r}{2T}}{(4r^2+\omega_n^2)}
=_{T=0} \frac{-m}{8\pi^2\gamma} \frac{\pi |\omega_n|}{4},
\\
\Pi(i\omega_n,{\bf q}) - \Pi(0,0)=  -\frac{\lambda^2 m}{16\pi \gamma} \int{d \theta} \sqrt{\omega_n^2+(v_F q_r\cos(\theta))^2+(\gamma q_z)^2},
\end{gathered}
\label{eq:Pi:nline}
\end{equation}
where $\Lambda$ is an upper cutoff of the order $k_F$. The integral can be evaluated analytically using elliptic functions resulting in Eq. (3) of the main text.

\subsection{Implications for QCP: Scaling Analysis}
One can see that the RPA correction dominates the bosonic propagator at low frequencies/momenta. We will now argue using scaling that, similar to Hertz theory, no further singular corrections from interactions are expected. Our aim is to find the scaling dimension of the Yukawa coupling including the anomalous dimension of the bosonic propagator found in RPA. First we need to fix the engineering dimensions of the fields.

We use a cylindrical momentum shell for fermions around the nodal line\cite{huh.2016} that yields $[\Psi] = -(d+1)/2$ for fermions (assuming nodal line plane and $d-2$ linearly dispersing momenta $L_\Psi \sim \int d \varepsilon k_F d^{d-1}k \Psi^\dagger \left(\varepsilon +\xi \sigma_3+ \sum_{i>2} \delta_i \sigma_2 \right)\Psi$) and a spherical one with the result $[\varphi]=-(d+2)/2$ for bosons ($L_\varphi\sim\int d \omega d^d q \varphi^* \sqrt{\omega^2+{\bf q }^2}\varphi $) [combined boson/fermion scaling of this type is also discussed in \cite{fitzpatrick.2013}]. The Yukawa coupling is then:
\[
\lambda\int d\omega d {\bf q} d\varepsilon k_F d^{d-1}k \Psi^\dagger\Psi\varphi,
\]
resulting in $[\lambda] = -(d+1+d - (d+1)-(d+2)/2) =(2-d)/2$, irrelevant for $d>2$.

Moreover, for the bosonic quartic term we have:
\[
u \left[\int d\omega d {\bf q}\right]^3 \varphi^4,
\]
and $[u] = -(3(d+1) - 2(d+2)) = 1-d$, irrelevant for $d>1$.

Other scaling schemes have been also applied to $q=0$ transitions, such as the nematic one \cite{metlitski.2010}. Here we show that the 'patching scheme' used there cannot be applied here. In the nematic case, the use of patching scheme has been motivated by arguing that a boson with momentum ${\bf q}$ is most strongly coupled to the regions on the Fermi to which ${\bf q}$ is tangent. For a single fermion patch the Lagrangian is:
\[
\int d k_x d^{d-1} k_y d\varepsilon \Psi^\dagger (i\varepsilon +v_Fk_x+\gamma k_y^2) \Psi
\]
while for bosons after the RPA correction from the full Fermi surface is included one has:
\[
\int d\omega d^d q \varphi^* \left(\frac{\omega}{q} +q_x^2+q_y^2\right) \varphi.
\]
Keeping the fermion action invariant requires $[k_x] = [k_y]^2$ and thus $q_x$ should be omitted from the bosonic propagator in the scaling limit for a single patch. Importantly, the polarization operator of a single patch indeed yields $\Pi\sim\frac{\omega}{q_y}$ \cite{sachdev.1999}, vindicating the argument that taking a single patch into account is sufficient to describe the nematic QCP in the scaling limit.

In our case, on the other hand, the RPA boson Lagrangian is (for momenta being in-plane)
\[
\int d\tau d^dx \varphi^*  F(i\omega, \sqrt{q_x^2+q_y^2}) \varphi;
\]
as before one should neglect $q_x$ in the scaling limit for the fermionic patch. However it can be seen from eq. (\ref{eq:Pi:nline}), that the contribution $f(i\omega, |q_y|)$ comes from $\theta\approx0$ region in the integral, i.e. from the patch with $v_F \parallel q_y$. Thus, contributions from other patches are necessary in the scaling limit, not allowing for self-consistent description of this QCP within a single patch. 

A key role in the scheme above is played by the assumption that the Fermi surface curvature dictates the scaling limit $[k_x] = [k_y]^2$. As it is shown above to lead to an inconsistency, one may use instead the scaling dimensions dictated by the bosonic propagator (i.e. $[k_x]=[k_y]=[\varepsilon]$) and ignore the Fermi surface curvature, which is however equivalent to the cylindrical scheme used above.

Another implementation of the cylindrical scaling scheme, where the momenta are scaled towards the nodal ring also for the vertex corrections, has been discussed \cite{uchoa2019}; while in the $N_f\to\infty$ limit the results agree with the ones presented here, there appear to be distinct $1/N_f$ corrections (see also an additional discussion below).

\subsection{Stability of Hertz-Millis Fixed point: Explicit Calculations of Lowest-order Diagrams}

To substantiate our statement of the stability of the Hertz-Millis fixed point we additionally provide an explicit calculation for the lowest-order diagrams (corresponding to $1/N_f$ corrections) in the large-$N_f$ limit beside the bosonic self-energy: fermionic self-energy and vertex correction. The former one is given by (see also Fig. \ref{fig:nlineseflen}):

\begin{figure}[h!]
	\includegraphics[width=0.5\linewidth]{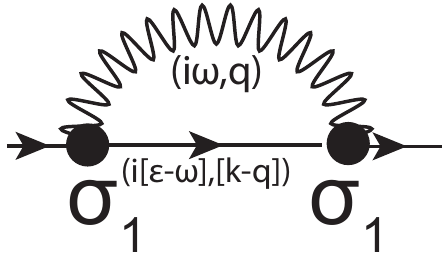}
	\caption{Feynman diagram for the lowest-order fermionic self-energy.}
	\label{fig:nlineseflen}
\end{figure}

\begin{gather*}
\Sigma (i\varepsilon, {\bf k}) = \lambda^2 \int \frac{d{\bf q} d\omega}{(2\pi)^{4}} D(i\omega,{\bf q}) \sigma_1 G(i(\varepsilon-\omega), {\bf k}-{\bf q})\sigma_1 
\approx
\\
\approx
-\lambda^2\int \frac{d{\bf q} d\omega}{(2\pi)^{4}}
D(i\omega,{\bf q})
\frac{
	i\varepsilon (-\omega^2+(v_F q \cos \varphi)^2+(\gamma q_z)^2)
	-
	\xi \sigma_3 (\omega^2-(v_F q \cos \varphi)^2+(\gamma q_z)^2)
	-\gamma k_z \sigma_2 (\omega^2+(v_F q \cos \varphi)^2-(\gamma q_z)^2)
}
{[\varepsilon^2+(v_f q \cos \varphi)^2+(\gamma q_z)^2]^2},
\end{gather*}
where one can see that an infrared singularity is located at $\varepsilon=0,{\bf q}=0$ for any value of $\varphi$. For the bare boson propagator $D_0(i\omega,{\bf q}) = (\omega^2+c^2q^2)^{-1}$ the singularity is logarithmic, however, for the renormalized one $D(i\omega,{\bf q}_r,q_z)|_{\omega,q\to0}\sim (\omega,q_r,q_z)^{-1}$ the integral is regular and the self-energy does not change the scaling properties of the fermionic propagator. The expression for the vertex correction (Fig. \ref{fig:nlinevc}) is:

\begin{figure}[h!]
	\includegraphics[width=0.5\linewidth]{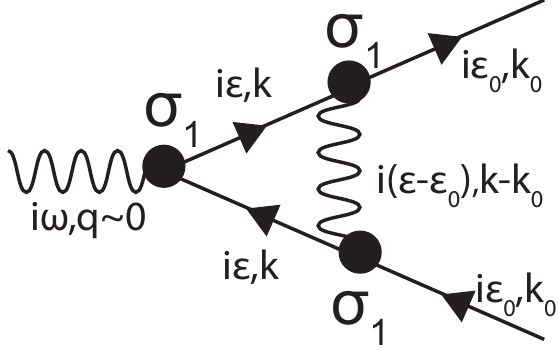}
	\caption{Feynman diagram for the lowest-order vertex correction.}
	\label{fig:nlinevc}
\end{figure}

\begin{gather*}
\lambda^3\int \frac{d\varepsilon d {\bf k}}{(2\pi)^4} 
\sigma_1G(i \varepsilon, {\bf k}) \sigma_1 G(i \varepsilon, {\bf k}) \sigma_1 D(i(\varepsilon-\varepsilon_0),|{\bf k_r}-{\bf k_r}_0|,k_z)
\approx
\\
\approx
-\sigma_1 \lambda^3
\int \frac{ k_F d\varepsilon dk_r d\varphi dk_z}{(2\pi)^4} 
\frac{1}{\varepsilon^2+v_F^2(k_r-k_F)^2+\gamma^2k_z^2}
D(i\varepsilon,2 k_F|\sin \frac{\varphi}{2}|,k_z),	
\end{gather*}
where one can see that an infrared singularity is located at $\varepsilon=0,\;k_r=k_F,\;k_z=0,\;\varphi=0$ (note that away from $\varphi=0$ the integral is regular even around $\varepsilon=0,\;k_r=k_F,\;k_z=0$). For the bare boson propagator $D_0(i\omega,{\bf q}) = (\omega^2+c^2q^2)^{-1}$ the singularity is logarithmic, however, for the renormalized one $D(i\omega,{\bf q}_r,q_z)|_{\omega,q\to0}\sim (\omega,q_r,q_z)^{-1}$ the integral is regular and thus the vertex correction can be incorporated perturbatively, not affecting the scaling properties. Note that a result consistent with a previous study \cite{uchoa2019}, where the vertex correction has an $1/x$ singularity for the bare propagator (the divergence in the fermion self-energy is consistent with the one presented here) can be obtained from the above if integration over $\varphi$ is disregarded and $\varphi$ is set to $0$.

\section{Nodal Point Systems}

Before we go into details of particular cases, let us start with scaling and power counting arguments with unrenormalized propagators. We assume Yukawa coupling between polar mode and fermions; the bare action for fermions is $\int \frac{d\varepsilon d^dk}{(2\pi)^{d+1}} \psi^\dagger(-i\varepsilon+{\bf k}{\bf \sigma})\psi$ and for bosons $\int \frac{d\omega d^dq}{2(2\pi)^{d+1}} \varphi_q(\omega^2+c_s q^2)\varphi_{-q}$. Unlike the nodal line case, the scaling scheme for the momenta can be taken to be the same for both fermionic and bosonic fields, i.e. we take a spherical shell around ${\bf k},{\bf q}=0$. One obtains the bare scaling dimensions $[\Psi] = -(d+2)/2,\;[\varphi] = -(d+3)/2$ resulting in $[\lambda] = -[2(d+1)-(d+3)/2-(d+2)] = (3-d)/2$. The fermion-boson interaction is then marginal in 3D and relevant in 2D.

Analyzing the divergencies of low-order diagrams we found that near the gaussian fixed point $\Pi,\;\Sigma,\;\Gamma$ are all expected to show the same behavior: $\log[1/E]$ in 3D and $1/E$ in 2D.

Additionally, the interboson interaction $g\int ( d \omega d^dq )^2 \varphi^4$ has the scaling dimension at the bare level $[g] = -( 3(d+1)-2(d+3)) = 3-d$. However, its effect in the lowest order is to renormalize the bosonic mass which has to be exactly zero at the QCP; higher-order effects that can lead to a bosonic self-energy will be ignored here, while the Yukawa coupling leads to a bosonic self-energy already at the lowest order.

\subsection{Details for the 2D Dirac Case}
In the absence of SOC Dirac points are guaranteed to exist in case the $\mathcal{PT}$ symmetry is intact. Some details of the results, however, may depend on the particular realization. Let us consider a two-band model. In the case $\mathcal{P}\sim\sigma_0$, the Hamiltonian takes the form $f_1({\bf k})\sigma_1+f_3({\bf k})\sigma_3$, where both $f_1({\bf k})$ and $f_3({\bf k})$ are even functions of ${\bf k}$. The ISB term can be only of the form $f_2({\bf k})\sigma_2$ with an odd $f_2({\bf k})$. In this case the band crossing $f_1=f_2=0$ may occur at a generic point in the BZ, but for linear dispersion it should not be a TRS-invariant point. It follows then that there has to be an even number of Dirac points, where $f_2({\bf k})\neq0$ resulting in a Yukawa-like coupling.

If we consider the case where $\mathcal{P}\sim\sigma_1$, the Hamiltonian is restricted to $f_1({\bf k})\sigma_1+f_2({\bf k})\sigma_2$, where $f_1({\bf k})$ is even and $f_2({\bf k})$ is odd. The ISB term is then $f_3({\bf k})\sigma_3$ with an odd $f_3({\bf k})$. Once again, the condition $f_1({\bf k})=0$ does not have to be satisfied at a TRS momentum and thus $f_3({\bf k})\neq 0$ at the Dirac node, but there necessarily will be an even number of them (e.g. as in graphene). It's important to check that the ISB term wouldn't vanish at the Dirac node, as otherwise the interaction would have the form $\sim \lambda k \varphi \psi^\dagger\psi$ and would have the scaling dimension $(1-d)/2$ making it irrelevant.

Now we focus on the case $\mathcal{P}\sim\sigma_1$; after a proper renaming of the axes the Hamiltonian for a single Dirac node can be written as
\begin{equation}
\begin{gathered}
H_{Dir} = v_F(k_x \sigma_x +k_y \sigma_y)
\\
H_{coupl} =
\lambda \sum_{{\bf q},{\bf k}}
{\varphi}_{\bf q}
c^\dagger_{{\bf k}+{\bf q}/2} \sigma_3 c_{{\bf k}-{\bf q}/2}.
\end{gathered}
\label{eq:ham:npoint}
\end{equation}
Note that in this case the fermions couple only to a single component of $\varphi$, suggesting that it is an Ising order parameter. This is easy to understand in the case of graphene lattice, where the electronic polar order parameter is just the charge imbalance between the two sublattices. The expression for the polar mode self-energy (Fig. \ref{fig:diagr} [I]) is:

\begin{figure}[h!]
	\includegraphics[width=0.8\linewidth]{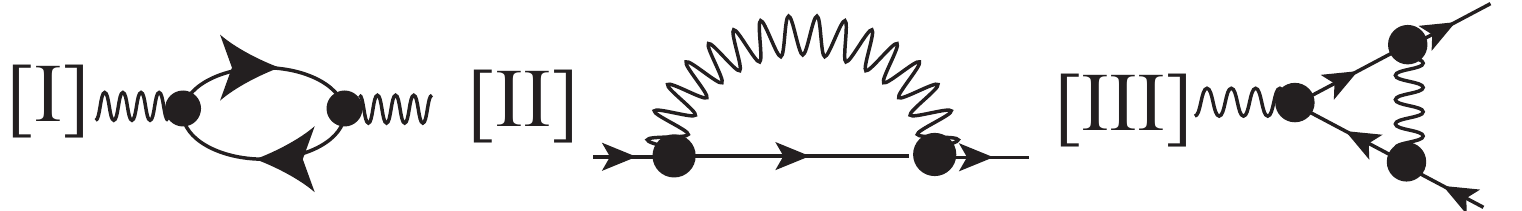}
	\caption{Lowest-order diagrams in Yukawa coupling.}
	\label{fig:diagr}
\end{figure}

\begin{equation}
\begin{gathered}
\Pi(i\omega,{\bf q}) = -\lambda^2T\sum_{\varepsilon_n}
\int\frac{d {\bf k}}{(2\pi)^3}
{\rm Tr} \left[\sigma_3G_0(i(\varepsilon_n+\omega_n/2), {\bf k}+{\bf q}/2)
\sigma_3G_0(i(\varepsilon_n-\omega_n/2), {\bf k}-{\bf q}/2)\right]
=
\\
=
-\lambda^2T\sum_{\varepsilon_n}
\int\frac{d {\bf k}}{(2\pi)^2}
\frac{
	{\rm Tr} \left[
	\sigma_3
	[i\varepsilon_+ + v_F (k_{x,+}\sigma_1+k_{y,+}\sigma_2)]
	\sigma_3
	[i\varepsilon_- + v_F (k_{x,-}\sigma_1+k_{y,-}\sigma_2)]
	\right]
}
{
	(\varepsilon_+^2 +v_F^2 k_+^2)
	(\varepsilon_-^2 + v_F^2 k_-^2)
}=
\\
=
2\lambda^2
T\sum_{\varepsilon_n}
\int\frac{d k_x d k_y}{(2\pi)^2}
\frac{
	\varepsilon_+\varepsilon_- +v_F^2 k_{x,+}k_{x,-}+v_F^2 k_{y,+}k_{y,-}
}
{
	(\varepsilon_+^2 +v_F^2 k_+^2)
	(\varepsilon_-^2 + v_F^2 k_-^2)
}
\end{gathered}
\label{eq:2dnp:PiGen}
\end{equation}
Using the previous results for the nodal line case, we get the following. For $T=0$ one has
\begin{equation}
\begin{gathered}
\Pi(i\omega,{\bf q}) =
\frac{\lambda^2}{2\pi v_F^2}\left(\Lambda-|\mu|-\frac{\sqrt{\omega^2+(v_Fq)^2}}{2}\left[\frac{\pi}{2}-\arctan \frac{2 |\mu|}{\sqrt{\omega^2+(v_Fq)^2}}\right]\right);
\\
\Pi(i\omega,{\bf q})_{\mu=0} - \Pi(0,0)_{\mu=0} =
-\frac{\lambda^2(\omega^2+(v_Fq)^2)}{8\pi v_F} \int_0^\infty \frac{dk}{(v_Fk)^2+(\omega^2+(v_Fq)^2)/4}
=
-
\frac{\lambda^2}{8 v_F^2}\sqrt{\omega^2+(v_F q)^2},
\end{gathered}
\label{eq:2dnp:pol}
\end{equation}
where $\mu$ is the chemical potential with respect to the Dirac point. Note that there is no additional integration over the angle $\theta$ as in the nodal line case, which was related in that case to the finite $k_F$ in the plane of the nodal line. Let us move onto the fermion self-energy. For the purpose of the RG calculation to follow we restrict ourselves to $T=0$ and expand the self-energy (Fig. \ref{fig:diagr} [II]) in momentum and frequency ${\bf k}$ and $\varepsilon$:
\begin{equation}
\begin{gathered}
\Sigma (i\varepsilon, {\bf k}) = \lambda^2 \int \frac{d^2q d\omega}{(2\pi)^{3}} D(-{\bf q},-\omega) \sigma_3 G(i(\varepsilon+\omega), {\bf k}+{\bf q})\sigma_3 =
\\
=
-\lambda^2\int \frac{d^2q d\omega}{(2\pi)^{3}}
D({\bf q},\omega)
\frac{i(\varepsilon+\omega)-v_F({\bf k}+{\bf q})\vec{\sigma}}
{(\varepsilon+\omega)^2+v_F^2({\bf k}+{\bf q})^2}
\approx
\\
\approx
-\lambda^2\int \frac{d^2q }{(2\pi)^2}
\int \frac{d\omega}{2\pi}
D({\bf q},\omega)\left(
\frac{i\varepsilon (-\omega^2+v_F^2q^2)}{(\omega^2+v_F^2q^2)^2}
+
\frac{-v_F({\bf k}\vec{\sigma})(\omega^2+v_F^2q^2) + v_F^3({\bf q}\vec{\sigma}) 2 ({\bf k q})}{(\omega^2+v_F^2q^2)^2}
\right)
=
\\
=
-\lambda^2\int \frac{q d q }{2\pi}
\int \frac{d\omega}{2\pi}
D(q,\omega)\left(
\frac{i\varepsilon (-\omega^2+v_F^2q^2)}{(\omega^2+v_F^2q^2)^2}
+
\frac{-v_F({\bf k}\vec{\sigma})\omega^2}{(\omega^2+v_F^2q^2)^2}
\right).
\end{gathered}
\label{eq:2dnp:selfen}
\end{equation}
It is evident that for $D(q,\omega)\sim(\omega^2+(cq)^2)$ the integral diverges as $1/q$, thus the fermion and boson self-energies need, in principle, to be considered on equal footing. Finally, we consider the lowest-order vertex correction in Fig. \ref{fig:diagr} [III] (for the RG purposes it is enough to take all the incoming momenta/frequencies to be 0):
\begin{equation}
\begin{gathered}
\Gamma =\lambda^3 \int \frac{d^2q d\omega}{(2\pi)^{3}}
D(i\omega,{\bf q})
\sigma_3 G(i\omega,{\bf q})\sigma_3G(i\omega,{\bf q})\sigma_3=
\\
=
-\lambda^3 \sigma_3 \int \frac{d^2q d\omega}{(2\pi)^{3}}
D(i\omega,{\bf q})
\frac{1}{\omega^2+v_F^2q^2}.
\end{gathered}
\label{eq:2dnp:vert}
\end{equation}

\subsubsection{RG: Cylindrical (Momentum) Shell in 2D}
We now take the second-order terms in $q,\omega$ in \eqref{eq:2dnp:PiGen}, first-order in $\varepsilon,p$ in \eqref{eq:2dnp:selfen}, and set the incoming momenta/frequencies to zero in \eqref{eq:2dnp:vert} and evaluate the integrals
on the momentum shell $k,q\in[\Lambda e^{-dl},\Lambda],\;\varepsilon,\omega\in [-\infty,\infty]$. Note also that we take the frequency/momenta of Green's functions in the polarization operator to be $\varepsilon+\omega,p+q$ and $\varepsilon,p$, unlike \eqref{eq:2dnp:PiGen}. We get
\begin{gather*}
\delta\Pi = -\frac{\lambda^2}{8\pi v_F^3} (\omega^2+(v_F q)^2/2) \frac{dl}{\Lambda}
\\
\delta\Sigma = -i\varepsilon \frac{\lambda^2}{4\pi c_s(c_s+v_F)^2}\frac{dl}{\Lambda}
+v_F({\bf k}\vec{\sigma})\frac{\lambda^2}{8\pi v_F(c_s+v_F)^2}\frac{dl}{\Lambda}
\\
\delta\Gamma(0,0) = -\sigma_3\lambda\frac{\lambda^2}{4\pi c_s v_F (c_s+v_F)}\frac{dl}{\Lambda}
\end{gather*}

We define now the renormalized propagators for fermions $G=(G_0^{-1}-\Sigma)^{-1}$ and bosons $D=(D_0^{-1}-\Pi)^{-1}$:
\[
D = \frac{1}{a_\varphi(l)^2\omega^2+(c_s(l))^2q^2},\;G=\frac{1}{ia_\psi(l)\varepsilon - v_F(l) {\bf k}\cdot\sigma}.
\]
Evaluating the same diagrams with these propagators we get:
\begin{gather*}
\delta\Pi = -\frac{\lambda^2/(a_\psi)^2}{8\pi (v_F/a_\psi)^3} (\omega^2+(v_F q/a_\psi)^2/2) \frac{dl}{\Lambda}
\\
\delta\Sigma = -i\varepsilon \frac{\lambda^2/(a_\psi a_\varphi^2)}{4\pi (c_s/a_\varphi)[c_s/a_\varphi+v_F/a_\psi]^2}\frac{dl}{\Lambda}
+(v_F/a_\psi)({\bf k}\vec{\sigma})
\frac{\lambda^2/(a_\psi a_\varphi^2)}
{8\pi (v_F/a_\psi)[c_s/a_\varphi+v_F/a_\psi]^2}\frac{dl}{\Lambda}
\\
\delta\Gamma(0,0) = -\sigma_3\lambda\frac{\lambda^2/(a_\psi^2 a_\varphi^2)}{4\pi (c_s/a_\varphi) (v_F/a_\psi) [c_s/a_\varphi+v_F/a_\psi]}\frac{dl}{\Lambda}
\end{gather*}
Introducing the dimensionless coupling constant $\alpha = \frac{\lambda^2}{8\pi v_F^3 \Lambda}$ and extending the previous results to the case of $N_f$ Dirac nodes, the RG equations can now be derived:
\begin{equation}
\begin{gathered}
\frac{d a_\varphi^2}{dl} = \alpha a_\psi N_f;
\\
\frac{d c_s^2}{dl} = N_f \alpha v_F^2/(2 a_\psi);
\\
\frac{d a_\psi}{dl} =\frac{2 \alpha v_F^3}{a_\psi a_\varphi c_s [c_s/a_\varphi+v_F/a_\psi]^2}
=\frac{2 \alpha a_\psi^2/a_\varphi^2}{\eta(1+\eta)^2};
\\
\frac{d v_F}{dl} = \frac{\alpha v_F a_\psi/a_\varphi^2}{(1+\eta)^2};
\\
\frac{d \alpha}{dl} =
\alpha+2\alpha \frac{1}{\lambda}\frac{d \lambda}{d l} - 3\alpha \frac{1}{v_F}\frac{d v_F}{d l}
=\alpha-4\frac{\alpha^2 a_\psi/a_\varphi^2}{\eta(1+\eta)}-3\frac{\alpha^2 a_\psi/a_\varphi^2}{(1+\eta)^2}
;
\\
\frac{d \eta}{dl} = \eta \left(
\frac{1}{c_s}\frac{d c_s}{d l}+\frac{1}{a_\psi}\frac{d a_\psi}{d l}-\frac{1}{v_F}\frac{d v_F}{d l}
-\frac{1}{a_\varphi}\frac{d a_\varphi}{d l}
\right)=
\\
\eta \left(
\frac{\alpha a_\psi/a_\varphi^2}{4\eta^2} N_f
+
\frac{2 \alpha a_\psi/a_\varphi^2}{\eta(1+\eta)^2}
-
\frac{\alpha a_\psi/a_\varphi^2}{(1+\eta)^2}
-
\alpha \frac{a_\psi}{2a_\varphi^2} N_f
\right)=
\\
=
\frac{\eta \alpha a_\psi N_f}{2a_\varphi^2} \left(\frac{1}{2\eta^2}-1\right)
+
\frac{\eta \alpha a_\psi}{a_\varphi^2(1+\eta)^2} \left(\frac{2}{\eta}-1\right)
\end{gathered}
\end{equation}
where $\eta=a_\psi c_s/(a_\varphi v_F)$. The RG flow is fully described by the four coupled equations for $a_\psi,\;a_\varphi,\;\alpha$ and $\eta$:
\begin{equation}
\begin{gathered}
\frac{d a_\varphi}{dl} = \frac{\alpha a_\psi N_f}{2 a_\varphi};
\\
\frac{d a_\psi}{dl} =\frac{2 \alpha a_\psi^2/a_\varphi^2}{\eta(1+\eta)^2};
\\
\frac{d \alpha}{dl}
=\alpha-4\frac{\alpha^2 a_\psi/a_\varphi^2}{\eta(1+\eta)}-3\frac{\alpha^2 a_\psi/a_\varphi^2}{(1+\eta)^2}
;
\\
\frac{d \eta}{dl} =
\frac{\eta \alpha a_\psi N_f}{2a_\varphi^2} \left(\frac{1}{2\eta^2}-1\right)
+
\frac{\eta \alpha a_\psi}{a_\varphi^2(1+\eta)^2} \left(\frac{2}{\eta}-1\right)
\end{gathered}
\end{equation}
It is furthermore useful to rewrite those using the coupling constant $\beta = \frac{\lambda^2 a_\psi}{8\pi v_F^3 a_\varphi^2 \Lambda}$:
\begin{equation}
\begin{gathered}
\frac{d a_\varphi}{dl} = \frac{\beta a_\varphi N_f}{2};
\\
\frac{d a_\psi}{dl} =\frac{2 \beta a_\psi}{\eta(1+\eta)^2};
\\
\frac{d \beta}{dl}
=\beta-4\frac{\beta^2}{\eta(1+\eta)}-3\frac{\beta^2}{(1+\eta)^2}
+\beta\left(\frac{2 \beta}{\eta(1+\eta)^2}-\beta N_f\right)
;
\\
\frac{d \eta}{dl} =
\frac{\eta \beta N_f}{2} \left(\frac{1}{2\eta^2}-1\right)
+
\frac{\eta \beta}{(1+\eta)^2} \left(\frac{2}{\eta}-1\right).
\end{gathered}
\end{equation}
from the equation for $\beta$ it is evident that $\beta=0$ fixed point is unstable; increasing $\beta$ drives the increase of $a_\psi$ and $a_\varphi$; in the limit $a_\varphi\to \infty$ (or, alternatively, $Z_\varphi\equiv 1/a_\varphi = 0$) there exists a fixed point at $\eta=\eta_0,\; \beta= \beta_0\;Z_\varphi,Z_\psi = 0$ (see an example of the flow in Fig. \ref{fig:RGNP}). For $N_f\gg1$ $\eta_0=1$ and $\beta_0 = N_f^{-1}$.

\begin{figure}[h!]
	\includegraphics[width=\linewidth]{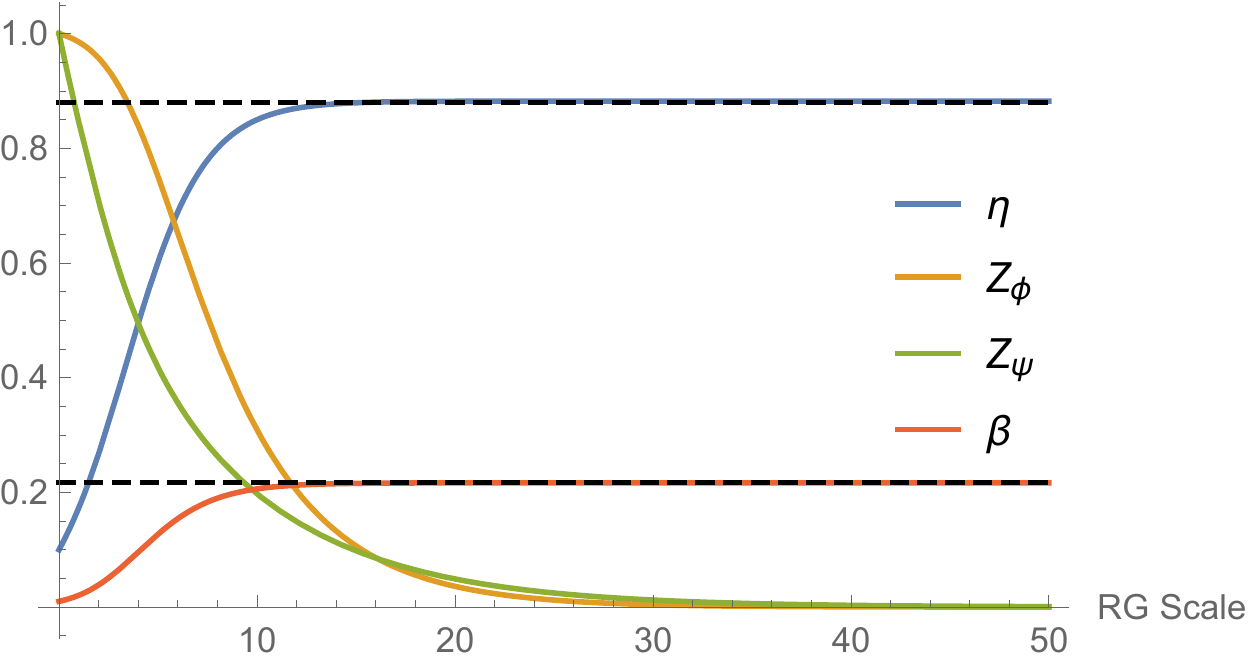}
	\caption{RG flow for $N_f=2;\beta(0)=0.01$;$\eta(0)=0.1,Z_\varphi(0)=Z_\psi(0)=1$; black dashed lines are fixed point values $\beta_0 \approx0.22$ and $\eta_0\approx0.88$.}
	\label{fig:RGNP}
\end{figure}

We can also estimate the anomalous scaling properties near the fixed point. One has then $a_\varphi = e^{l \beta_0 N_f/2};\;a_\psi\sim e^{l 2\beta_0/\eta_0/(1+\eta_0)^2};\;v_F\sim e^{l\beta_0/(1+\eta_0)^2}$ and $c_s\sim \exp \beta_0 l[N_f/2+(1-2/\eta_0)/(1+\eta_0)^2]$. Note that there is a correction to the dynamical critical exponent as the momentum and frequency scale differently: $z \neq 1$, but this deviation is the same for both bosons and fermions. As $l\sim -\log(k)$ we get that $\varepsilon, k \to \varepsilon^{1-2\beta_0/\eta_0/(1+\eta_0)^2}, k^{1-\beta_0/(1+\eta_0)^2}$ for fermions and $\omega, q \to \omega^{1-\beta_0 N_f/2}, q^{1-\beta_0 [N_f/2+(1-2/\eta_0)/(1+\eta_0)^2}$ for bosons. For $N_f\to \infty$ we recover the RPA result: $\omega^2\to\omega$ and $q^2\to q$. One can see that $N_f$ can be used to control the fixed point value of the coupling constant allowing for a possible justification of the weak-coupling RG.

\subsubsection{$\varepsilon$-Expansion}
A different way to arrange an RG scheme in 2D is to expand in space dimensionality around the upper critical dimension, i.e. $d=3$. The integration is performed then on the momentum shell in 3D, and since all integrals are logarithmically divergent, it is sufficient for the leading order in $d-3$ to perform the angular averaging may be performed in 3D \cite{goldefeld1992}. Additionally, $\sigma$ matrices are to be replaced with $\gamma$ matrices, such that a mass term anticommuting with the non-interacting Hamiltonian is allowed. The results are:

\begin{gather*}
\delta\Pi = -\frac{\lambda^2}{4\pi^2 v_F^3} (\omega^2+(v_F q)^2) \frac{dl}{\Lambda^{3-d}},
\\
\delta\Sigma = -i\varepsilon \frac{\lambda^2}{4\pi^2 c_s(c_s+v_F)^2}\frac{dl}{\Lambda^{3-d}}
+v_F({\bf k}\vec{\sigma})\frac{\lambda^2}{4\pi^2}\frac{2 c_s+v_F}{3 v_F c_s(c_s+v_F)^2}\frac{dl}{\Lambda^{3-d}},
\\
\delta\Gamma(0,0) = - \gamma_m \lambda\frac{\lambda^2}{4\pi^2 c_s v_F (c_s+v_F)}\frac{dl}{\Lambda^{3-d}}.
\end{gather*}
Defining the coupling constant $\alpha=\frac{\lambda^2}{4\pi^2 v_F^3 \Lambda^{3-d}}$ and reevaluating the contributions above with renormalized Green's functions we get the RG equations:
\begin{equation}
\begin{gathered}
\frac{d a_\varphi^2}{dl} = \alpha a_\psi N_f;
\\
\frac{d c_s^2}{dl} = N_f \alpha v_F^2/a_\psi;
\\
\frac{d a_\psi}{dl} =\frac{\alpha a_\psi^2/a_\varphi^2}{\eta(1+\eta)^2};
\\
\frac{d v_F}{dl} = \frac{\alpha v_F a_\psi/a_\varphi^2}{(1+\eta)^2}\frac{2\eta+1}{3\eta};
\\
\frac{d \alpha}{dl} =(3-d)\alpha-2\frac{\alpha^2 a_\psi/a_\varphi^2}{\eta(1+\eta)}-\frac{\alpha^2 a_\psi/a_\varphi^2(2\eta+1)}{\eta(1+\eta)^2}
;
\\
\frac{d \eta}{dl}
=
\frac{\eta \alpha a_\psi N_f}{2a_\varphi^2} \left(\frac{1}{\eta^2}-1\right)
+
\frac{\eta \alpha a_\psi}{a_\varphi^2(1+\eta)^2} \frac{2(1-\eta)}{3\eta}
\end{gathered}
\end{equation}
where $\eta=a_\psi c_s/(a_\varphi v_F)$. The RG flow is fully described by the four coupled equations for $a_\psi,\;a_\varphi,\;\eta$ and $\beta\equiv \frac{\lambda^2 a_\psi}{4\pi^2 v_F^3 \Lambda^{3-d} a_\varphi^2} = \alpha a_\psi/a_\varphi^2$:
\begin{equation}
\begin{gathered}
\frac{d a_\varphi}{dl} = \frac{\beta N_f a_\varphi}{2};
\\
\frac{d a_\psi}{dl} =\frac{\beta a_\psi}{\eta(1+\eta)^2};
\\
\frac{d \beta}{dl}
=(3-d)\beta-2\frac{\beta^2}{\eta(1+\eta)}-\frac{\beta^2(2\eta+1)}{(1+\eta)^2}
+\frac{\beta^2}{\eta(1+\eta)^2} - \beta^2 N_f
\\
\frac{d \eta}{dl}
=
\frac{\eta \beta N_f}{2} \left(\frac{1}{\eta^2}-1\right)
+
\frac{\eta \beta }{(1+\eta)^2} \frac{2(1-\eta)}{3\eta}.
\end{gathered}
\end{equation}
The fixed point of these equations is at $\eta=1$, from which it follows that at fixed point $\beta = \frac{3-d}{N_f+3/2}$. One gets then $a_\psi\sim e^{\frac{\beta}{4}l}$ and (see equations above) $v_F\sim e^{\frac{\beta}{(1+\eta)^2}l} = e^{\frac{\beta}{4}l}$. For bosons $a_\varphi\sim e^{\frac{\beta N_f}{2}l}$ and consequently $c_s\sim \eta a_\varphi v_F/a_\psi\sim e^{\frac{\beta N_f}{2}l}$. We thus recover that the Non-Fermi liquid here has time and space (momentum and frequency) scaling in the same way at the fixed point - i.e. emergent Lorentz invariance, as is known to be the case in this model \cite{mihaila.2017,lang.2018}.

\begin{figure}[h!]
	\includegraphics[width=\linewidth]{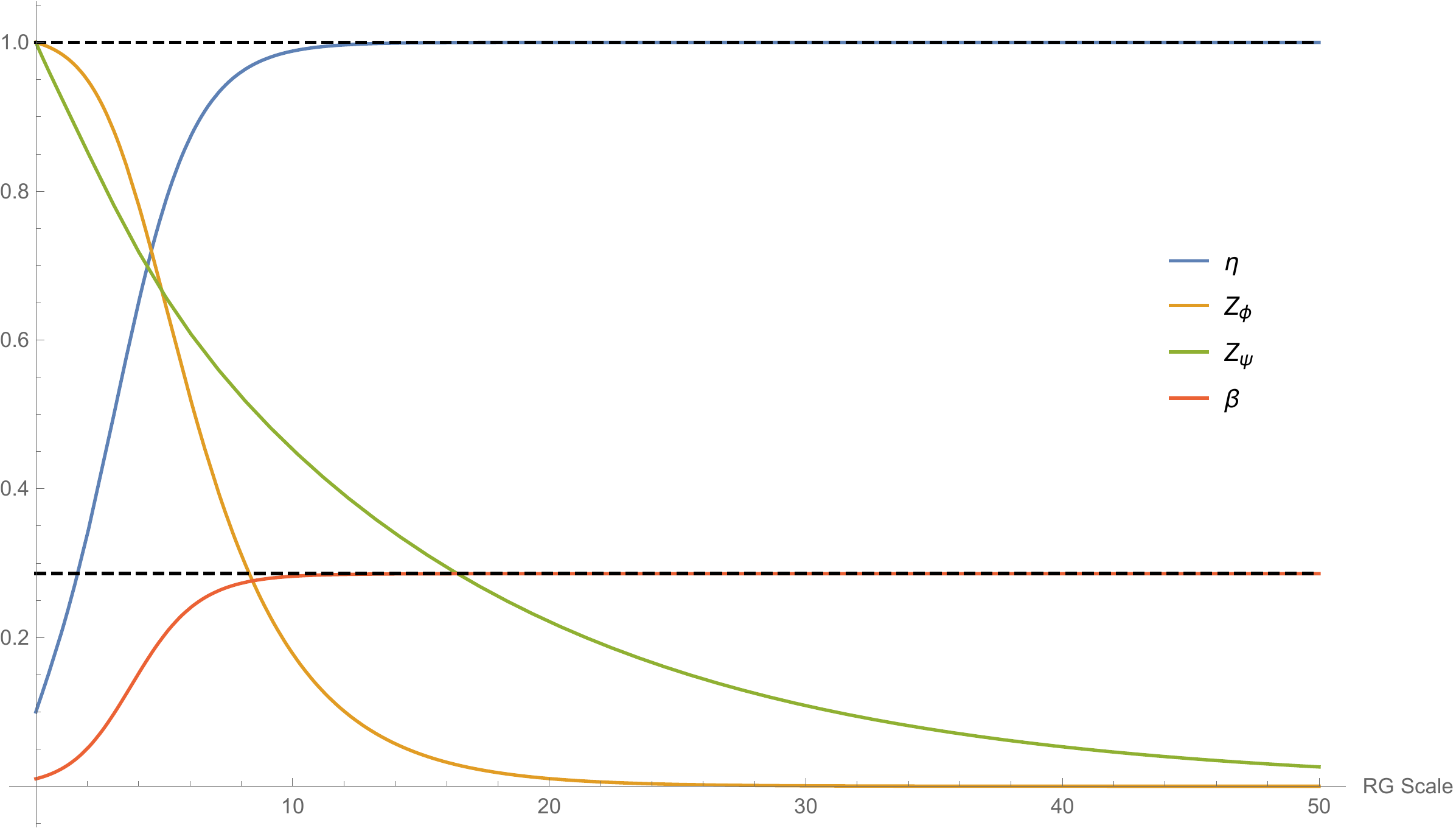}
	\caption{RG flow for $N_f=2;\beta(0)=0.01$;$\eta(0)=0.1,Z_\varphi(0)=Z_\psi(0)=1$; black dashed lines are fixed point values $\beta_0 \approx0.286$ and $\eta_0=1$.}
	\label{fig:RGNPeps}
\end{figure}

\subsection{Weyl Nodal Point in 3D}
Another important case is the Weyl point in 3D. These should be generic to systems with either broken $\mathcal{T}$ or $\mathcal{P}$; as the latter corresponds to the QCP we study we consider a polar QCP in a Weyl system with no TRS. If we neglect possible anisotropies, the Hamiltonian takes the form
\begin{equation}
\begin{gathered}
H_{W} = v_F \vec{k}\cdot\vec{\sigma}
\\
H_{coupl} =
\lambda \sum_{{\bf q},{\bf k}}
\vec{\varphi}_{\bf q}
c^\dagger_{{\bf k}+{\bf q}/2} \vec{\sigma} c_{{\bf k}-{\bf q}/2}.
\end{gathered}
\label{eq:ham:weyl}
\end{equation}
Let us consider now the boson self-energy (note the indices due to vector coupling, diagrams here in what follows look equivalent to those in Fig. \ref{fig:diagr}):
\begin{gather*}
\Pi_{\alpha\alpha}(i\omega,{\bf q}) = -\lambda^2
\int\frac{d \varepsilon d {\bf k}}{(2\pi)^4}
{\rm Tr} \left[\sigma_\alpha G_0(i(\varepsilon_n+\omega_n/2), {\bf k}+{\bf q}/2)
\sigma_\alpha G_0(i(\varepsilon_n-\omega_n/2), {\bf k}-{\bf q}/2)\right]
=
\\
=
2\lambda^2
\int\frac{d \varepsilon d {\bf k}}{(2\pi)^4}
\frac{
	\varepsilon_+\varepsilon_- - v_Fk_{\alpha,+}v_Fk_{\alpha,-}+\sum_{\beta\neq \alpha }v_F k_{\beta,+} v_F k_{\beta,-}
}
{
	(\varepsilon_+^2 +v_F^2 k_+^2)
	(\varepsilon_-^2 + v_F^2 k_-^2)
}
\end{gather*}
Evaluating the integrals even for $T=0$ is quite complicated and the existing results are specifically for real frequencies\cite{Thakur2018,Zhou2018}, so we limit ourselves to calculations on the momentum shell $k\in[\Lambda e^{-dl},\Lambda],\;\omega\in[-\infty,\infty]$. One gets
\[
\delta \Pi_{\alpha \alpha} = - \frac{\lambda^2}{\pi^2v_F^3} \frac{\omega^2+\sum_{\beta\neq \alpha } v_F^2q_\beta^2}{12} dl.
\]
Off-diagonal components are also non-zero:
\begin{gather*}
\Pi_{\alpha\beta}(i\omega,{\bf q}) = -\lambda^2
\int\frac{d \varepsilon d {\bf k}}{(2\pi)^4}
{\rm Tr} \left[\sigma_\alpha G_0(i(\varepsilon_n+\omega_n/2), {\bf k}+{\bf q}/2)
\sigma_\beta G_0(i(\varepsilon_n-\omega_n/2), {\bf k}-{\bf q}/2)\right]
=
\\
=
-\lambda^2
\int\frac{d \varepsilon d {\bf k}}{(2\pi)^4}
\frac{
	2(v_F k_{\alpha,+} v_F k_{\beta,-}+v_F k_{\beta,+} v_F k_{\alpha,-})
	+
	i v_F {\rm Tr} [\sigma_\alpha\sigma_\gamma\sigma_\beta]
	(\omega k_\gamma-\varepsilon q_\gamma)
}
{
	(\varepsilon_+^2 +v_F^2 k_+^2)
	(\varepsilon_-^2 + v_F^2 k_-^2)
},
\end{gather*}
where $\gamma\neq\alpha,\beta$. Evaluating the integrals for the momentum shell we get (note that the second term in the numerator does not contribute a logarithmically divergent term\cite{Zhou2018}):
\[
\delta \Pi_{\alpha \beta} =  \frac{\lambda^2}{\pi^2 v_F^3} \frac{v_F^2q_\alpha q_\beta}{12} dl.
\]
One can combine both contributions in the following simple form:
\begin{equation}
\delta \Pi_{\alpha \beta} =  - \frac{\lambda^2}{\pi^2 v_F^3} \frac{\omega^2 +v_F^2(q^2\delta_{\alpha\beta} - q_\alpha q_\beta)}{12} dl.
\end{equation}
The momentum-dependent part of this self-energy is non-zero for transverse, but not longitudinal modes; hence we are led to necessity of considering the difference between longitudinal and transverse phonon velocities. We consider the following form of the bare bosonic propagator:
\begin{equation}
\begin{gathered}
D^0_{\alpha\beta} = ((\omega^2+c_T^2 q^2)\delta_{\alpha\beta}+ (c_L^2-c_T^2)q_\alpha q_\beta)^{-1}=
\frac{\delta_{\alpha\beta} - q_\alpha q_\beta/q^2}{\omega^2+c_T^2 q^2}
+\frac{q_\alpha q_\beta/q^2}{\omega^2+c_L^2 q^2}\equiv
\\
\equiv
A(\omega,q)\delta_{\alpha\beta}+B(\omega,q)q_\alpha q_\beta,
\end{gathered}
\end{equation}
where $A(\omega,q) = (\omega^2+c_T^2 q^2)^{-1}$ and $q^2 B(\omega,q) = -(c_L^2-c_T^2)(\omega^2+c_T^2 q^2)^{-1}(\omega^2+c_L^2 q^2)^{-1}$. With this we can move to the fermionic self-energy:
\begin{equation}
\begin{gathered}
\Sigma (i\varepsilon, {\bf k}) = \lambda^2 \int \frac{d^3q d\omega}{(2\pi)^{4}} D^0_{\alpha\beta}(-{\bf q},-\omega) \sigma_\alpha G(i(\varepsilon+\omega),{\bf k}+{\bf q})\sigma_\beta =
\\
=
-\lambda^2 \int \frac{d^3q d\omega}{(2\pi)^{4}} D^0_{\alpha\beta}({\bf q},\omega)
\sigma_\alpha
\frac{i(\varepsilon+\omega)+v_F({\bf k}+{\bf q})\vec{\sigma}}
{(\varepsilon+\omega)^2+v_F^2({\bf k}+{\bf q})^2}
\sigma_\beta
\approx
\\
\approx
-\lambda^2 \int \frac{d^3q d\omega}{(2\pi)^{4}}
\frac{i\varepsilon (-\omega^2+v_F^2q^2)}{(\omega^2+v_F^2q^2)^2}
D^0_{\alpha\beta}({\bf q},\omega)
\sigma_\alpha\sigma_\beta
+
\\
+
\left(
\frac{\sigma_\alpha ({\bf k}\cdot \vec{\sigma})\sigma_\beta}{\omega^2+v_F^2q^2}
-
\frac{2 \left(\sum_a q_a k_a\right)\sigma_\alpha ({\bf q}\cdot \vec{\sigma}) \sigma_\beta}
{(\omega^2+v_F^2q^2)^2}
\right)D^0_{\alpha\beta}({\bf q},\omega)=
\\
-\lambda^2 \int \frac{d^3q d\omega}{(2\pi)^{4}}
\frac{i\varepsilon (-\omega^2+v_F^2q^2)}{(\omega^2+v_F^2q^2)^2}
\left(\frac{2}{\omega^2+c_T^2 q^2}+\frac{1}{\omega^2+c_L^2 q^2}\right)
-
\frac{2{\bf k}\cdot \vec{\sigma}}{3}\frac{\omega^2-v_F^2q^2}{(\omega^2+v_F^2q^2)^2(\omega^2+c_T^2 q^2)}
\\
-
\frac{{\bf k}\cdot \vec{\sigma}}{3}
\frac{\omega^2+3v_F^2q^2}{(\omega^2+v_F^2q^2)^2(\omega^2+c_L^2 q^2)},
\end{gathered}
\label{eq:3dwp:selfen}
\end{equation}
resulting in
\begin{equation}
\delta \Sigma = -i\varepsilon \frac{\lambda^2}{4\pi^2}\left(\frac{2}{c_T(c_T+v_F)^2}+\frac{1}{c_L(c_L+v_F)^2}\right)dl
+v_F\vec{k}\cdot\vec{\sigma} \frac{\lambda^2}{4\pi^2}\left(-\frac{2}{3c_T(c_T+v_F)^2}+\frac{3v_F+2c_L}{3c_Lv_F(c_L+v_F)^2}\right)dl.
\end{equation}
Finally, the vertex correction is:
\begin{equation}
\begin{gathered}
\delta \Gamma_i = 
\lambda^3 \int \frac{d^3q d\omega}{(2\pi)^{4}}
D_{\alpha\beta}(-i\omega,-{\bf q})
\sigma_\alpha G(i\omega,{\bf q})\sigma_i G(i\omega,{\bf q})\sigma_\beta=
\\
=
\lambda^3 \int \frac{d^3q d\omega}{(2\pi)^{4}}
\frac{D_{\alpha\beta}(-i\omega,-{\bf q})}{(\omega^2+v_F^2q^2)^2}
\sigma_\alpha (-\omega^2\sigma_i+({\bf q}\cdot \vec{\sigma})\sigma_i ({\bf q}\cdot \vec{\sigma})\sigma_\beta
\\
=
\lambda^3\sigma_i \int \frac{d^3q d\omega}{(2\pi)^{4}}
\frac{2}{3} \frac{\omega^2-v_F^2q^2}{(\omega^2+c_T^2 q^2)(\omega^2+v_F^2q^2)^2}
+
\frac{\omega^2+3v_F^2q^2}{3(\omega^2+c_L^2 q^2)(\omega^2+v_F^2q^2)^2}
=
\\
=
\sigma_i \frac{\lambda^3}{6\pi^2}
\left(-\frac{1}{c_T(c_T+v_F)^2}+\frac{3v_F+2c_L}{2c_Lv_F(c_L+v_F)^2}\right)dl
\end{gathered}
\end{equation}
Analogously to the 2D Dirac case we proceed
\begin{equation}
\begin{gathered}
\frac{d a_\varphi^2}{dl} = N_f \frac{\lambda^2}{12 \pi^2 v_F^3} a_\psi
\\
\frac{d c_T^2}{dl} = N_f \frac{\lambda^2}{12 \pi^2 v_F a_\psi}
\\
\frac{d c_L^2}{dl} = 0
\\
\frac{d a_\psi}{dl} = \frac{\lambda^2}{4 \pi^2 a_\psi a_\varphi^2}
\left(
\frac{2}{(c_T/a_\varphi) (c_T/a_\varphi+v_F/a_\psi)^2}
+\frac{1}{(c_L/a_\varphi) (c_L/a_\varphi+v_F/a_\psi)^2}
\right)
\\
\frac{d v_F}{dl} = \frac{\lambda^2 v_F/a_\psi}{4 \pi^2 a_\psi a_\varphi^2}
\left(
-\frac{2}{3(c_T/a_\varphi) (c_T/a_\varphi+v_F/a_\psi)^2}
+\frac{3(v_F/a_\psi)+2(c_L/a_\varphi)}{3(c_L/a_\varphi)(v_F/a_\psi) (c_L/a_\varphi+v_F/a_\psi)^2}
\right)
\\
\frac{d\lambda}{dl} = \frac{\lambda^3 }{6 \pi^2 a_\psi^2 a_\varphi^2}
\left(-
\frac{1}{(c_T/a_\varphi) (c_T/a_\varphi+v_F/a_\psi)^2}
+
\frac{3(v_F/a_\psi)+2(c_L/a_\varphi)}{2(c_L/a_\varphi)(v_F/a_\psi) (c_L/a_\varphi+v_F/a_\psi)^2}
\right)
\end{gathered}
\end{equation}

Defining the coupling constant $\alpha= \frac{\lambda^2 a_\psi}{12 \pi^2 a_\varphi^2 v_F^3}$ we get:
\begin{equation}
\begin{gathered}
\frac{d a_\varphi}{dl} =\frac{N_f \alpha a_\varphi}{2};
\\
\frac{d a_\psi}{dl} = 3\alpha a_\psi
\left(
\frac{2}{\eta_T (1+\eta_T)^2}
+\frac{1}{\eta_L(1+\eta_L)^2}
\right);
\\
\frac{d \alpha}{dl} =
2\alpha \frac{1}{\lambda}\frac{d \lambda}{d l} - 3\alpha \frac{1}{v_F}\frac{d v_F}{d l}
+\alpha \frac{1}{a_\psi}\frac{d a_\psi}{d l}
-2\alpha \frac{1}{a_\varphi}\frac{d a_\varphi}{d l}
=
\\
=\alpha^2\left(
-N_f+\frac{6(1+1)-2}{\eta_T(1+\eta_T)^2}
+\frac{3-3(3+2\eta_L)+3+2\eta_L}{\eta_L(1+\eta_L)^2}
\right)=
\\
=\alpha^2\left(
-N_f+\frac{10}{\eta_T(1+\eta_T)^2}
-\frac{3+4\eta_L}{\eta_L(1+\eta_L)^2}
\right);
\\
\frac{d \eta_T}{dl} = \eta_T \left(
\frac{1}{c_T}\frac{d c_T}{d l}+\frac{1}{a_\psi}\frac{d a_\psi}{d l}-\frac{1}{v_F}\frac{d v_F}{d l}
-\frac{1}{a_\varphi}\frac{d a_\varphi}{d l}
\right)=
\\
=\frac{\eta_T N_f \alpha}{2} \left(\frac{1}{\eta_T^2} -1\right)
+
\alpha \eta_T
\left(
\frac{8}{\eta_T(1+\eta_T)^2}
-
\frac{2}{(1+\eta_L)^2}
\right);
\\
\frac{d \eta_L}{dl} = \eta_L \left(
\frac{1}{c_L}\frac{d c_L}{d l}+\frac{1}{a_\psi}\frac{d a_\psi}{d l}-\frac{1}{v_F}\frac{d v_F}{d l}
-\frac{1}{a_\varphi}\frac{d a_\varphi}{d l}
\right)=
\\
=-\frac{\eta_L N_f \alpha}{2}
+
\alpha \eta_L
\left(
\frac{8}{\eta_T(1+\eta_T)^2}
-
\frac{2}{(1+\eta_L)^2}
\right);
\end{gathered}
\end{equation}

	\begin{figure}
	\centering
	\begin{minipage}{0.4\linewidth}
		\centering
		\includegraphics[width=\linewidth]{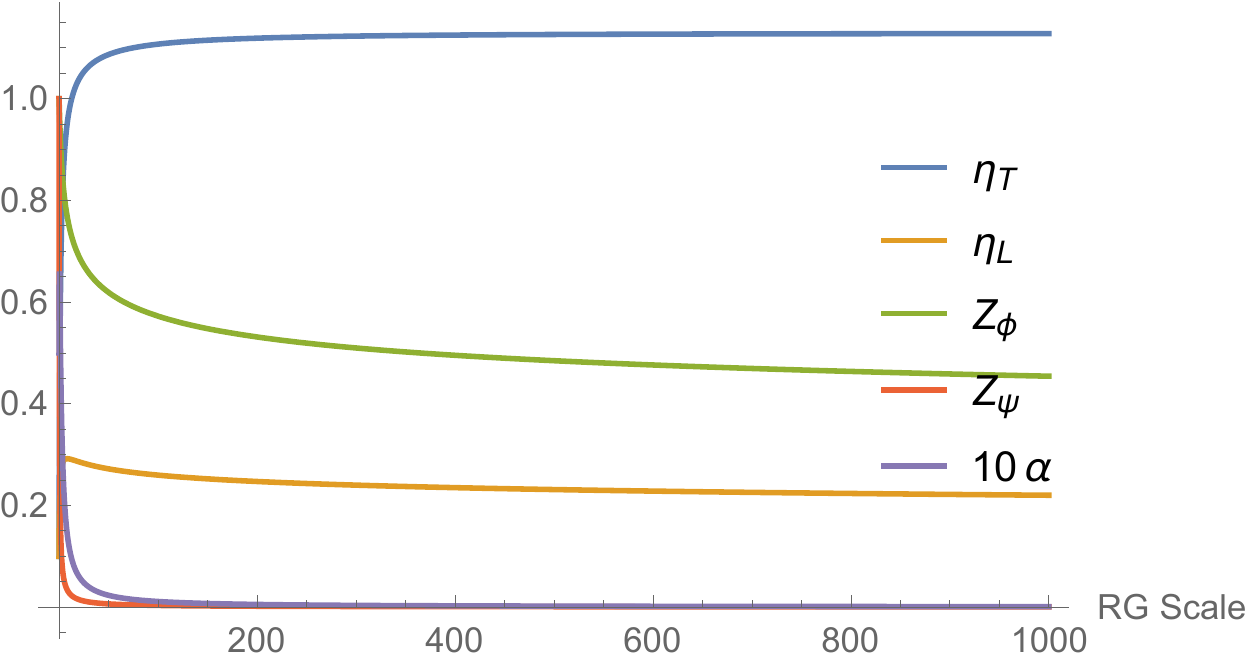}
		\caption{RG flow for $N_f=2;\beta(0)=0.05$;$\eta_T(0)=0.1$; $ \eta_L(0)=0.2;Z_\varphi(0)=Z_\psi(0)=1$;}
	\end{minipage}
	\begin{minipage}{0.2\linewidth}
	\end{minipage}%
	\begin{minipage}{0.4\linewidth}
		\centering
		\includegraphics[width=\linewidth]{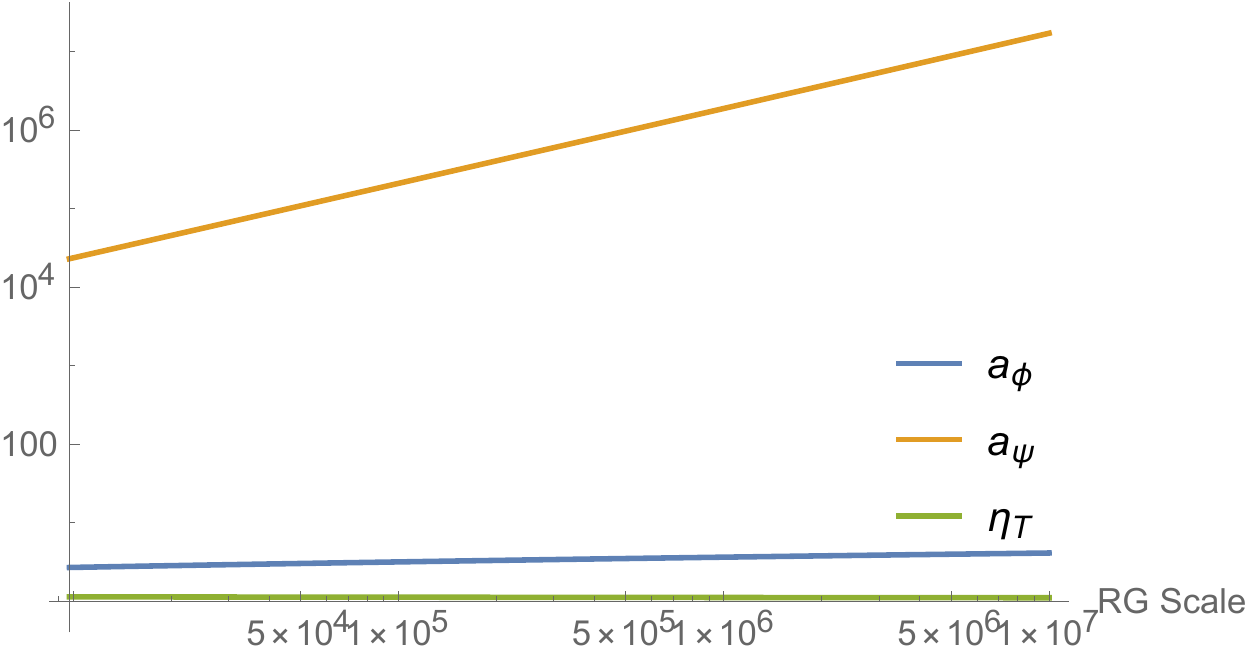}
		\caption{Details of RG flow for large RG scales for $\eta_T, a_{\psi,\varphi} = Z_{\psi,\varphi}^{-1}$}
		\label{fig:sub2}
	\end{minipage}
	\label{fig:RGWeyl}
\end{figure}

The numerical solution of RG equations show that $Z_\psi\equiv 1/a_\psi$ and $\alpha$ flow to zero first; moreover at larger scales that $\eta_L$ exhibits slow decrease. This allows one to extract the large-$l$ asymptotic analytically in the limit $\eta_L\to0\;a_\psi\gg1$. In this limit, $\eta_T$ reaches a fixed-point value of $1$. The remaining equations are simplified as follows:
\begin{equation}
\begin{gathered}
\frac{d a_\varphi}{dl} =\frac{N_f \alpha a_\varphi}{2};
\;
\frac{d a_\psi}{dl} = 3\alpha\frac{a_\psi}{\eta_L};
\\
\frac{d \alpha}{dl} = - \frac{3\alpha^2}{\eta_L};
\\
\frac{d \eta_L}{dl} =
-\frac{\eta_L N_f \alpha}{2};
\end{gathered}
\end{equation}
One can combine the above equations to obtain an equation for $\beta\equiv \frac{\alpha}{\eta_L}$:
\[
\frac{d \beta}{dl} = -3\beta^2+\frac{N_f}{2}\beta^2\eta_L\approx-3\beta^2,
\]
resulting in the large-$l$ asymptotic $\beta \to 1/(3 l)$. Using this, we obtain:
\begin{equation}
\begin{gathered}
\eta_L\to \frac{6}{N_f\log{l}};
\;
\beta\approx \frac{1+(\log l)^{-1}+O((\log l)^{-2})}{3l};
\;
\alpha\to \frac{2}{N_f l \log l};
\;
a_\psi\to {\rm const}\cdot l;
\\
a_\varphi\to{\rm const} \cdot \log l;
\;
\eta_T =1.
\end{gathered}
\end{equation}
Thus, we find that $\eta_L/\eta_T=c_L/c_T\to0$ indicating that the transverse phonon velocity hardens, while the fermionic quasiparticle residue vanishes as $1/l$ taking into account that $l=\log(k^{-1},\omega^{-1})$ that means that there are logarithmic non-Fermi liquid corrections. On the other hand, the bosonic quasiparticle residue vanishes only as $1/\log{l}$, i.e. the bosons receive only loglog corrections and are relatively well-defined at the QCP.

Let us also evaluate the asymptotic behavior of the other observables:
\begin{equation}
\begin{gathered}
\frac{d c_T^2}{dl}  = N_f \alpha c_T^2/\eta_T^2 \to c_T^2 = const (\log l)^2,
\\
\frac{d v_F}{dl} \approx 3 \alpha v_F /\eta_L = v_F/l \to v_F= const \cdot l.
\end{gathered}
\end{equation}
The asymptotic form of the propagators is thus
\begin{equation}
\begin{gathered}
D_T(i\omega, {\bf q}) \sim \frac{1}{a(\omega \log \omega^{-1})^2+b(q \log q^{-1})^2};
\\
D_L(i\omega, {\bf q}) \sim \frac{1}{(a\omega \log \omega^{-1})^2+c_L^2 q^2};
\\
G(i\varepsilon, {\bf k}) \sim \frac{1}{i a\varepsilon\log\varepsilon^{-1}+b\log k^{-1} {\bf k}\cdot\vec{\sigma}} .
\end{gathered}
\end{equation}

\section{Nodal Surface Systems}
Results for the case of a nodal surface are actually easily obtained from above by taking $\gamma=0$ in Eq. (2) of the main text and generalizing $\frac{k_x^2+k_y^2-k_F^2}{2m}\to\varepsilon({\bf k})$ and then $\varepsilon({\bf k}+{\bf q}/2) \approx v_F(\theta,\varphi)|{\bf k}-{\bf k}_F(\theta,\varphi)|+ {\bf v}_F(\theta,\varphi){\bf q}/2$. After similar steps we get
\begin{gather*}
\Pi(i\omega_n,{\bf q})
=
-2\int \frac{d\varphi \sin \theta d\theta}{v_F(\theta,\varphi)} g(\sqrt{\omega_n^2+({\bf v}_F(\theta,\varphi){\bf q})^2})
\\
g(\omega_n) =
\int_{-\infty}^\infty\frac{ d \xi }{(2\pi)^3}
\frac{\xi \tanh \frac{\xi+\mu}{2T}}{(4\xi^2+\omega_n^2)}
+
\frac{\xi\tanh \frac{\xi-\mu}{2T}}{(4\xi^2+\omega_n^2)}.
\end{gather*}
Note that there is an infrared log-divergence for $\mu=0,\;T=0$ that actually signifies that a polar instability will occur for arbitrary weak coupling with the conduction electrons and boson mass as is expected from a situation with perfect nesting. Thus a QCP may only occur for finite $\mu$, driven by the initial boson mass. At this QCP one has
\[
g(\omega_n) =
\int_{\mu}^\infty\frac{ d \xi }{\pi^2} \frac{\xi}{(4\xi^2+\omega_n^2)}=
\frac{\log \frac{\Lambda}{\mu^2+\omega_n^2}}{8\pi^2},
\]
where $\Lambda$ is the UV cutoff. One can see that the dynamical critical exponent here remains the same and thus one expects only log-corrections due to quartic interboson interactions.

\section{Effect of Coulomb Interactions}

Let us now consider the effects of Coulomb interactions:
\begin{equation}
\begin{gathered}
\int d^d r d \tau \sum_{\alpha,\beta} \frac{e^2}{|r-r'|}
\left[\Psi^\dagger_\alpha({\bf r},\tau)\Psi_\alpha({\bf r},\tau) + Q_0 \nabla \cdot \vec{\varphi}({\bf r},\tau) \right]
\cdot
\\
\cdot
\left[\Psi^\dagger_\beta({\bf r}',\tau)\Psi_\beta({\bf r}',\tau)+Q_0 \nabla \cdot \vec{\varphi}({\bf r}',\tau)\right]=
\\
=
\int d^d q d \omega V_{Coul}({\bf q})
\left[
\left(\sum_{\alpha} d^d k d \varepsilon \Psi^\dagger_\alpha({\bf k}+{\bf q},\varepsilon+\omega)\Psi_\alpha({\bf k},\varepsilon)\right)
+ iQ_0 {\bf q} \cdot \vec{\varphi}({\bf q},\omega) \right]
\cdot
\\
\cdot
\left[
d^d k' d \varepsilon'
\left(\sum_{\beta} \Psi^\dagger_\beta({\bf k}'-{\bf q},\varepsilon'-\omega)\Psi_\beta({\bf k}',\varepsilon')\right)
-iQ_0 {\bf q} \cdot \vec{\varphi}(-{\bf q},-\omega)\right]
\end{gathered}
\label{eq:Coul}
\end{equation}
where $V_{Coul}({\bf q})$ is the Fourier transform of $1/r$ in $d$ dimensions, $\alpha$ are the internal fermionic degrees of freedom (i.e., spin and band) and $Q_0$ is proportional to the effective dipole moment of the optical phonon displacement. To perform preliminary scaling analysis we rewrite this term using an auxiliary field $\Phi({\bf r},{\tau})$:
\begin{equation}
\begin{gathered}
S_{Coul} =
\int d^d q d \omega  \Phi({\bf q},\omega)V_{Coul}^{-1}({\bf q}) \Phi(-{\bf q},-\omega)+
\\
+
i e\int d^d q d \omega d^d k d \varepsilon
\Phi({\bf q},\omega)
\Psi^\dagger_\alpha({\bf k}+{\bf q},\varepsilon+\omega)\Psi_\alpha({\bf k},\varepsilon)
+
\\
+
Q_0 \int d^d q d \omega
\left[{\bf q} \cdot \vec{\varphi}({\bf q},\omega)\right]
\Phi(-{\bf q},-\omega),
\end{gathered}
\label{eq:CoulLagr}
\end{equation}
where integrating over $\Phi({\bf q},\omega)$ results in the usual form of the Coulomb interaction.

\subsection{Nodal line in 3D}
The way to understand this case is to consider first separately the fermion-Coulomb and fermion-polar boson problems independently (as if there were no Coulomb coupling between bosons and fermions and no effect of Coulomb on bosons) and then include the remaining effects.

As is shown above, the RPA-like fixed point is stable for the fermion-boson and the fermion-Coulomb \cite{huh.2016} problems; thus we can start by replacing the bare bosonic propagator $D_0\to(-\Pi_{\sigma_3}(i\omega,{\bf q}))^{-1}$ and the Coulomb interaction $V_{Coul}\to (-\Pi(i\omega,{\bf q}))^{-1}$, where both polarization operators are linear in $q$ at low $q$.

We can now find the scaling dimensions of two remaining terms that emerge due to Coulomb interaction. The fermion-boson interaction is
\[
\int d^d q d \omega d^d k d \varepsilon \sum_{\alpha,\beta} V_{Coul}({\bf q})
i Q_0 {\bf q} \cdot \vec{\varphi}({\bf q},\omega)
\Psi^\dagger_\alpha({\bf k}+{\bf q},\varepsilon+\omega)\Psi_\alpha({\bf k},\varepsilon),
\]
As $[V_{Coul}]=-1$ the scaling dimension of this term is same as Yukawa coupling and it's irrelevant. The boson-boson term is \[
\int d^d q d \omega V_{Coul}({\bf q})
Q_0^2 [{\bf q} \cdot \vec{\varphi}({\bf q},\omega)][{\bf q} \cdot \vec{\varphi}(-{\bf q},-\omega)],
\]
has the scaling dimension $-[d+1-1+2(1-(d+2)/2)] = 0$ and is marginal. However, its effect can be easily included as it is only a quadratic term:
\[
D_{Coul}(i\omega,{\bf q}) = (-\Pi_{\sigma_3}(i\omega,{\bf q}) + V_{Coul}({\bf q}) Q_0^2 q^2 \cos^2 \eta),
\]
where $\eta$ is the angle between ${\bf q}$ and the polarization of $\varphi$.

\subsection{Nodal Points}
The first observation one can make is that the coupling $Q_0$ in \eqref{eq:CoulLagr} has the scaling dimension $-[d+2 -d-(d+3)/2] = (d-1)/2$ ($[\Phi] = -(d+1+[d-1])/2= -d$, $[\varphi] = - (d+3)/2$) and is relevant for $d>1$. Furthermore, it receives no perturbative corrections and, consequently, always flows to strong coupling. On the other hand, the coupling to fermions is always marginal $[e] = -(2(d+1) - (d+2) - d)=0$. Thus, we take the boson-Coulomb coupling into account first, which can be done exactly. Note that this is in principle equivalent to solving for the eigenmodes of the Coulomb-polar mode system in each step of RG: since there's only a quadratic coupling between them, fields with momenta belonging to different shells do not mix. It results in renormalized bosonic propagator and Coulomb interaction:

\begin{gather*}
V_{Coul}(i\omega, {\bf q}) = \frac{1}{A_d q^{d-1}+Q_0^2 \sum_{ab} q_a q_b D_0^{ab} (i\omega, {\bf q})},
\\
D^{ab}(i\omega, {\bf q}) = \frac{1}{[D_0^{ab} (i\omega, {\bf q})]^{-1}+\frac{Q_0^2}{A_d} q_a q_b q^{1-d}},
\end{gather*}
where $A_d = 1/(4\pi)$ in 3D and $1/(2\pi)$ in 2D, respectively. 

\subsubsection{Weyl point 3D}
In the case of 3D Weyl point we take the isotropic system $D_0^{ab} (i\omega, {\bf q}) = \delta_{ab}(\omega^2+c^2 q^2)^{-1}$ to get the Coulomb interaction to be
\[
V_{Coul}(i\omega, {\bf q}) = \frac{4\pi (\omega^2+c^2 q^2)}{q^2(\omega^2+c^2 q^2)+4\pi Q_0^2 q^2}.
\]
Clearly, such a renormalization changes the scaling dimension of the field $\Phi$. Under the assumption of $z=1$ we get $[\Phi] = -(d+1)/2$ and $[e] = (1-d)/2$, irrelevant for $d>1$. Thus, the polar mode at the QCP screens the Coulomb interaction with fermions out and we can neglect it in the RG analysis. On the other hand, the bosonic propagator is now:
\[
D^{ab}(i\omega, {\bf q}) = \frac{1}{(\omega^2+c^2 q^2)\delta_{ab}+4\pi Q_0^2 q_a q_b /q^2} 
=
\frac{\delta_{\alpha\beta} - q_\alpha q_\beta/q^2}{\omega^2+c^2 q^2}
+\frac{q_\alpha q_\beta/q^2}{\omega^2+c^2q^2+4\pi Q_0^2},
\]
i.e. the longitudinal mode is gapped and absent from the scaling limit. Thus, the RG equations for this case can be obtained by simply taking the limit $c_L\to \infty$ :
\begin{equation}
\begin{gathered}
\frac{d a_\varphi}{dl} =\frac{N_f \alpha a_\varphi}{2};
\\
\frac{d a_\psi}{dl} = 
\frac{6\alpha a_\psi}{\eta_T (1+\eta_T)^2};
\\
\frac{d \alpha}{dl} 
=\alpha^2\left(
-N_f+\frac{10}{\eta_T(1+\eta_T)^2}
\right);
\\
\frac{d \eta_T}{dl} 
=\frac{\eta_T N_f \alpha}{2} \left(\frac{1}{\eta_T^2} -1\right)
+
\frac{8\alpha}{(1+\eta_T)^2}.
\end{gathered}
\end{equation}
In this case $\eta_T$ still exhibits a stable fixed point at $\eta_T=\eta_T^0$, however it is now not equal to $1$:
\[
\frac{\eta_T^0 N_f}{2} \left(\frac{1}{(\eta_T^0)^2} -1\right)
+
\frac{8}{(1+\eta_T^0)^2}=0.
\]
It is important that: a) $\eta_T^0>1$; b) for $N_f= 2$ numerical solution is $\eta_T^0\approx1.69$. Using this estimates one can show that the r.h.s. of the equation for $\alpha$ is always negative at this fixed point value and thus the system once again flows to weak coupling. The asymptotic behavior of the RG equations solution is given by:
\begin{gather*}
\alpha\approx 1/(\kappa_0 l);
\;
a_\varphi\to l^{N_f/(2\kappa_0)};
\;
a_\psi\to l^{6/(\eta_T^0(1+\eta_T^0)^2\kappa_0)};
\\
\eta_T^0\approx|_{N_f\to \infty} =1+ \frac{2}{N_f};
\;
\kappa_0 = -\left(
-N_f+\frac{10}{\eta_T^0(1+\eta_T^0)^2}
\right)
\approx|_{N_f\to \infty} =N_f,
\end{gather*}
where one notes that the bosonic quasiparticle residue now also vanishes as a power-law in $l$, i.e. as a power-law in logarithm of the frequency/momentum. Also
\[
\frac{d c_T^2}{dl} =\frac{N_f \alpha c_T^2}{\eta_T^2} \to c_T^2 = const\cdot l^{N_f/(\kappa_0(\eta_T^0)^2)}
;\;
\frac{d v_F}{dl} =-\frac{2 \alpha v_F}{\eta_T^0(1+\eta_T^0)^2} \to v_F = const \cdot l^{-2/(\eta_T^0(1+\eta_T^0)^2\kappa_0)}
\]

The form of the critical propagators is:

\begin{equation}
\begin{gathered}
D_L(i\omega, {\bf q}) \sim \frac{1}{a \omega^2(\log \omega^{-1})^{N_f/\kappa_0}+(c_L q)^2};
\\
D_T(i\omega, {\bf q}) \sim \frac{1}{a \omega^2(\log \omega^{-1})^{N_f/\kappa_0}+bq^2 (\log q^{-1})^{N_f/(\kappa_0(\eta_T^0)^2)}};
\\
G(i\varepsilon, {\bf k}) \sim \frac{1}{i a\varepsilon(\log\varepsilon^{-1})^{3N_f/(8\kappa_0)(1-(\eta_T^0)^{-2})}+b(\log k^{-1})^{-N_f/(8\kappa_0)(1-(\eta_T^0)^{-2})} {\bf k}\cdot\vec{\sigma}} .
\end{gathered}
\end{equation}

\subsubsection{Dirac point 2D}
For the 2D Dirac case we've found that Yukawa coupling occurs for a single component of the polar order only. Thus, we get the Coulomb interaction to be screened by this mode as:
\[
V_{Coul}(i\omega, {\bf q}) = \frac{2\pi (\omega^2+c^2 q^2)}{q(\omega^2+c^2 q^2)+2\pi Q_0^2 q_x^2}.
\]
This renormalization changes the scaling dimension of the field $\Phi$. However, if $q_x=0$ one seems to recover the bare Coulomb interaction. Thus, we are forced to consider the anisotropy in an essential way. Namely, the angle with the $x$-axis should be also considered as an additional coordinate. Namely, $q_x = q \cos \theta\approx q \theta$ for $\theta\approx \pi/2$. Assuming $z=1$ to find the bare scaling dimension we find that $[\theta] = [q,\omega]/2$. The scaling dimesnion of the momentum/frequency integral thus becomes $d+1+1/2$.

The bosonic propagator is now:
\[
D(i\omega, {\bf q}) = \frac{1}{(\omega^2+c^2 q^2)+2\pi Q_0^2 q_x^2/q} 
\approx
\frac{1}{(\omega^2+c^2 q^2)+2\pi Q_0^2 q \theta^2} ,
\]

We are now in position to determine the bare scaling dimensions of the Coulomb coupling. We get $[\Phi] = -(2d+0.5)/2,\; [\varphi] = -(d+3.5)/2,\; [\Psi] = -(d+2.5)/2$. We get $[e] = -1/4$, making it irrelevant. On the other hand, $[\lambda] = (2.5-d)/2$, relevant for 2D. We now derive the RG equations in the 2D cylindrical scheme for the case. We will also perform the angular integration approximately around $\theta=\pi/2$.

Also note that one has to allow for anisotropy when calculating the diagrams. The fermionic self-energy is:
\begin{gather*}
\Sigma (i\varepsilon, {\bf k}) = \lambda^2 \int \frac{d^2q d\omega}{(2\pi)^{3}} D({\bf q},\omega) \sigma_3 G(i\varepsilon)\sigma_3 =
\\
\approx
-\lambda^2\int \frac{q dq d \theta }{(2\pi)^2}
\int \frac{d\omega}{2\pi}
\frac{1}{\omega^2+c^2 q^2+2\pi Q_0^2 q \cos^2\theta}
\left(
\frac{i\varepsilon (-\omega^2+v_{Fx}^2q_x^2+v_{Fy}^2q_y^2)}
{(\omega^2+v_{Fx}^2q_x^2+v_{Fy}^2q_y^2)^2}
+
\right.
\\
\left.
+
\frac{-(v_{Fx} k_x \sigma_x+v_{Fy} k_y \sigma_y)(\omega^2+v_{Fx}^2q_x^2+v_{Fy}^2q_y^2) 
	+ 
	2(v_{Fx} q_x \sigma_x+v_{Fy} q_y \sigma_x)
	(v_{Fx}^2 k_x q_x + v_{Fy}^2 k_y q_y)}
{(\omega^2+v_{Fx}^2q_x^2+v_{Fy}^2q_y^2)^2}
\right).
\end{gather*}
The angular integral can be evaluated approximately for $q, \omega \to 0$. The strongest singularity is then contained in the bosonic propagator for $\theta\approx \pi/2$; moreover, as in the integral the relevant range is $\cos\theta \sim q$ we can neglect $q_x$ everywhere except for the bosonic propagator. The result is:

\begin{equation}
\begin{gathered}
\delta\Sigma (i\varepsilon, {\bf k})
\approx
-
\frac{\lambda^2}{2 \pi^2 v_{Fy}^2\sqrt{2 \pi Q_0^2}}
\frac{dl}{\sqrt{\Lambda}}
\left[
i\varepsilon F_1(c/v_{Fy})
-
v_{Fx} k_x \sigma_x F_2 (c/v_{Fy})
+
v_{Fy} k_y \sigma_y F_1 (c/v_{Fy})
\right],
\\
F_1(x) = \frac{x^2-1+\sqrt{1-x^2} \log \left[\frac{1+\sqrt{1-x^2}}{x}\right]}{(1-x^2)^2},
\\
F_2(x) = \frac{\log \left[\frac{1+\sqrt{1-x^2}}{x}\right] }{\sqrt{1-x^2}}.
\end{gathered}
\label{eq:2dnpCoul:selfen}
\end{equation}
The vertex correction is evaluated analogously:
\begin{equation}
\begin{gathered}
\delta\Gamma =
-\lambda^3 \sigma_3 \int \frac{d^2q d\omega}{(2\pi)^{3}}
D(i\omega,{\bf q})
\frac{1}{\omega^2+v_{Fx}^2q_x^2+v_{Fy}^2q_y^2}
\approx
\\
\approx
-
\frac{\lambda^3}{2 \pi^2 v_{Fy}^2\sqrt{2 \pi Q_0^2}}
\frac{dl}{\sqrt{\Lambda}}
F_2 (c/v_{Fy})
\end{gathered}
\label{eq:2dnpCoul:vert}
\end{equation}
Finally, let us evaluate the correction due to the polarization operator (boson self-energy):
\begin{equation}
\begin{gathered}
\delta\Pi(i\omega,{\bf q}) = 
2\lambda^2
T\sum_{\varepsilon_n}
\int\frac{d k_x d k_y}{(2\pi)^2}
\frac{
	(\varepsilon+\omega)\varepsilon +v_{Fx}^2 (k_x+q_x)k_x+v_{Fy}^2 (k_y+q_y)k_y
}
{
	(\varepsilon+\omega)^2 +v_{Fx}^2 (k_x+q_x)^2+v_{Fy}^2 (k_y+q_y)^2)^2
	(\varepsilon^2 + v_{Fx}^2 k_x^2+v_{Fy}^2 k_y^2)^2
}
\\
=
-
\frac{\lambda^2}{2 \pi^2 v_{Fy}^3}
\frac{d l}{\Lambda}
\left\{
F_\omega(v_{Fx}/v_{Fy})
\omega^2
+
v_{Fy}^2 F_{qx}(v_{Fx}/v_{Fy})q_x^2
+
v_{Fy}^2 F_{qy}(v_{Fy}/v_{Fx})q_y^2
\right\},
\\
F_\omega(x) = \frac{E[1 - x^2]}{2 x^2},
\\
F_{qx}(x) = x \frac{x^2E[1-1/x^2]+x E[1-x^2]-xK[1-x^2]-K[1 - 1/x^2]}{4 (1- x^2)},
\\
F_{qy}(x) = \frac{E[1-x^2]+x E[1-1/x^2]-xK[1-1/x^2]-x^2K[1 - x^2]}{4 x^2 (x^2- 1)},
\end{gathered}
\label{eq:2dnpCoul:pol}
\end{equation}

Evaluating the same diagrams with renormalized propagators and introducing $N_f$ fermion flavors we get:
\begin{gather*}
\delta\Pi = -
\frac{N_f \lambda^2/ a_\psi^2}{2 \pi^2 (v_{Fy}/a_\psi)^3}
\frac{d l}{\Lambda}
\left\{
F_\omega(v_{Fx}/v_{Fy})
\omega^2
+
(v_{Fy}/a_\psi)^2 F_{qx}(v_{Fx}/v_{Fy})q_x^2
+
(v_{Fy}/a_\psi)^2 F_{qy}(v_{Fy}/v_{Fx})q_y^2
\right\}
\\
\delta\Sigma = -
\frac{\lambda^2/a_\psi/a_\varphi^2}
{2 \pi^2 (v_{Fy}/a_\psi)^2\sqrt{2 \pi Q_0^2/a_\varphi^2}}
\frac{dl}{\sqrt{\Lambda}}
\cdot
\\
\cdot
\left[
i\varepsilon F_1(c_y a_\psi/(a_\varphi v_{Fy}))
-
(v_{Fx}/a_\psi) k_x \sigma_x F_2 (c_y a_\psi/(a_\varphi v_{Fy}))
+
(v_{Fy}/a_\psi) k_y \sigma_y F_1 (c_y a_\psi/(a_\varphi v_{Fy}))
\right]
\\
\delta\Gamma = -
\sigma_3
\frac{\lambda^3/(a_\psi^2 a_\varphi^2)}{2 \pi^2 (v_{Fy}/a_\psi)^2\sqrt{2 \pi Q_0^2/a_\varphi^2}}
\frac{dl}{\sqrt{\Lambda}}
F_2 (c_y a_\psi/(a_\varphi v_{Fy}))
\end{gather*}

Defining the coupling constants $\beta_\varphi = \frac{\lambda^2 a_\psi}{2\pi^2v_{Fy}^3 a_\varphi^2 \Lambda},\; 
\beta_\psi = \frac{\lambda^2 }{2\pi^2v_{Fy}^2 a_\varphi \sqrt{2 \pi Q_0^2 \Lambda}}$ we get:

\begin{gather*}
\frac{d a_\varphi}{d l} = \frac{N_f \beta_\varphi a_\varphi}{2} F_\omega(\eta_f);
\;
\frac{d c_x^2}{d l} = 
N_f \beta_\varphi (v_{Fy}/a_\psi)^2 a_\varphi^2 F_{qx}(\eta_f);
\;
\frac{d c_y^2}{d l} = 
N_f \beta_\varphi (v_{Fy}/a_\psi)^2 a_\varphi^2 F_{qy}(\eta_f);
\\
\frac{d a_\psi}{d l} = \beta_\psi a_\psi F_1(\eta);
\;
\frac{d v_{Fy}}{d l} = -v_{Fy} \beta_\psi F_1(\eta);
\;
\frac{d v_{Fx}}{d l} = v_{Fx} \beta_\psi F_2(\eta);
\\
\frac{d \lambda}{d l} =-\lambda \beta_\psi
F_2 (\eta);
\\
\frac{d \beta_\varphi}{d l} =
\beta_\varphi
-
2\frac{\beta_\varphi}{a_\varphi}\frac{d a_\varphi}{d l}
+
2 \frac{\beta_\varphi}{\lambda}\frac{d \lambda}{d l}
+
\frac{\beta_\varphi}{a_\psi}\frac{d a_\psi}{d l}
-
3 \frac{\beta_\varphi}{v_{Fy}}\frac{d v_{Fy}}{d l}
=
\\
=\beta_\varphi
-
N_f \beta_\varphi^2 F_\omega(\eta_f)
-
2\beta_\psi\beta_\varphi F_2(\eta)
+
4\beta_\psi\beta_\varphi F_1(\eta);
\\
\frac{d \beta_\psi}{d l} =
\frac{\beta_\psi}{2}
-
\frac{\beta_\psi}{a_\varphi}\frac{d a_\varphi}{d l}
+
2 \frac{\beta_\psi}{\lambda}\frac{d \lambda}{d l}
-
2 \frac{\beta_\psi}{v_{Fy}}\frac{d v_{Fy}}{d l} =
\\
=
\frac{\beta_\psi}{2}
-
\frac{N_f \beta_\varphi\beta_\psi F_\omega(\eta_f)}{2}
-
2\beta_\psi^2 F_2(\eta)
+
2\beta_\psi^2 F_1(\eta);
\\
\frac{d \eta_f}{d l}
=
\eta_f\left\{\frac{1}{v_{Fx}}\frac{d v_{Fx}}{dl} - \frac{1}{v_{Fy}}\frac{d v_{Fy}}{dl}\right\}
=
\eta_f \beta_\psi [F_1(\eta)+F_2(\eta)];
\\
\frac{d \eta}{d l}
=
\eta\left\{
\frac{1}{a_\psi}\frac{d a_\psi}{dl}
+\frac{1}{c_y}\frac{d c_y}{dl}
- \frac{1}{v_{Fy}}\frac{d v_{Fy}}{dl}
- \frac{1}{a_\varphi}\frac{d a_\varphi}{dl}
\right\}=
\\
=
2\eta \beta_\psi F_1(\eta)
+\frac{\eta N_f\beta_\varphi}{2}
[
F_{qy}(\eta_f)/\eta^2
-
F_\omega(\eta_f)
].
\end{gather*}
where $\eta_f = v_{Fx}/v_{Fy}$ and $\eta = a_\psi c_y/(a_\varphi v_{Fy})$.
\begin{figure}[h!]
	\includegraphics[width=\linewidth]{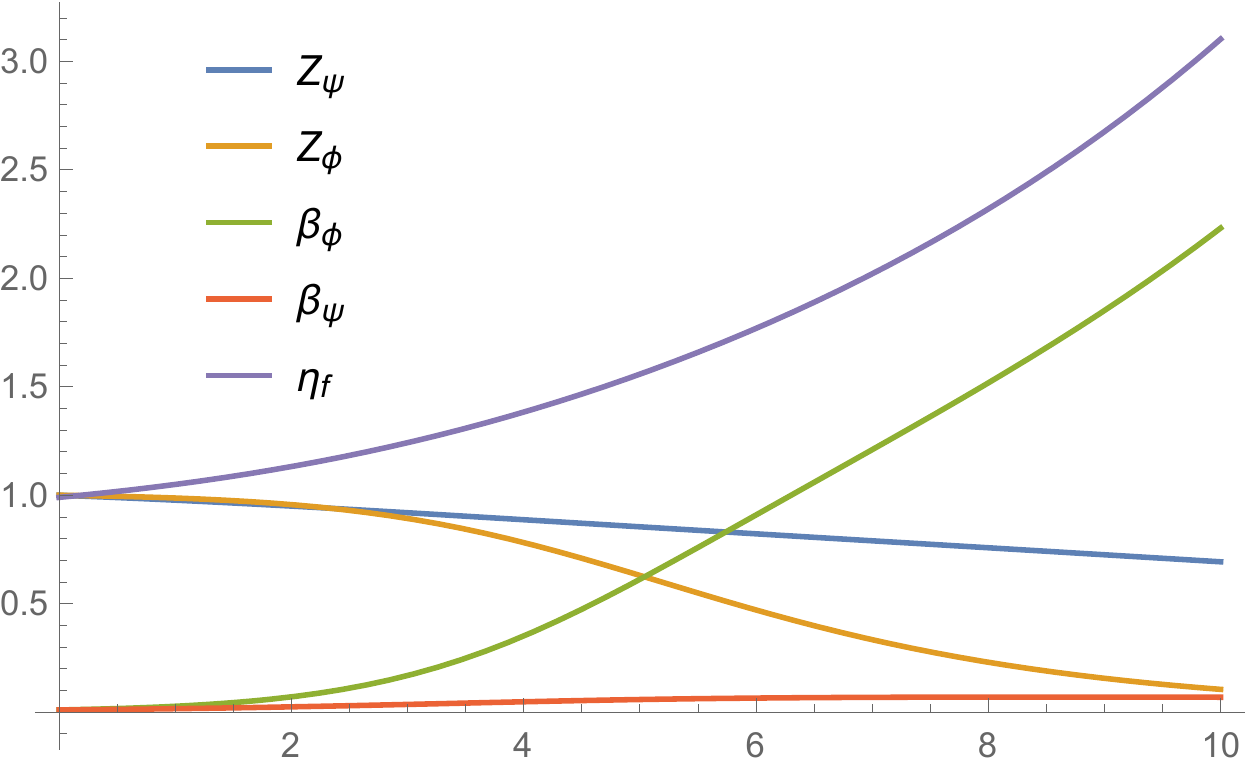}
	\caption{RG flow for $N_f=2;\beta_\varphi(0)=\beta_\psi(0)=0.01$;$\eta(0)=0.1,\eta_f(0)=0.99,Z_\varphi(0)=Z_\psi(0)=1$.}
	\label{fig:RGDPCoul}
\end{figure}

The numerical solution of the equations has the property that $\eta_f$ and $\beta_\varphi$ grow unbounded.

Indeed, the Gaussian fixed point $\beta_\psi=\beta_\varphi$ is evidently unstable due to the linear terms in the r.h.s. At the interacting fixed point one finds combining the equations for $\beta_\psi$ and $\beta_\varphi$ that $\beta_\psi F_2(\eta)$ has to vanish. Using $F_2(\eta)>F_1(\eta)$, both being positive, (as follows from the numerical solutions) one concludes that $\beta_\psi F_1(\eta)=0$ too. Next, one observes that $N_f \beta_\varphi F_\omega(\eta_f)=1$ at the fixed point. 

Then there are two options: either $\eta,\; \eta_f$ and $\beta_\varphi$ are finite and $\beta_\psi=0$ or $\eta_f$ grows to infinity. The former case turns out to be unstable. Indeed, expanding the r.h.s. of the equations near the fixed point we get:

\begin{gather*}
\frac{d\delta \beta_\varphi}{d l} = -\delta \beta_\varphi
+(-2 F_2(\eta)+4 F_1(\eta))\beta_\varphi^0\delta\beta_\psi
-N_f(\beta_\varphi^0)^2F'_\omega(\eta_f^0)\delta\eta_f;
\\
\frac{d(\delta \beta_\psi)}{dl} \approx 
-2 (F_2(\eta^0)-F_1(\eta^0)) \delta \beta_\psi^2
-\frac{N_f\beta_\varphi^0 F_\omega'(\eta_f^0)}{2}\delta\beta_\psi \delta\eta_f
-\frac{N_f F_\omega(\eta_f^0)}{2}\delta \beta_\psi\delta \beta_\varphi;
\\
\frac{d(\delta \eta_f)}{dl} \approx
\eta_f^0(F_1(\eta^0)+F_2(\eta^0))\delta \beta_\psi
\\
\frac{d(\delta \eta)}{dl} \approx
-\delta \eta
+2 \eta_0 F_1(\eta_0) \delta \beta_\psi
+\frac{\eta_0 N_f\beta_\varphi^0}{2}\left(\frac{F_{qy}'(\eta_f^0)}{\eta_0^2} - F'_\omega(\eta_f^0) \right) \delta \eta_f.
\end{gather*}
One notes that in the linear order, $\delta \eta_f$ is unstable: for nonzero $\delta\beta_\psi$ in the initial conditions it will move away from the fixed-point value.

Let us now consider the case when $\eta_f$ grows to infinity. The asymptotic expressions for $F_\omega,\; F_{qy}$ are $F_\omega(\eta_f)\to \frac{1}{2\eta_f},\; F_{qy}\to\frac{\log \eta_f}{2 \eta_f^3}$ for this case. As $N_f \beta_\varphi F_\omega(\eta_f)=1$, it follows that $\beta_\varphi$ should also grow to infinity in this case. In the large $\eta_f$ limit this leads to decreasing $\eta$, which eventually goes to zero in the numerical calculations. In this limit one can simplify the equations (using $F_2-F_1\approx1,\;F_1\sim\log (1/\eta),\;F_2\sim\log (1/\eta)+1$):

\begin{gather*}
\frac{d \beta_\varphi}{d l}
=
\beta_\varphi \left( 1- \frac{N_f\beta_\varphi}{2\eta_f}\right)
+
2\beta_\varphi\beta_\psi [\log(1/\eta)-1]
\\
\frac{d \beta_\psi}{d l} =
\frac{\beta_\psi}{2} \left( 1- \frac{N_f\beta_\varphi}{2\eta_f}\right)
-2\beta_\psi^2;
\\
\frac{d \eta_f}{d l}
=
\eta_f\beta_\psi[2\log(1/\eta)+1]
\\
\frac{d \eta}{d l}
=
2\eta\beta_\psi\log(1/\eta)
+
\frac{\eta N_f\beta_\varphi}{2}
\left(
\frac{\log \eta_f}{2\eta^2\eta_f^3}
-
\frac{1}{2\eta_f}
\right)
\end{gather*}

The equations for $x = N_f\beta_\varphi/(2\eta_f)$ and $\beta_\psi$ are decoupled from the others:
\begin{gather*}
\frac{dx}{dl} =  x(1-x)-3\beta_\psi x
\\
\frac{d \beta_\psi}{d l} =
\frac{\beta_\psi}{2} (1-x)-2\beta_\psi^2.
\end{gather*}
In the limit $l\to \infty$ one can check that
\[
x(l) \equiv N_f\beta_\varphi/(2\eta_f) = 1-\frac{6}{l};\;\beta_\psi=\frac{2}{l},
\]
indeed satisfy the equations above. Next we solve the equation for $\eta$ with the exponential ansatz $\eta\sim e^{-\alpha l}$ dropping the second term (we show that it is small afterwards) to find $\alpha=1/10$ resulting in:
\[
\eta \to e^{-l/10};\; \eta_f\to e^{2l/5} ;\; \beta_\varphi \to  e^{2l/5}.
\]

Inserting these into the equations for other parameters one gets (we use $F_{qx}(\eta_f)\to \eta_f/2$):

\begin{gather*}
\frac{d a_\varphi}{d l} = \frac{a_\varphi}{2};
\\
\frac{d a_\psi}{d l} =0.2  a_\psi;
\;
\frac{d v_{Fy}}{d l} = -v_{Fy} \beta_\psi F_1(\eta) =  -0.2v_{Fy};
\;
\frac{d v_{Fx}}{d l} = v_{Fx} \beta_\psi F_2(\eta) = 0.2v_{Fx} ;
\end{gather*}
resulting in
\[
a_\varphi\sim e^{l/2};\;a_\psi\sim e^{0.2l};\;v_{Fy}\sim e^{-0.2l};\;v_{Fx}\sim e^{0.2 l}
\]
and
\begin{gather*}
\frac{d c_x^2}{d l} = 
N_f \beta_\varphi (v_{Fy}/a_\psi)^2 a_\varphi^2 F_{qx}(\eta_f)
\sim e^{l};
\\
\frac{d c_y^2}{d l} = 
N_f \beta_\varphi (v_{Fy}/a_\psi)^2 a_\varphi^2 F_{qy}(\eta_f)
\sim l e^{-0.6l}
.
\end{gather*}
that yields:
\[
c_x^2 \sim e^{l};\;c_y^2\sim const.
\]

One can note that the main physical effects are: 1) strong renormalization of bosons, which become incoherent 2) strongly enhanced anisotropy of the fermionic dispersion. Taking $l\sim \log(k^{-1},\omega^{-1})$ we Eq. (6) of the main text.

\section{Hertz-Millis estimated for the scattering rates and conductivity/resistivity}

Here we generalize the expression used in \cite{schofield1999} to estimate the scattering rates near a QCP.

Let us start with the expression for the scattering rate at $T=0$ of a particle at an energy $\varepsilon_{\bf p}$ above the Fermi level and momentum ${\bf p}$ due to creation of particle-hole pairs according to the Fermi's golden rule:

\begin{gather*}
\Gamma = \frac{2 \pi}{\hbar}\int d \rho_f |W|^2 
\sim 
\\
\sim
\int d^d q d^d p 
\delta(\varepsilon_{\bf p-q} +\varepsilon_{\bf p_h+q} - \varepsilon_{\bf p} - \varepsilon_{\bf p_h})
|V_{\bf q}|^2 
\theta(|{\bf p}|-p_F)
\theta(|{\bf p-q}|-p_F)
\theta(|{\bf p}_h+{\bf q}|-p_F)
\theta(-|{\bf p}_h|+p_F),
\end{gather*}
where $V_{\bf q}$ is the Fourier transform of the interaction potential, ${\bf q}$ is the transferred momentum, and the $\theta-$functions are due to the Pauli principle. We first change the integration variables in the integral over ${\bf q}$. Namely, using $\omega \equiv\varepsilon_{\bf p} - \varepsilon_{\bf p-q}  \approx v_F q \cos \theta_{pq}$ we can rewrite $d \cos \theta_{pq} = \frac{d\omega}{v_F q}$ with the integration limits being from zero to $\varepsilon_{\bf p}$.

Note that from the above it follows that as $q\to0$, $\omega\approx v_F q \cos \theta_{pq}< v_F q$ and thus one has the lower limit for the radial part of the $q$ integration to be $\omega/v_F$. 
The upper limit for this integral comes from requirement that $\varepsilon_{\bf p-q}<\varepsilon_{\bf p}$: assuming a spherical Fermi surface this is only possible if $q<2 p\approx2 p_F$ (the diameter of the isoenergy sphere with energy $\varepsilon_p$).

What remains is to perform the integration over ${\bf p}_h$. We choose the $z$ axis of the spherical coordinates to be along ${\bf q}$. The requirements $\varepsilon_{\bf p_h}<0,\;\varepsilon_{\bf p_h+q}>0$ result in the integration limits $p_F-\omega/v_F<p_h<p_F$. Furthermore, for $q\ll p_F$ the integration over the angle can be performed as follows:

\[
\int d \cos\theta_h \delta(\varepsilon_{\bf p_h+q} - \varepsilon_{\bf p_h}-\omega) 
\approx
\int d \cos\theta_h \delta(v_F q \cos\theta_h-\omega)  = \frac{1}{v_F q},
\]
where one sees explicitly that the argument of the delta function can be always set to zero for a certain $\theta_h$ as $\omega<q v_F$. The integral over $p_h$ yields:
\begin{equation}
\int_{p_F-\omega/v_F}^{p_F} p_h^{d-1} d p_h \approx \nu_F \omega,
\label{eq:phint}
\end{equation}
where $\nu_F$ is the density of states at the Fermi level. Collecting the expressions above results in:
\begin{equation}
\Gamma \sim \int_0^{\varepsilon_p} \nu_F \omega d \omega \int_{\omega/v_F}^{2 p_F} 
\frac{|V_{\bf q}|^2}{(v_Fq)^2}  q^{d-1} dq
\end{equation}

Note that for the purposes of resistivity estimates one should use the transport scattering rate supplemented by a factor of $1-\cos \theta_{pq}$ in the integral above; for $q\ll p_F$ it is of the order $q^2/p_F^2$:
\begin{equation}
\Gamma_{tr} \sim \int_0^{\varepsilon_p} \nu_F \omega d \omega \int_{\omega/v_F}^{2 p_F} 
\frac{|V_{\bf q}|^2}{(v_Fq)^2} \frac{q^2}{p_F^2} q^{d-1} dq
\end{equation}

Finally, at finite temperatures, the estimates for the transport lifetimes may be obtained by assuming the quasiparticle energy to be of the order of $T$, i.e. $\varepsilon_p\to T$ in the above equations.

\subsection{Quantum critical systems}
Moving to quantum critical systems, one needs to take into account that the interaction is mediated by a (damped) bosonic mode leading to the transition amplitude being dependent on both momentum and frequency/energy transfer, i.e. $V_q\to D_{q,\omega}\sim \frac{1}{i \gamma \frac{\omega}{q^{z-2}} +q^2}$, where $z$ is the dynamical critical exponent. 

One can now separate the integral over $q$ into two regions. For $q\ll (\gamma \omega)^{1/z}$ one has  $|D_{q,\omega}|^2\sim \frac{q^{2(z-2)}}{\omega^2} $. For $z \geq 2$ the $q$  integral accumulates at the upper limit  $q \approx (\gamma \omega)^{1/z}$. In the opposite case,  $q\gg (\gamma \omega)^{1/z}$, $|D_{q,\omega}|^2\sim \frac{1}{q^4} $ and we expect the main contribution to come from the lower limit $q \approx (\gamma \omega)^{1/z}$. As both contributions yield answers of the same form, we can use only the second one to get:

\begin{equation}
\Gamma_{tr} \sim \int_0^{\varepsilon_p} \nu_F \omega d \omega \int_{(\gamma \omega)^{1/z}}^{2 p_F} 
\frac{1}{(v_Fq)^2} \frac{1}{q^4} \frac{q^2}{p_F^2} q^{d-1} dq.
\end{equation}

\subsection{Nodal semimetals}
For the case of nodal semimetals several adjustments need to be made. First, there are two kinds of particle-hole pair creation processes possible: inter-band and intra-band. In the latter case, one has $\omega<v_Fq$ for the scattering of the quasiparticle (we assume that the lower band is occupied, while the upper one is empty), but $\omega>v_F p_h$ for the interband process, thus $q>p_h$. The energy conservation condition $v_F|{\bf p}_h+{\bf q}|+v_Fp_h=\omega$ requires that $(q-p_h+p_h)v_F = q v_F \leq\omega$, which is only achieved if ${\bf p}_h$ and ${\bf q}$ are antiparallel and $\omega=q v_F$. This means that the integral over $q$ is of measure zero in this case.

For the intraband scattering the issue is that there are no quasiparticles in the upper band. However, at finite temperatures one expects a density of thermally excited quasiparticles ($\sim T^d$ for nodal points $\sim T^2$ for 3D nodal line) to be present. To emulate this in our $T=0$ estimates we assume a finite chemical potential $\mu$ to be present and set its scale to $T$ at the end of the calculation.

The calculation is then similar to the case of a usual Fermi surface above with the following adjustments. For nodal points, the integration over $p_h$ in \ref{eq:phint} results in an overall factor $\omega \mu^{d-1}$ instead of $\omega$. Secondly, the upper limit of the $q$ integration  - $2 k_F$ is actually of the order $\mu/v_F\sim T/v_F$, which does not result in a strong suppression of the critical boson propagator (unlike the case $k_F\gg (\varepsilon,T)/v_F$). Thus, backscattering is not suppressed and the factor $q^2/p_F^2$ in the transport scattering rate can be omitted.

In 3D, up to logarithmic correction, the scattering rate is:
\begin{equation}
\begin{gathered}
\Gamma_{WP3D} \sim \mu^2 \int_0^{\varepsilon_p} \omega d\omega \int_{\omega/v_F}^{2\mu/v_F} 
\frac{1}{(v_Fq)^2} \frac{1}{q^4} q^2 dq\sim|_{\mu\to T} T,
\\
\sigma\sim \frac{n(T) }{\Gamma}\sim \frac{T^2}{T} \sim T,
\\
\rho = \sigma^{-1} \sim T^{-1},
\end{gathered}
\end{equation}
agreeing with the known result \cite{armitage.2018}.

In 2D, within Hertz-Millis approach, we take $D_q\sim \frac{1}{q}$ to get:
\begin{equation}
\begin{gathered}
\Gamma_{DP2D} \sim \mu \int_0^{\varepsilon_p} \omega d\omega \int_{\omega/v_F}^{2\mu/v_F} 
\frac{1}{(v_Fq)^2} \frac{1}{q^2} q dq\sim|_{\mu\to T} T,
\\
\sigma\sim \frac{n(T)}{\Gamma}\sim \frac{T}{T} \sim const,
\\
\rho = \sigma^{-1} \sim const.
\end{gathered}
\end{equation}

For the nodal line case the same argument applies for the interband scattering as for nodal lines. For intraband one, in-plane momentum transfers of the order $k_F\gg T/v_F$ are possible; we will show now that the transport scattering rate for such momentum transfers and the low momentum transfers (where the transport scattering prefactor $q^2/k_F^2$ applies) give the same temperature dependence:
\begin{equation}
\begin{gathered}
\Gamma^{tr,q_r \sim 2k_F}_{NL3D} \sim 
\mu \int_0^{\varepsilon_p} \omega d\omega 
\int_{\sim k_F}^{2k_F} 
q_r dq_r 
\int_{\sqrt{\omega^2-(v_F q_r)^2}/\gamma}^{2 \omega/\gamma} d q_z
\frac{1}{(v_Fq)^2} \frac{1}{(2 k_F)^2}\sim|_{\mu\to T} T^4,
\\
\Gamma^{tr,q_r \to 0}_{NL3D} \sim 
\mu \int_0^{\varepsilon_p} \omega d\omega 
\int_{\omega/v_F}^{\sim\omega/v_F} 
q_r dq_r 
\int_{\sqrt{\omega^2-(v_F q_r)^2}/\gamma}^{2 \omega/\gamma} d q_z
\frac{1}{(v_Fq)^2} \frac{1}{q^2} \frac{q^2}{k_F^2}\sim|_{\mu\to T} T^4,
\\
\sigma\sim \frac{n(T)}{\Gamma}\sim \frac{T}{T^4} \sim T^{-3},
\\
\rho = \sigma^{-1} \sim T^{3}.
\end{gathered}
\end{equation}

\section{Specific heat for the Hertz-Millis case}

Here we consider a simple estimate for the specific heat capacity of a system within the Hertz-Millis approach above the upper critical dimension. In that case, one can obtain the effective free energy by integrating the fermionic fields out, re-exponentiating the result and expanding the fermionic logarithm to the second order in bosonic field. This procedure yields:
\begin{gather*}
S_{eff}[\varphi]
=
-T \log
[
G_0^-1
]
+
\varphi(\chi^{-1})\varphi,
\\
\chi^{-1} = D_0^{-1} - \Pi .
\end{gather*}
Integrating over the bosonic fields we get the following free energy:
\[
F(T) = F_{fer}^0(T)+ 
T\sum_{\omega_n} \int\frac{d^d q}{(2\pi)^d} 
\log \chi^{-1} (i \omega_n, {\bf q}).
\]
where the first term is the free energy of non-interacting fermions arising from the first term in the effective action. Evaluating the Matsubara sum (deforming the contour assuming a branch cut along the real axis) results in :
\begin{gather*}
\int \frac{d^d q}{(2\pi)^d}  \frac{d \nu}{2\pi}
2 n_B(\nu)
\arctan 
\left[
\frac{{\rm Im}\chi (\nu, {\bf q})}{{\rm Re}\chi (\nu, {\bf q})}
\right]
\approx
\int \frac{d^d q}{(2\pi)^d}  \frac{d \nu}{\pi}
n_B(\nu)
\arctan \left[ C \frac{\nu}{q^z}\right]
.
\end{gather*}
Making the integral dimensionless results in:
\begin{equation}
F_{bos}(T)
=
T^{d/z}T
\int \frac{d^d x}{(2\pi)^d}  \frac{d x_0}{\pi}
n_B(x_0)
\arctan \left[ C \frac{x_0}{x^z}\right].
\end{equation}
Finally, it follows that the bosonic contribution to specific heat $C = - T \frac{\partial^2 F_{bos}}{\partial T^2}$ scales with temperature as:

\begin{equation}
C_{bos}(T)
\sim
T^{d/z}
\end{equation}

\end{widetext}


\begin{thebibliography}{80}%
		\makeatletter
		\providecommand \@ifxundefined [1]{%
			\@ifx{#1\undefined}
		}%
		\providecommand \@ifnum [1]{%
			\ifnum #1\expandafter \@firstoftwo
			\else \expandafter \@secondoftwo
			\fi
		}%
		\providecommand \@ifx [1]{%
			\ifx #1\expandafter \@firstoftwo
			\else \expandafter \@secondoftwo
			\fi
		}%
		\providecommand \natexlab [1]{#1}%
		\providecommand \enquote  [1]{``#1''}%
		\providecommand \bibnamefont  [1]{#1}%
		\providecommand \bibfnamefont [1]{#1}%
		\providecommand \citenamefont [1]{#1}%
		\providecommand \href@noop [0]{\@secondoftwo}%
		\providecommand \href [0]{\begingroup \@sanitize@url \@href}%
		\providecommand \@href[1]{\@@startlink{#1}\@@href}%
		\providecommand \@@href[1]{\endgroup#1\@@endlink}%
		\providecommand \@sanitize@url [0]{\catcode `\\12\catcode `\$12\catcode
			`\&12\catcode `\#12\catcode `\^12\catcode `\_12\catcode `\%12\relax}%
		\providecommand \@@startlink[1]{}%
		\providecommand \@@endlink[0]{}%
		\providecommand \url  [0]{\begingroup\@sanitize@url \@url }%
		\providecommand \@url [1]{\endgroup\@href {#1}{\urlprefix }}%
		\providecommand \urlprefix  [0]{URL }%
		\providecommand \Eprint [0]{\href }%
		\providecommand \doibase [0]{http://dx.doi.org/}%
		\providecommand \selectlanguage [0]{\@gobble}%
		\providecommand \bibinfo  [0]{\@secondoftwo}%
		\providecommand \bibfield  [0]{\@secondoftwo}%
		\providecommand \translation [1]{[#1]}%
		\providecommand \BibitemOpen [0]{}%
		\providecommand \bibitemStop [0]{}%
		\providecommand \bibitemNoStop [0]{.\EOS\space}%
		\providecommand \EOS [0]{\spacefactor3000\relax}%
		\providecommand \BibitemShut  [1]{\csname bibitem#1\endcsname}%
		\let\auto@bib@innerbib\@empty
		%</preamble>
		\bibitem [{\citenamefont {Sachdev}(1999)}]{sachdev.1999}%
		\BibitemOpen
		\bibfield  {author} {\bibinfo {author} {\bibfnamefont {S.}~\bibnamefont
				{Sachdev}},\ }\href@noop {} {\emph {\bibinfo {title} {Quantum Phase
					Transitions}}}\ (\bibinfo  {publisher} {Cambridge Universtity Press},\
		\bibinfo {year} {1999})\BibitemShut {NoStop}%
		\bibitem [{\citenamefont {Scalapino}(2012)}]{scalapino2012}%
		\BibitemOpen
		\bibfield  {author} {\bibinfo {author} {\bibfnamefont {D.~J.}\ \bibnamefont
				{Scalapino}},\ }\bibfield  {title} {\enquote {\bibinfo {title} {A common
					thread: The pairing interaction for unconventional superconductors},}\ }\href
		{\doibase 10.1103/RevModPhys.84.1383} {\bibfield  {journal} {\bibinfo
				{journal} {Rev. Mod. Phys.}\ }\textbf {\bibinfo {volume} {84}},\ \bibinfo
			{pages} {1383--1417} (\bibinfo {year} {2012})}\BibitemShut {NoStop}%
		\bibitem [{\citenamefont {Abanov}\ \emph {et~al.}(2003)\citenamefont {Abanov},
			\citenamefont {Chubukov},\ and\ \citenamefont {Schmalian}}]{abanov2003}%
		\BibitemOpen
		\bibfield  {author} {\bibinfo {author} {\bibfnamefont {Ar.}\ \bibnamefont
				{Abanov}}, \bibinfo {author} {\bibfnamefont {Andrey~V.}\ \bibnamefont
				{Chubukov}}, \ and\ \bibinfo {author} {\bibfnamefont {J.}~\bibnamefont
				{Schmalian}},\ }\bibfield  {title} {\enquote {\bibinfo {title}
				{Quantum-critical theory of the spin-fermion model and its application to
					cuprates: Normal state analysis},}\ }\href {\doibase
			10.1080/0001873021000057123} {\bibfield  {journal} {\bibinfo  {journal}
				{Advances in Physics}\ }\textbf {\bibinfo {volume} {52}},\ \bibinfo {pages}
			{119--218} (\bibinfo {year} {2003})},\ \Eprint
		{http://arxiv.org/abs/https://doi.org/10.1080/0001873021000057123}
		{https://doi.org/10.1080/0001873021000057123} \BibitemShut {NoStop}%
		\bibitem [{\citenamefont {L\"ohneysen}\ \emph {et~al.}(2007)\citenamefont
			{L\"ohneysen}, \citenamefont {Rosch}, \citenamefont {Vojta},\ and\
			\citenamefont {W\"olfle}}]{lohneysen.2007}%
		\BibitemOpen
		\bibfield  {author} {\bibinfo {author} {\bibfnamefont {Hilbert~v.}\
				\bibnamefont {L\"ohneysen}}, \bibinfo {author} {\bibfnamefont {Achim}\
				\bibnamefont {Rosch}}, \bibinfo {author} {\bibfnamefont {Matthias}\
				\bibnamefont {Vojta}}, \ and\ \bibinfo {author} {\bibfnamefont {Peter}\
				\bibnamefont {W\"olfle}},\ }\bibfield  {title} {\enquote {\bibinfo {title}
				{Fermi-liquid instabilities at magnetic quantum phase transitions},}\ }\href
		{\doibase 10.1103/RevModPhys.79.1015} {\bibfield  {journal} {\bibinfo
				{journal} {Rev. Mod. Phys.}\ }\textbf {\bibinfo {volume} {79}},\ \bibinfo
			{pages} {1015--1075} (\bibinfo {year} {2007})}\BibitemShut {NoStop}%
		\bibitem [{\citenamefont {Brando}\ \emph {et~al.}(2016)\citenamefont {Brando},
			\citenamefont {Belitz}, \citenamefont {Grosche},\ and\ \citenamefont
			{Kirkpatrick}}]{brando.2016}%
		\BibitemOpen
		\bibfield  {author} {\bibinfo {author} {\bibfnamefont {M.}~\bibnamefont
				{Brando}}, \bibinfo {author} {\bibfnamefont {D.}~\bibnamefont {Belitz}},
			\bibinfo {author} {\bibfnamefont {F.~M.}\ \bibnamefont {Grosche}}, \ and\
			\bibinfo {author} {\bibfnamefont {T.~R.}\ \bibnamefont {Kirkpatrick}},\
		}\bibfield  {title} {\enquote {\bibinfo {title} {Metallic quantum
					ferromagnets},}\ }\href {\doibase 10.1103/RevModPhys.88.025006} {\bibfield
			{journal} {\bibinfo  {journal} {Rev. Mod. Phys.}\ }\textbf {\bibinfo {volume}
				{88}},\ \bibinfo {pages} {025006} (\bibinfo {year} {2016})}\BibitemShut
		{NoStop}%
		\bibitem [{\citenamefont {Shibauchi}\ \emph {et~al.}(2014)\citenamefont
			{Shibauchi}, \citenamefont {Carrington},\ and\ \citenamefont
			{Matsuda}}]{shibauchi.2014}%
		\BibitemOpen
		\bibfield  {author} {\bibinfo {author} {\bibfnamefont {T.}~\bibnamefont
				{Shibauchi}}, \bibinfo {author} {\bibfnamefont {A.}~\bibnamefont
				{Carrington}}, \ and\ \bibinfo {author} {\bibfnamefont {Y.}~\bibnamefont
				{Matsuda}},\ }\bibfield  {title} {\enquote {\bibinfo {title} {A quantum
					critical point lying beneath the superconducting dome in iron pnictides},}\
		}\href {\doibase 10.1146/annurev-conmatphys-031113-133921} {\bibfield
			{journal} {\bibinfo  {journal} {Annual Review of Condensed Matter Physics}\
			}\textbf {\bibinfo {volume} {5}},\ \bibinfo {pages} {113--135} (\bibinfo
			{year} {2014})},\ \Eprint
		{http://arxiv.org/abs/https://doi.org/10.1146/annurev-conmatphys-031113-133921}
		{https://doi.org/10.1146/annurev-conmatphys-031113-133921} \BibitemShut
		{NoStop}%
		\bibitem [{\citenamefont {Kolodiazhnyi}\ \emph {et~al.}(2010)\citenamefont
			{Kolodiazhnyi}, \citenamefont {Tachibana}, \citenamefont {Kawaji},
			\citenamefont {Hwang},\ and\ \citenamefont
			{Takayama-Muromachi}}]{kolod.2010}%
		\BibitemOpen
		\bibfield  {author} {\bibinfo {author} {\bibfnamefont {T.}~\bibnamefont
				{Kolodiazhnyi}}, \bibinfo {author} {\bibfnamefont {M.}~\bibnamefont
				{Tachibana}}, \bibinfo {author} {\bibfnamefont {H.}~\bibnamefont {Kawaji}},
			\bibinfo {author} {\bibfnamefont {J.}~\bibnamefont {Hwang}}, \ and\ \bibinfo
			{author} {\bibfnamefont {E.}~\bibnamefont {Takayama-Muromachi}},\ }\bibfield
		{title} {\enquote {\bibinfo {title} {{Persistence of Ferroelectricity in
						${\mathrm{BaTiO}}_{3}$ through the Insulator-Metal Transition}},}\ }\href
		{\doibase 10.1103/PhysRevLett.104.147602} {\bibfield  {journal} {\bibinfo
				{journal} {Phys. Rev. Lett.}\ }\textbf {\bibinfo {volume} {104}},\ \bibinfo
			{pages} {147602} (\bibinfo {year} {2010})}\BibitemShut {NoStop}%
		\bibitem [{\citenamefont {Shi}\ \emph {et~al.}(2013)\citenamefont {Shi},
			\citenamefont {Guo}, \citenamefont {Wang}, \citenamefont {Princep},
			\citenamefont {Khalyavin}, \citenamefont {Manuel}, \citenamefont {Michiue},
			\citenamefont {Sato}, \citenamefont {Tsuda}, \citenamefont {Yu},
			\citenamefont {Arai}, \citenamefont {Shirako}, \citenamefont {Akaogi},
			\citenamefont {Wang}, \citenamefont {Yamaura},\ and\ \citenamefont
			{Boothroyd}}]{shi2013}%
		\BibitemOpen
		\bibfield  {author} {\bibinfo {author} {\bibfnamefont {Youguo}\ \bibnamefont
				{Shi}}, \bibinfo {author} {\bibfnamefont {Yanfeng}\ \bibnamefont {Guo}},
			\bibinfo {author} {\bibfnamefont {Xia}\ \bibnamefont {Wang}}, \bibinfo
			{author} {\bibfnamefont {Andrew~J.}\ \bibnamefont {Princep}}, \bibinfo
			{author} {\bibfnamefont {Dmitry}\ \bibnamefont {Khalyavin}}, \bibinfo
			{author} {\bibfnamefont {Pascal}\ \bibnamefont {Manuel}}, \bibinfo {author}
			{\bibfnamefont {Yuichi}\ \bibnamefont {Michiue}}, \bibinfo {author}
			{\bibfnamefont {Akira}\ \bibnamefont {Sato}}, \bibinfo {author}
			{\bibfnamefont {Kenji}\ \bibnamefont {Tsuda}}, \bibinfo {author}
			{\bibfnamefont {Shan}\ \bibnamefont {Yu}}, \bibinfo {author} {\bibfnamefont
				{Masao}\ \bibnamefont {Arai}}, \bibinfo {author} {\bibfnamefont {Yuichi}\
				\bibnamefont {Shirako}}, \bibinfo {author} {\bibfnamefont {Masaki}\
				\bibnamefont {Akaogi}}, \bibinfo {author} {\bibfnamefont {Nanlin}\
				\bibnamefont {Wang}}, \bibinfo {author} {\bibfnamefont {Kazunari}\
				\bibnamefont {Yamaura}}, \ and\ \bibinfo {author} {\bibfnamefont {Andrew~T.}\
				\bibnamefont {Boothroyd}},\ }\bibfield  {title} {\enquote {\bibinfo {title}
				{A ferroelectric-like structural transition in a metal},}\ }\href
		{https://doi.org/10.1038/nmat3754} {\bibfield  {journal} {\bibinfo  {journal}
				{Nature Materials}\ }\textbf {\bibinfo {volume} {12}},\ \bibinfo {pages}
			{1024} (\bibinfo {year} {2013})}\BibitemShut {NoStop}%
		\bibitem [{\citenamefont {Rischau}\ \emph {et~al.}(2017)\citenamefont
			{Rischau}, \citenamefont {Lin}, \citenamefont {Grams}, \citenamefont {Finck},
			\citenamefont {Harms}, \citenamefont {Engelmayer}, \citenamefont {Lorenz},
			\citenamefont {Gallais}, \citenamefont {Fauque}, \citenamefont {Hemberger},\
			and\ \citenamefont {Behnia}}]{rischau.2017}%
		\BibitemOpen
		\bibfield  {author} {\bibinfo {author} {\bibfnamefont {Carl~Willem}\
				\bibnamefont {Rischau}}, \bibinfo {author} {\bibfnamefont {Xiao}\
				\bibnamefont {Lin}}, \bibinfo {author} {\bibfnamefont {Christoph~P.}\
				\bibnamefont {Grams}}, \bibinfo {author} {\bibfnamefont {Dennis}\
				\bibnamefont {Finck}}, \bibinfo {author} {\bibfnamefont {Steffen}\
				\bibnamefont {Harms}}, \bibinfo {author} {\bibfnamefont {Johannes}\
				\bibnamefont {Engelmayer}}, \bibinfo {author} {\bibfnamefont {Thomas}\
				\bibnamefont {Lorenz}}, \bibinfo {author} {\bibfnamefont {Yann}\ \bibnamefont
				{Gallais}}, \bibinfo {author} {\bibfnamefont {Benoit}\ \bibnamefont
				{Fauque}}, \bibinfo {author} {\bibfnamefont {Joachim}\ \bibnamefont
				{Hemberger}}, \ and\ \bibinfo {author} {\bibfnamefont {Kamran}\ \bibnamefont
				{Behnia}},\ }\bibfield  {title} {\enquote {\bibinfo {title} {{A ferroelectric
						quantum phase transition inside the superconducting dome of
						Sr$_{1-x}$Ca$_x$TiO$_{3-\delta}$}},}\ }\href@noop {} {\bibfield  {journal}
			{\bibinfo  {journal} {Nature Physics}\ }\textbf {\bibinfo {volume} {13}},\
			\bibinfo {pages} {643} (\bibinfo {year} {2017})}\BibitemShut {NoStop}%
		\bibitem [{\citenamefont {Jiang}\ \emph {et~al.}(2017)\citenamefont {Jiang},
			\citenamefont {Liu}, \citenamefont {Sun}, \citenamefont {Yang}, \citenamefont
			{Rajamathi}, \citenamefont {Qi}, \citenamefont {Yang}, \citenamefont {Chen},
			\citenamefont {Peng}, \citenamefont {Hwang} \emph {et~al.}}]{jiang2017}%
		\BibitemOpen
		\bibfield  {author} {\bibinfo {author} {\bibfnamefont {Juan}\ \bibnamefont
				{Jiang}}, \bibinfo {author} {\bibfnamefont {ZK}~\bibnamefont {Liu}}, \bibinfo
			{author} {\bibfnamefont {Y}~\bibnamefont {Sun}}, \bibinfo {author}
			{\bibfnamefont {HF}~\bibnamefont {Yang}}, \bibinfo {author} {\bibfnamefont
				{CR}~\bibnamefont {Rajamathi}}, \bibinfo {author} {\bibfnamefont
				{YP}~\bibnamefont {Qi}}, \bibinfo {author} {\bibfnamefont {LX}~\bibnamefont
				{Yang}}, \bibinfo {author} {\bibfnamefont {C}~\bibnamefont {Chen}}, \bibinfo
			{author} {\bibfnamefont {H}~\bibnamefont {Peng}}, \bibinfo {author}
			{\bibfnamefont {CC}~\bibnamefont {Hwang}},  \emph {et~al.},\ }\bibfield
		{title} {\enquote {\bibinfo {title} {{Signature of type-II Weyl semimetal
						phase in MoTe${}_2$}},}\ }\href@noop {} {\bibfield  {journal} {\bibinfo
				{journal} {Nature communications}\ }\textbf {\bibinfo {volume} {8}},\
			\bibinfo {pages} {13973} (\bibinfo {year} {2017})}\BibitemShut {NoStop}%
		\bibitem [{\citenamefont {Yu}\ \emph {et~al.}(2018)\citenamefont {Yu},
			\citenamefont {Zhou}, \citenamefont {Chuang}, \citenamefont {Yang},
			\citenamefont {Lin},\ and\ \citenamefont {Bansil}}]{yu2018}%
		\BibitemOpen
		\bibfield  {author} {\bibinfo {author} {\bibfnamefont {Wing~Chi}\
				\bibnamefont {Yu}}, \bibinfo {author} {\bibfnamefont {Xiaoting}\ \bibnamefont
				{Zhou}}, \bibinfo {author} {\bibfnamefont {Feng-Chuan}\ \bibnamefont
				{Chuang}}, \bibinfo {author} {\bibfnamefont {Shengyuan~A.}\ \bibnamefont
				{Yang}}, \bibinfo {author} {\bibfnamefont {Hsin}\ \bibnamefont {Lin}}, \ and\
			\bibinfo {author} {\bibfnamefont {Arun}\ \bibnamefont {Bansil}},\ }\bibfield
		{title} {\enquote {\bibinfo {title} {Nonsymmorphic cubic dirac point and
					crossed nodal rings across the ferroelectric phase transition in
					${\mathrm{lioso}}_{3}$},}\ }\href {\doibase
			10.1103/PhysRevMaterials.2.051201} {\bibfield  {journal} {\bibinfo  {journal}
				{Phys. Rev. Materials}\ }\textbf {\bibinfo {volume} {2}},\ \bibinfo {pages}
			{051201} (\bibinfo {year} {2018})}\BibitemShut {NoStop}%
		\bibitem [{\citenamefont {Cao}\ \emph {et~al.}(2018)\citenamefont {Cao},
			\citenamefont {Wang}, \citenamefont {Park}, \citenamefont {Yuan},
			\citenamefont {Liu}, \citenamefont {Nikitin}, \citenamefont {Akamatsu},
			\citenamefont {Kareev}, \citenamefont {Middey}, \citenamefont {Meyers},
			\citenamefont {Thompson}, \citenamefont {Ryan}, \citenamefont {Shafer},
			\citenamefont {N'Diaye}, \citenamefont {Arenholz}, \citenamefont {Gopalan},
			\citenamefont {Zhu}, \citenamefont {Rabe},\ and\ \citenamefont
			{Chakhalian}}]{cao2018}%
		\BibitemOpen
		\bibfield  {author} {\bibinfo {author} {\bibfnamefont {Yanwei}\ \bibnamefont
				{Cao}}, \bibinfo {author} {\bibfnamefont {Zhen}\ \bibnamefont {Wang}},
			\bibinfo {author} {\bibfnamefont {Se~Young}\ \bibnamefont {Park}}, \bibinfo
			{author} {\bibfnamefont {Yakun}\ \bibnamefont {Yuan}}, \bibinfo {author}
			{\bibfnamefont {Xiaoran}\ \bibnamefont {Liu}}, \bibinfo {author}
			{\bibfnamefont {Sergey~M.}\ \bibnamefont {Nikitin}}, \bibinfo {author}
			{\bibfnamefont {Hirofumi}\ \bibnamefont {Akamatsu}}, \bibinfo {author}
			{\bibfnamefont {M.}~\bibnamefont {Kareev}}, \bibinfo {author} {\bibfnamefont
				{S.}~\bibnamefont {Middey}}, \bibinfo {author} {\bibfnamefont
				{D.}~\bibnamefont {Meyers}}, \bibinfo {author} {\bibfnamefont
				{P.}~\bibnamefont {Thompson}}, \bibinfo {author} {\bibfnamefont {P.J.}\
				\bibnamefont {Ryan}}, \bibinfo {author} {\bibfnamefont {Padraic}\
				\bibnamefont {Shafer}}, \bibinfo {author} {\bibfnamefont {A.}~\bibnamefont
				{N'Diaye}}, \bibinfo {author} {\bibfnamefont {E.}~\bibnamefont {Arenholz}},
			\bibinfo {author} {\bibfnamefont {Venkatraman}\ \bibnamefont {Gopalan}},
			\bibinfo {author} {\bibfnamefont {Yimei}\ \bibnamefont {Zhu}}, \bibinfo
			{author} {\bibfnamefont {Karin~M.}\ \bibnamefont {Rabe}}, \ and\ \bibinfo
			{author} {\bibfnamefont {J.}~\bibnamefont {Chakhalian}},\ }\bibfield  {title}
		{\enquote {\bibinfo {title} {Artificial two-dimensional polar metal at room
					temperature},}\ }\href {\doibase 10.1038/s41467-018-03964-9} {\bibfield
			{journal} {\bibinfo  {journal} {Nature Communications}\ }\textbf {\bibinfo
				{volume} {9}},\ \bibinfo {pages} {1547} (\bibinfo {year} {2018})}\BibitemShut
		{NoStop}%
		\bibitem [{\citenamefont {Fei}\ \emph {et~al.}(2018)\citenamefont {Fei},
			\citenamefont {Zhao}, \citenamefont {Palomaki}, \citenamefont {Sun},
			\citenamefont {Miller}, \citenamefont {Zhao}, \citenamefont {Yan},
			\citenamefont {Xu},\ and\ \citenamefont {Cobden}}]{fei2018}%
		\BibitemOpen
		\bibfield  {author} {\bibinfo {author} {\bibfnamefont {Zaiyao}\ \bibnamefont
				{Fei}}, \bibinfo {author} {\bibfnamefont {Wenjin}\ \bibnamefont {Zhao}},
			\bibinfo {author} {\bibfnamefont {Tauno~A}\ \bibnamefont {Palomaki}},
			\bibinfo {author} {\bibfnamefont {Bosong}\ \bibnamefont {Sun}}, \bibinfo
			{author} {\bibfnamefont {Moira~K}\ \bibnamefont {Miller}}, \bibinfo {author}
			{\bibfnamefont {Zhiying}\ \bibnamefont {Zhao}}, \bibinfo {author}
			{\bibfnamefont {Jiaqiang}\ \bibnamefont {Yan}}, \bibinfo {author}
			{\bibfnamefont {Xiaodong}\ \bibnamefont {Xu}}, \ and\ \bibinfo {author}
			{\bibfnamefont {David~H}\ \bibnamefont {Cobden}},\ }\bibfield  {title}
		{\enquote {\bibinfo {title} {Ferroelectric switching of a two-dimensional
					metal},}\ }\href@noop {} {\bibfield  {journal} {\bibinfo  {journal} {Nature}\
			}\textbf {\bibinfo {volume} {560}},\ \bibinfo {pages} {336--339} (\bibinfo
			{year} {2018})}\BibitemShut {NoStop}%
		\bibitem [{\citenamefont {Shirodkar}\ and\ \citenamefont
			{Waghmare}(2014)}]{shirodkar2014}%
		\BibitemOpen
		\bibfield  {author} {\bibinfo {author} {\bibfnamefont {Sharmila~N.}\
				\bibnamefont {Shirodkar}}\ and\ \bibinfo {author} {\bibfnamefont {Umesh~V.}\
				\bibnamefont {Waghmare}},\ }\bibfield  {title} {\enquote {\bibinfo {title}
				{{Emergence of Ferroelectricity at a Metal-Semiconductor Transition in a $1T$
						Monolayer of ${\mathrm{MoS}}_{2}$}},}\ }\href {\doibase
			10.1103/PhysRevLett.112.157601} {\bibfield  {journal} {\bibinfo  {journal}
				{Phys. Rev. Lett.}\ }\textbf {\bibinfo {volume} {112}},\ \bibinfo {pages}
			{157601} (\bibinfo {year} {2014})}\BibitemShut {NoStop}%
		\bibitem [{\citenamefont {Fei}\ \emph {et~al.}(2016)\citenamefont {Fei},
			\citenamefont {Kang},\ and\ \citenamefont {Yang}}]{fei2016}%
		\BibitemOpen
		\bibfield  {author} {\bibinfo {author} {\bibfnamefont {Ruixiang}\
				\bibnamefont {Fei}}, \bibinfo {author} {\bibfnamefont {Wei}\ \bibnamefont
				{Kang}}, \ and\ \bibinfo {author} {\bibfnamefont {Li}~\bibnamefont {Yang}},\
		}\bibfield  {title} {\enquote {\bibinfo {title} {{Ferroelectricity and Phase
						Transitions in Monolayer Group-$IV$ Monochalcogenides}},}\ }\href {\doibase
			10.1103/PhysRevLett.117.097601} {\bibfield  {journal} {\bibinfo  {journal}
				{Phys. Rev. Lett.}\ }\textbf {\bibinfo {volume} {117}},\ \bibinfo {pages}
			{097601} (\bibinfo {year} {2016})}\BibitemShut {NoStop}%
		\bibitem [{\citenamefont {Ding}\ \emph {et~al.}(2017)\citenamefont {Ding},
			\citenamefont {Zhu}, \citenamefont {Wang}, \citenamefont {Gao}, \citenamefont
			{Xiao}, \citenamefont {Gu}, \citenamefont {Zhang},\ and\ \citenamefont
			{Zhu}}]{ding2017}%
		\BibitemOpen
		\bibfield  {author} {\bibinfo {author} {\bibfnamefont {Wenjun}\ \bibnamefont
				{Ding}}, \bibinfo {author} {\bibfnamefont {Jianbao}\ \bibnamefont {Zhu}},
			\bibinfo {author} {\bibfnamefont {Zhe}\ \bibnamefont {Wang}}, \bibinfo
			{author} {\bibfnamefont {Yanfei}\ \bibnamefont {Gao}}, \bibinfo {author}
			{\bibfnamefont {Di}~\bibnamefont {Xiao}}, \bibinfo {author} {\bibfnamefont
				{Yi}~\bibnamefont {Gu}}, \bibinfo {author} {\bibfnamefont {Zhenyu}\
				\bibnamefont {Zhang}}, \ and\ \bibinfo {author} {\bibfnamefont {Wenguang}\
				\bibnamefont {Zhu}},\ }\bibfield  {title} {\enquote {\bibinfo {title}
				{{Prediction of intrinsic two-dimensional ferroelectrics in
						${\mathrm{In}}_{2}{\mathrm{Se}}_{3}$ and other ${\mathrm
							{III}}_{2}$-${\mathrm{VI}}_{3}$ van der Waals materials}},}\ }\href {\doibase
			10.1038/ncomms14956} {\bibfield  {journal} {\bibinfo  {journal} {Nature
					Communications}\ }\textbf {\bibinfo {volume} {8}},\ \bibinfo {pages} {14956}
			(\bibinfo {year} {2017})}\BibitemShut {NoStop}%
		\bibitem [{\citenamefont {Benedek}\ and\ \citenamefont
			{Birol}(2016)}]{benedek2016}%
		\BibitemOpen
		\bibfield  {author} {\bibinfo {author} {\bibfnamefont {Nicole~A.}\
				\bibnamefont {Benedek}}\ and\ \bibinfo {author} {\bibfnamefont {Turan}\
				\bibnamefont {Birol}},\ }\bibfield  {title} {\enquote {\bibinfo {title}
				{?ferroelectric? metals reexamined: fundamental mechanisms and design
					considerations for new materials},}\ }\href {\doibase 10.1039/C5TC03856A}
		{\bibfield  {journal} {\bibinfo  {journal} {J. Mater. Chem. C}\ }\textbf
			{\bibinfo {volume} {4}},\ \bibinfo {pages} {4000--4015} (\bibinfo {year}
			{2016})}\BibitemShut {NoStop}%
		\bibitem [{\citenamefont {Anderson}\ and\ \citenamefont
			{Blount}(1965)}]{anderson.1965}%
		\BibitemOpen
		\bibfield  {author} {\bibinfo {author} {\bibfnamefont {P.~W.}\ \bibnamefont
				{Anderson}}\ and\ \bibinfo {author} {\bibfnamefont {E.~I.}\ \bibnamefont
				{Blount}},\ }\bibfield  {title} {\enquote {\bibinfo {title} {Symmetry
					considerations on martensitic transformations: "ferroelectric" metals?}}\
		}\href {\doibase 10.1103/PhysRevLett.14.217} {\bibfield  {journal} {\bibinfo
				{journal} {Phys. Rev. Lett.}\ }\textbf {\bibinfo {volume} {14}},\ \bibinfo
			{pages} {217--219} (\bibinfo {year} {1965})}\BibitemShut {NoStop}%
		\bibitem [{\citenamefont {Shneerson}(1973)}]{khmelnitskii1973}%
		\BibitemOpen
		\bibfield  {author} {\bibinfo {author} {\bibfnamefont {VL}~\bibnamefont
				{Shneerson}},\ }\bibfield  {title} {\enquote {\bibinfo {title}
				{Phase-transition of displacement type in crystals at very low
					temperatures},}\ }\href@noop {} {\bibfield  {journal} {\bibinfo  {journal}
				{Zh. Eksp. Teor. Fiz}\ }\textbf {\bibinfo {volume} {64}},\ \bibinfo {pages}
			{316--330} (\bibinfo {year} {1973})}\BibitemShut {NoStop}%
		\bibitem [{\citenamefont {Roussev}\ and\ \citenamefont
			{Millis}(2003)}]{roussev.2003}%
		\BibitemOpen
		\bibfield  {author} {\bibinfo {author} {\bibfnamefont {R.}~\bibnamefont
				{Roussev}}\ and\ \bibinfo {author} {\bibfnamefont {A.~J.}\ \bibnamefont
				{Millis}},\ }\bibfield  {title} {\enquote {\bibinfo {title} {Theory of the
					quantum paraelectric-ferroelectric transition},}\ }\href {\doibase
			10.1103/PhysRevB.67.014105} {\bibfield  {journal} {\bibinfo  {journal} {Phys.
					Rev. B}\ }\textbf {\bibinfo {volume} {67}},\ \bibinfo {pages} {014105}
			(\bibinfo {year} {2003})}\BibitemShut {NoStop}%
		\bibitem [{\citenamefont {Rowley}\ \emph {et~al.}(2014)\citenamefont {Rowley},
			\citenamefont {Spalek}, \citenamefont {Smith}, \citenamefont {Dean},
			\citenamefont {Itoh}, \citenamefont {Scott}, \citenamefont {Lonzarich},\ and\
			\citenamefont {Saxena}}]{Rowley2014}%
		\BibitemOpen
		\bibfield  {author} {\bibinfo {author} {\bibfnamefont {S.~E.}\ \bibnamefont
				{Rowley}}, \bibinfo {author} {\bibfnamefont {L.~J.}\ \bibnamefont {Spalek}},
			\bibinfo {author} {\bibfnamefont {R.~P.}\ \bibnamefont {Smith}}, \bibinfo
			{author} {\bibfnamefont {M.~P.~M.}\ \bibnamefont {Dean}}, \bibinfo {author}
			{\bibfnamefont {M.}~\bibnamefont {Itoh}}, \bibinfo {author} {\bibfnamefont
				{J.~F.}\ \bibnamefont {Scott}}, \bibinfo {author} {\bibfnamefont {G.~G.}\
				\bibnamefont {Lonzarich}}, \ and\ \bibinfo {author} {\bibfnamefont {S.~S.}\
				\bibnamefont {Saxena}},\ }\bibfield  {title} {\enquote {\bibinfo {title}
				{Ferroelectric quantum criticality},}\ }\href
		{https://doi.org/10.1038/nphys2924} {\bibfield  {journal} {\bibinfo
				{journal} {Nature Physics}\ }\textbf {\bibinfo {volume} {10}},\ \bibinfo
			{pages} {367} (\bibinfo {year} {2014})}\BibitemShut {NoStop}%
		\bibitem [{\citenamefont {Chandra}\ \emph {et~al.}(2017)\citenamefont
			{Chandra}, \citenamefont {Lonzarich}, \citenamefont {Rowley},\ and\
			\citenamefont {Scott}}]{chandra2017}%
		\BibitemOpen
		\bibfield  {author} {\bibinfo {author} {\bibfnamefont {P}~\bibnamefont
				{Chandra}}, \bibinfo {author} {\bibfnamefont {G~G}\ \bibnamefont
				{Lonzarich}}, \bibinfo {author} {\bibfnamefont {S~E}\ \bibnamefont {Rowley}},
			\ and\ \bibinfo {author} {\bibfnamefont {J~F}\ \bibnamefont {Scott}},\
		}\bibfield  {title} {\enquote {\bibinfo {title} {Prospects and applications
					near ferroelectric quantum phase transitions: a key issues review},}\ }\href
		{\doibase 10.1088/1361-6633/aa82d2} {\bibfield  {journal} {\bibinfo
				{journal} {Reports on Progress in Physics}\ }\textbf {\bibinfo {volume}
				{80}},\ \bibinfo {pages} {112502} (\bibinfo {year} {2017})}\BibitemShut
		{NoStop}%
		\bibitem [{\citenamefont {Chandra}\ \emph {et~al.}()\citenamefont {Chandra},
			\citenamefont {Coleman}, \citenamefont {Continentino},\ and\ \citenamefont
			{Lonzarich}}]{chandra2018}%
		\BibitemOpen
		\bibfield  {author} {\bibinfo {author} {\bibfnamefont {Premala}\ \bibnamefont
				{Chandra}}, \bibinfo {author} {\bibfnamefont {Piers}\ \bibnamefont
				{Coleman}}, \bibinfo {author} {\bibfnamefont {Mucio~A.}\ \bibnamefont
				{Continentino}}, \ and\ \bibinfo {author} {\bibfnamefont {Gilbert~G.}\
				\bibnamefont {Lonzarich}},\ }\href@noop {} {\ }\Eprint
		{http://arxiv.org/abs/1805.11771} {arXiv:1805.11771} \BibitemShut {NoStop}%
		\bibitem [{\citenamefont {Narayan}\ \emph {et~al.}(2019)\citenamefont
			{Narayan}, \citenamefont {Cano}, \citenamefont {Balatsky},\ and\
			\citenamefont {Spaldin}}]{narayan2019}%
		\BibitemOpen
		\bibfield  {author} {\bibinfo {author} {\bibfnamefont {Awadhesh}\
				\bibnamefont {Narayan}}, \bibinfo {author} {\bibfnamefont {Andr{\'e}s}\
				\bibnamefont {Cano}}, \bibinfo {author} {\bibfnamefont {Alexander~V.}\
				\bibnamefont {Balatsky}}, \ and\ \bibinfo {author} {\bibfnamefont
				{Nicola~A.}\ \bibnamefont {Spaldin}},\ }\bibfield  {title} {\enquote
			{\bibinfo {title} {Multiferroic quantum criticality},}\ }\href {\doibase
			10.1038/s41563-018-0255-6} {\bibfield  {journal} {\bibinfo  {journal} {Nature
					Materials}\ }\textbf {\bibinfo {volume} {18}},\ \bibinfo {pages} {223--228}
			(\bibinfo {year} {2019})}\BibitemShut {NoStop}%
		\bibitem [{\citenamefont {Yoshida}\ \emph {et~al.}(2005)\citenamefont
			{Yoshida}, \citenamefont {Ikeda}, \citenamefont {Matsuhata}, \citenamefont
			{Shirakawa}, \citenamefont {Lee},\ and\ \citenamefont
			{Katano}}]{yoshida.2005}%
		\BibitemOpen
		\bibfield  {author} {\bibinfo {author} {\bibfnamefont {Yoshiyuki}\
				\bibnamefont {Yoshida}}, \bibinfo {author} {\bibfnamefont {Shin-Ichi}\
				\bibnamefont {Ikeda}}, \bibinfo {author} {\bibfnamefont {Hirofumi}\
				\bibnamefont {Matsuhata}}, \bibinfo {author} {\bibfnamefont {Naoki}\
				\bibnamefont {Shirakawa}}, \bibinfo {author} {\bibfnamefont {C.~H.}\
				\bibnamefont {Lee}}, \ and\ \bibinfo {author} {\bibfnamefont {Susumu}\
				\bibnamefont {Katano}},\ }\bibfield  {title} {\enquote {\bibinfo {title}
				{{Crystal and magnetic structure of
						${\mathrm{Ca}}_{3}{\mathrm{Ru}}_{2}{\mathrm{O}}_{7}$}},}\ }\href {\doibase
			10.1103/PhysRevB.72.054412} {\bibfield  {journal} {\bibinfo  {journal} {Phys.
					Rev. B}\ }\textbf {\bibinfo {volume} {72}},\ \bibinfo {pages} {054412}
			(\bibinfo {year} {2005})}\BibitemShut {NoStop}%
		\bibitem [{\citenamefont {Lei}\ \emph {et~al.}(2018)\citenamefont {Lei},
			\citenamefont {Gu}, \citenamefont {Puggioni}, \citenamefont {Stone},
			\citenamefont {Peng}, \citenamefont {Ge}, \citenamefont {Wang}, \citenamefont
			{Wang}, \citenamefont {Yuan}, \citenamefont {Wang}, \citenamefont {Mao},
			\citenamefont {Rondinelli},\ and\ \citenamefont {Gopalan}}]{lei.2018}%
		\BibitemOpen
		\bibfield  {author} {\bibinfo {author} {\bibfnamefont {Shiming}\ \bibnamefont
				{Lei}}, \bibinfo {author} {\bibfnamefont {Mingqiang}\ \bibnamefont {Gu}},
			\bibinfo {author} {\bibfnamefont {Danilo}\ \bibnamefont {Puggioni}}, \bibinfo
			{author} {\bibfnamefont {Greg}\ \bibnamefont {Stone}}, \bibinfo {author}
			{\bibfnamefont {Jin}\ \bibnamefont {Peng}}, \bibinfo {author} {\bibfnamefont
				{Jianjian}\ \bibnamefont {Ge}}, \bibinfo {author} {\bibfnamefont
				{Yu}~\bibnamefont {Wang}}, \bibinfo {author} {\bibfnamefont {Baoming}\
				\bibnamefont {Wang}}, \bibinfo {author} {\bibfnamefont {Yakun}\ \bibnamefont
				{Yuan}}, \bibinfo {author} {\bibfnamefont {Ke}~\bibnamefont {Wang}}, \bibinfo
			{author} {\bibfnamefont {Zhiqiang}\ \bibnamefont {Mao}}, \bibinfo {author}
			{\bibfnamefont {James~M.}\ \bibnamefont {Rondinelli}}, \ and\ \bibinfo
			{author} {\bibfnamefont {Venkatraman}\ \bibnamefont {Gopalan}},\ }\bibfield
		{title} {\enquote {\bibinfo {title} {{Observation of Quasi-Two-Dimensional
						Polar Domains and Ferroelastic Switching in a Metal,
						Ca${}_3$Ru${}_2$O${}_7$}},}\ }\href {\doibase 10.1021/acs.nanolett.8b00633}
		{\bibfield  {journal} {\bibinfo  {journal} {Nano Letters}\ }\textbf {\bibinfo
				{volume} {18}},\ \bibinfo {pages} {3088--3095} (\bibinfo {year} {2018})},\
		\Eprint {http://arxiv.org/abs/https://doi.org/10.1021/acs.nanolett.8b00633}
		{https://doi.org/10.1021/acs.nanolett.8b00633} \BibitemShut {NoStop}%
		\bibitem [{\citenamefont {Kim}\ \emph {et~al.}(2016)\citenamefont {Kim},
			\citenamefont {Puggioni}, \citenamefont {Yuan}, \citenamefont {Xie},
			\citenamefont {Zhou}, \citenamefont {Campbell}, \citenamefont {Ryan},
			\citenamefont {Choi}, \citenamefont {Kim}, \citenamefont {Patzner} \emph
			{et~al.}}]{kim2016polar}%
		\BibitemOpen
		\bibfield  {author} {\bibinfo {author} {\bibfnamefont {TH}~\bibnamefont
				{Kim}}, \bibinfo {author} {\bibfnamefont {Danilo}\ \bibnamefont {Puggioni}},
			\bibinfo {author} {\bibfnamefont {Y}~\bibnamefont {Yuan}}, \bibinfo {author}
			{\bibfnamefont {L}~\bibnamefont {Xie}}, \bibinfo {author} {\bibfnamefont
				{H}~\bibnamefont {Zhou}}, \bibinfo {author} {\bibfnamefont {N}~\bibnamefont
				{Campbell}}, \bibinfo {author} {\bibfnamefont {PJ}~\bibnamefont {Ryan}},
			\bibinfo {author} {\bibfnamefont {Y}~\bibnamefont {Choi}}, \bibinfo {author}
			{\bibfnamefont {J-W}\ \bibnamefont {Kim}}, \bibinfo {author} {\bibfnamefont
				{JR}~\bibnamefont {Patzner}},  \emph {et~al.},\ }\bibfield  {title} {\enquote
			{\bibinfo {title} {Polar metals by geometric design},}\ }\href@noop {}
		{\bibfield  {journal} {\bibinfo  {journal} {Nature}\ }\textbf {\bibinfo
				{volume} {533}},\ \bibinfo {pages} {68--72} (\bibinfo {year}
			{2016})}\BibitemShut {NoStop}%
		\bibitem [{\citenamefont {Nukala}\ \emph {et~al.}(2017)\citenamefont {Nukala},
			\citenamefont {Ren}, \citenamefont {Agarwal}, \citenamefont {Berger},
			\citenamefont {Liu}, \citenamefont {Johnson},\ and\ \citenamefont
			{Agarwal}}]{nukala2017}%
		\BibitemOpen
		\bibfield  {author} {\bibinfo {author} {\bibfnamefont {Pavan}\ \bibnamefont
				{Nukala}}, \bibinfo {author} {\bibfnamefont {Mingliang}\ \bibnamefont {Ren}},
			\bibinfo {author} {\bibfnamefont {Rahul}\ \bibnamefont {Agarwal}}, \bibinfo
			{author} {\bibfnamefont {Jacob}\ \bibnamefont {Berger}}, \bibinfo {author}
			{\bibfnamefont {Gerui}\ \bibnamefont {Liu}}, \bibinfo {author} {\bibfnamefont
				{AT~Charlie}\ \bibnamefont {Johnson}}, \ and\ \bibinfo {author}
			{\bibfnamefont {Ritesh}\ \bibnamefont {Agarwal}},\ }\bibfield  {title}
		{\enquote {\bibinfo {title} {Inverting polar domains via electrical pulsing
					in metallic germanium telluride},}\ }\href@noop {} {\bibfield  {journal}
			{\bibinfo  {journal} {Nature communications}\ }\textbf {\bibinfo {volume}
				{8}},\ \bibinfo {pages} {1--9} (\bibinfo {year} {2017})}\BibitemShut
		{NoStop}%
		\bibitem [{\citenamefont {Sharma}\ \emph {et~al.}(2019)\citenamefont {Sharma},
			\citenamefont {Xiang}, \citenamefont {Shao}, \citenamefont {Zhang},
			\citenamefont {Tsymbal}, \citenamefont {Hamilton},\ and\ \citenamefont
			{Seidel}}]{sharma2019}%
		\BibitemOpen
		\bibfield  {author} {\bibinfo {author} {\bibfnamefont {Pankaj}\ \bibnamefont
				{Sharma}}, \bibinfo {author} {\bibfnamefont {Fei-Xiang}\ \bibnamefont
				{Xiang}}, \bibinfo {author} {\bibfnamefont {Ding-Fu}\ \bibnamefont {Shao}},
			\bibinfo {author} {\bibfnamefont {Dawei}\ \bibnamefont {Zhang}}, \bibinfo
			{author} {\bibfnamefont {Evgeny~Y.}\ \bibnamefont {Tsymbal}}, \bibinfo
			{author} {\bibfnamefont {Alex~R.}\ \bibnamefont {Hamilton}}, \ and\ \bibinfo
			{author} {\bibfnamefont {Jan}\ \bibnamefont {Seidel}},\ }\bibfield  {title}
		{\enquote {\bibinfo {title} {A room-temperature ferroelectric semimetal},}\
		}\href {\doibase 10.1126/sciadv.aax5080} {\bibfield  {journal} {\bibinfo
				{journal} {Science Advances}\ }\textbf {\bibinfo {volume} {5}} (\bibinfo
			{year} {2019}),\ 10.1126/sciadv.aax5080},\ \Eprint
		{http://arxiv.org/abs/https://advances.sciencemag.org/content/5/7/eaax5080.full.pdf}
		{https://advances.sciencemag.org/content/5/7/eaax5080.full.pdf} \BibitemShut
		{NoStop}%
		\bibitem [{\citenamefont {Hasan}\ \emph {et~al.}(2017)\citenamefont {Hasan},
			\citenamefont {Xu}, \citenamefont {Belopolski},\ and\ \citenamefont
			{Huang}}]{hasan2017}%
		\BibitemOpen
		\bibfield  {author} {\bibinfo {author} {\bibfnamefont {M.~Zahid}\
				\bibnamefont {Hasan}}, \bibinfo {author} {\bibfnamefont {Su-Yang}\
				\bibnamefont {Xu}}, \bibinfo {author} {\bibfnamefont {Ilya}\ \bibnamefont
				{Belopolski}}, \ and\ \bibinfo {author} {\bibfnamefont {Shin-Ming}\
				\bibnamefont {Huang}},\ }\bibfield  {title} {\enquote {\bibinfo {title}
				{Discovery of weyl fermion semimetals and topological fermi arc states},}\
		}\href {\doibase 10.1146/annurev-conmatphys-031016-025225} {\bibfield
			{journal} {\bibinfo  {journal} {Annual Review of Condensed Matter Physics}\
			}\textbf {\bibinfo {volume} {8}},\ \bibinfo {pages} {289--309} (\bibinfo
			{year} {2017})},\ \Eprint
		{http://arxiv.org/abs/https://doi.org/10.1146/annurev-conmatphys-031016-025225}
		{https://doi.org/10.1146/annurev-conmatphys-031016-025225} \BibitemShut
		{NoStop}%
		\bibitem [{\citenamefont {Sakai}\ \emph {et~al.}(2016)\citenamefont {Sakai},
			\citenamefont {Ikeura}, \citenamefont {Bahramy}, \citenamefont {Ogawa},
			\citenamefont {Hashizume}, \citenamefont {Fujioka}, \citenamefont {Tokura},\
			and\ \citenamefont {Ishiwata}}]{sakai2016critical}%
		\BibitemOpen
		\bibfield  {author} {\bibinfo {author} {\bibfnamefont {Hideaki}\ \bibnamefont
				{Sakai}}, \bibinfo {author} {\bibfnamefont {Koji}\ \bibnamefont {Ikeura}},
			\bibinfo {author} {\bibfnamefont {Mohammad~Saeed}\ \bibnamefont {Bahramy}},
			\bibinfo {author} {\bibfnamefont {Naoki}\ \bibnamefont {Ogawa}}, \bibinfo
			{author} {\bibfnamefont {Daisuke}\ \bibnamefont {Hashizume}}, \bibinfo
			{author} {\bibfnamefont {Jun}\ \bibnamefont {Fujioka}}, \bibinfo {author}
			{\bibfnamefont {Yoshinori}\ \bibnamefont {Tokura}}, \ and\ \bibinfo {author}
			{\bibfnamefont {Shintaro}\ \bibnamefont {Ishiwata}},\ }\bibfield  {title}
		{\enquote {\bibinfo {title} {{Critical enhancement of thermopower in a
						chemically tuned polar semimetal MoTe$_2$}},}\ }\href@noop {} {\bibfield
			{journal} {\bibinfo  {journal} {Science advances}\ }\textbf {\bibinfo
				{volume} {2}},\ \bibinfo {pages} {e1601378} (\bibinfo {year}
			{2016})}\BibitemShut {NoStop}%
		\bibitem [{\citenamefont {Barraza-Lopez}\ \emph {et~al.}(2018)\citenamefont
			{Barraza-Lopez}, \citenamefont {Kaloni}, \citenamefont {Poudel},\ and\
			\citenamefont {Kumar}}]{barraza2018}%
		\BibitemOpen
		\bibfield  {author} {\bibinfo {author} {\bibfnamefont {Salvador}\
				\bibnamefont {Barraza-Lopez}}, \bibinfo {author} {\bibfnamefont
				{Thaneshwor~P.}\ \bibnamefont {Kaloni}}, \bibinfo {author} {\bibfnamefont
				{Shiva~P.}\ \bibnamefont {Poudel}}, \ and\ \bibinfo {author} {\bibfnamefont
				{Pradeep}\ \bibnamefont {Kumar}},\ }\bibfield  {title} {\enquote {\bibinfo
				{title} {{Tuning the ferroelectric-to-paraelectric transition temperature and
						dipole orientation of group-IV monochalcogenide monolayers}},}\ }\href
		{\doibase 10.1103/PhysRevB.97.024110} {\bibfield  {journal} {\bibinfo
				{journal} {Phys. Rev. B}\ }\textbf {\bibinfo {volume} {97}},\ \bibinfo
			{pages} {024110} (\bibinfo {year} {2018})}\BibitemShut {NoStop}%
		\bibitem [{\citenamefont {Narayan}(2019)}]{narayan_2019}%
		\BibitemOpen
		\bibfield  {author} {\bibinfo {author} {\bibfnamefont {Awadhesh}\
				\bibnamefont {Narayan}},\ }\bibfield  {title} {\enquote {\bibinfo {title}
				{Effect of strain and doping on the polar metal phase in {LiOsO}3},}\ }\href
		{\doibase 10.1088/1361-648x/ab5a10} {\bibfield  {journal} {\bibinfo
				{journal} {Journal of Physics: Condensed Matter}\ }\textbf {\bibinfo {volume}
				{32}},\ \bibinfo {pages} {125501} (\bibinfo {year} {2019})}\BibitemShut
		{NoStop}%
		\bibitem [{\citenamefont {Cochran}(1960)}]{cochran.1960}%
		\BibitemOpen
		\bibfield  {author} {\bibinfo {author} {\bibfnamefont {W.}~\bibnamefont
				{Cochran}},\ }\bibfield  {title} {\enquote {\bibinfo {title} {Crystal
					stability and the theory of ferroelectricity},}\ }\href {\doibase
			10.1080/00018736000101229} {\bibfield  {journal} {\bibinfo  {journal}
				{Advances in Physics}\ }\textbf {\bibinfo {volume} {9}},\ \bibinfo {pages}
			{387--423} (\bibinfo {year} {1960})},\ \Eprint
		{http://arxiv.org/abs/https://doi.org/10.1080/00018736000101229}
		{https://doi.org/10.1080/00018736000101229} \BibitemShut {NoStop}%
		\bibitem [{\citenamefont {Gor{\textquoteright}kov}(2016)}]{Gorkov2016}%
		\BibitemOpen
		\bibfield  {author} {\bibinfo {author} {\bibfnamefont {Lev~P.}\ \bibnamefont
				{Gor{\textquoteright}kov}},\ }\bibfield  {title} {\enquote {\bibinfo {title}
				{{Phonon mechanism in the most dilute superconductor n-type SrTiO$_3$}},}\
		}\href {\doibase 10.1073/pnas.1604145113} {\bibfield  {journal} {\bibinfo
				{journal} {Proceedings of the National Academy of Sciences}\ }\textbf
			{\bibinfo {volume} {113}},\ \bibinfo {pages} {4646--4651} (\bibinfo {year}
			{2016})},\ \Eprint
		{http://arxiv.org/abs/https://www.pnas.org/content/113/17/4646.full.pdf}
		{https://www.pnas.org/content/113/17/4646.full.pdf} \BibitemShut {NoStop}%
		\bibitem [{\citenamefont {Ruhman}\ and\ \citenamefont
			{Lee}(2016)}]{ruhman.2016}%
		\BibitemOpen
		\bibfield  {author} {\bibinfo {author} {\bibfnamefont {Jonathan}\
				\bibnamefont {Ruhman}}\ and\ \bibinfo {author} {\bibfnamefont {Patrick~A.}\
				\bibnamefont {Lee}},\ }\bibfield  {title} {\enquote {\bibinfo {title}
				{Superconductivity at very low density: The case of strontium titanate},}\
		}\href {\doibase 10.1103/PhysRevB.94.224515} {\bibfield  {journal} {\bibinfo
				{journal} {Phys. Rev. B}\ }\textbf {\bibinfo {volume} {94}},\ \bibinfo
			{pages} {224515} (\bibinfo {year} {2016})}\BibitemShut {NoStop}%
		\bibitem [{\citenamefont {W\"olfle}\ and\ \citenamefont
			{Balatsky}(2018)}]{wolfle.2018}%
		\BibitemOpen
		\bibfield  {author} {\bibinfo {author} {\bibfnamefont {Peter}\ \bibnamefont
				{W\"olfle}}\ and\ \bibinfo {author} {\bibfnamefont {Alexander~V.}\
				\bibnamefont {Balatsky}},\ }\bibfield  {title} {\enquote {\bibinfo {title}
				{{Superconductivity at low density near a ferroelectric quantum critical
						point: Doped ${\mathrm{SrTiO}}_{3}$}},}\ }\href {\doibase
			10.1103/PhysRevB.98.104505} {\bibfield  {journal} {\bibinfo  {journal} {Phys.
					Rev. B}\ }\textbf {\bibinfo {volume} {98}},\ \bibinfo {pages} {104505}
			(\bibinfo {year} {2018})}\BibitemShut {NoStop}%
		\bibitem [{\citenamefont {Ruhman}\ and\ \citenamefont
			{Lee}(2019)}]{wolfle.2018com}%
		\BibitemOpen
		\bibfield  {author} {\bibinfo {author} {\bibfnamefont {Jonathan}\
				\bibnamefont {Ruhman}}\ and\ \bibinfo {author} {\bibfnamefont {Patrick~A.}\
				\bibnamefont {Lee}},\ }\bibfield  {title} {\enquote {\bibinfo {title}
				{{Comment on ``Superconductivity at low density near a ferroelectric quantum
						critical point: Doped ${\mathrm{SrTiO}}_{3}$''}},}\ }\href {\doibase
			10.1103/PhysRevB.100.226501} {\bibfield  {journal} {\bibinfo  {journal}
				{Phys. Rev. B}\ }\textbf {\bibinfo {volume} {100}},\ \bibinfo {pages}
			{226501} (\bibinfo {year} {2019})}\BibitemShut {NoStop}%
		\bibitem [{\citenamefont {W\"olfle}\ and\ \citenamefont
			{Balatsky}(2019)}]{wolfle.2018repl}%
		\BibitemOpen
		\bibfield  {author} {\bibinfo {author} {\bibfnamefont {Peter}\ \bibnamefont
				{W\"olfle}}\ and\ \bibinfo {author} {\bibfnamefont {Alexander~V.}\
				\bibnamefont {Balatsky}},\ }\bibfield  {title} {\enquote {\bibinfo {title}
				{{Reply to ``Comment on `Superconductivity at low density near a
						ferroelectric quantum critical point: Doped ${\mathrm{SrTiO}}_{3}$'''}},}\
		}\href {\doibase 10.1103/PhysRevB.100.226502} {\bibfield  {journal} {\bibinfo
				{journal} {Phys. Rev. B}\ }\textbf {\bibinfo {volume} {100}},\ \bibinfo
			{pages} {226502} (\bibinfo {year} {2019})}\BibitemShut {NoStop}%
		\bibitem [{\citenamefont {van~der Marel}\ \emph {et~al.}(2019)\citenamefont
			{van~der Marel}, \citenamefont {Barantani},\ and\ \citenamefont
			{Rischau}}]{marel2019}%
		\BibitemOpen
		\bibfield  {author} {\bibinfo {author} {\bibfnamefont {D.}~\bibnamefont
				{van~der Marel}}, \bibinfo {author} {\bibfnamefont {F.}~\bibnamefont
				{Barantani}}, \ and\ \bibinfo {author} {\bibfnamefont {C.~W.}\ \bibnamefont
				{Rischau}},\ }\bibfield  {title} {\enquote {\bibinfo {title} {{Possible
						mechanism for superconductivity in doped ${\mathrm{SrTiO}}_{3}$}},}\ }\href
		{\doibase 10.1103/PhysRevResearch.1.013003} {\bibfield  {journal} {\bibinfo
				{journal} {Phys. Rev. Research}\ }\textbf {\bibinfo {volume} {1}},\ \bibinfo
			{pages} {013003} (\bibinfo {year} {2019})}\BibitemShut {NoStop}%
		\bibitem [{\citenamefont {Mooradian}\ and\ \citenamefont
			{Wright}(1966)}]{mooradian.1966}%
		\BibitemOpen
		\bibfield  {author} {\bibinfo {author} {\bibfnamefont {A.}~\bibnamefont
				{Mooradian}}\ and\ \bibinfo {author} {\bibfnamefont {G.~B.}\ \bibnamefont
				{Wright}},\ }\bibfield  {title} {\enquote {\bibinfo {title} {{Observation of
						the Interaction of Plasmons with Longitudinal Optical Phonons in GaAs}},}\
		}\href {\doibase 10.1103/PhysRevLett.16.999} {\bibfield  {journal} {\bibinfo
				{journal} {Phys. Rev. Lett.}\ }\textbf {\bibinfo {volume} {16}},\ \bibinfo
			{pages} {999--1001} (\bibinfo {year} {1966})}\BibitemShut {NoStop}%
		\bibitem [{\citenamefont {LARKIN}\ and\ \citenamefont
			{KHMEL'NITSKll}(1969)}]{larkin1969}%
		\BibitemOpen
		\bibfield  {author} {\bibinfo {author} {\bibfnamefont {AI}~\bibnamefont
				{LARKIN}}\ and\ \bibinfo {author} {\bibfnamefont {DE}~\bibnamefont
				{KHMEL'NITSKll}},\ }\bibfield  {title} {\enquote {\bibinfo {title} {Phase
					transition in uniaxial ferroelectrics},}\ }\href@noop {} {\bibfield
			{journal} {\bibinfo  {journal} {SOVIET PHYSICS JETP}\ }\textbf {\bibinfo
				{volume} {29}} (\bibinfo {year} {1969})}\BibitemShut {NoStop}%
		\bibitem [{\citenamefont {Garrity}\ \emph {et~al.}(2014)\citenamefont
			{Garrity}, \citenamefont {Rabe},\ and\ \citenamefont
			{Vanderbilt}}]{garrity2014}%
		\BibitemOpen
		\bibfield  {author} {\bibinfo {author} {\bibfnamefont {Kevin~F.}\
				\bibnamefont {Garrity}}, \bibinfo {author} {\bibfnamefont {Karin~M.}\
				\bibnamefont {Rabe}}, \ and\ \bibinfo {author} {\bibfnamefont {David}\
				\bibnamefont {Vanderbilt}},\ }\bibfield  {title} {\enquote {\bibinfo {title}
				{Hyperferroelectrics: Proper ferroelectrics with persistent polarization},}\
		}\href {\doibase 10.1103/PhysRevLett.112.127601} {\bibfield  {journal}
			{\bibinfo  {journal} {Phys. Rev. Lett.}\ }\textbf {\bibinfo {volume} {112}},\
			\bibinfo {pages} {127601} (\bibinfo {year} {2014})}\BibitemShut {NoStop}%
		\bibitem [{\citenamefont {Tagantsev}(1988)}]{tagantsev1988}%
		\BibitemOpen
		\bibfield  {author} {\bibinfo {author} {\bibfnamefont {Alexander~K.}\
				\bibnamefont {Tagantsev}},\ }\bibfield  {title} {\enquote {\bibinfo {title}
				{Weak ferroelectrics},}\ }\href {\doibase 10.1080/00150198808229398}
		{\bibfield  {journal} {\bibinfo  {journal} {Ferroelectrics}\ }\textbf
			{\bibinfo {volume} {79}},\ \bibinfo {pages} {57--60} (\bibinfo {year}
			{1988})},\ \Eprint
		{http://arxiv.org/abs/https://www.tandfonline.com/doi/pdf/10.1080/00150198808229398}
		{https://www.tandfonline.com/doi/pdf/10.1080/00150198808229398} \BibitemShut
		{NoStop}%
		\bibitem [{\citenamefont {Kozii}\ and\ \citenamefont {Fu}(2015)}]{kozii.2015}%
		\BibitemOpen
		\bibfield  {author} {\bibinfo {author} {\bibfnamefont {Vladyslav}\
				\bibnamefont {Kozii}}\ and\ \bibinfo {author} {\bibfnamefont {Liang}\
				\bibnamefont {Fu}},\ }\bibfield  {title} {\enquote {\bibinfo {title}
				{Odd-parity superconductivity in the vicinity of inversion symmetry breaking
					in spin-orbit-coupled systems},}\ }\href {\doibase
			10.1103/PhysRevLett.115.207002} {\bibfield  {journal} {\bibinfo  {journal}
				{Phys. Rev. Lett.}\ }\textbf {\bibinfo {volume} {115}},\ \bibinfo {pages}
			{207002} (\bibinfo {year} {2015})}\BibitemShut {NoStop}%
		\bibitem [{\citenamefont {Wu}\ and\ \citenamefont {Martin}(2017)}]{wu2017}%
		\BibitemOpen
		\bibfield  {author} {\bibinfo {author} {\bibfnamefont {Fengcheng}\
				\bibnamefont {Wu}}\ and\ \bibinfo {author} {\bibfnamefont {Ivar}\
				\bibnamefont {Martin}},\ }\bibfield  {title} {\enquote {\bibinfo {title}
				{Nematic and chiral superconductivity induced by odd-parity fluctuations},}\
		}\href {\doibase 10.1103/PhysRevB.96.144504} {\bibfield  {journal} {\bibinfo
				{journal} {Phys. Rev. B}\ }\textbf {\bibinfo {volume} {96}},\ \bibinfo
			{pages} {144504} (\bibinfo {year} {2017})}\BibitemShut {NoStop}%
		\bibitem [{\citenamefont {Kanasugi}\ and\ \citenamefont
			{Yanase}(2018)}]{yanase2018}%
		\BibitemOpen
		\bibfield  {author} {\bibinfo {author} {\bibfnamefont {Shota}\ \bibnamefont
				{Kanasugi}}\ and\ \bibinfo {author} {\bibfnamefont {Youichi}\ \bibnamefont
				{Yanase}},\ }\bibfield  {title} {\enquote {\bibinfo {title}
				{Spin-orbit-coupled ferroelectric superconductivity},}\ }\href {\doibase
			10.1103/PhysRevB.98.024521} {\bibfield  {journal} {\bibinfo  {journal} {Phys.
					Rev. B}\ }\textbf {\bibinfo {volume} {98}},\ \bibinfo {pages} {024521}
			(\bibinfo {year} {2018})}\BibitemShut {NoStop}%
		\bibitem [{\citenamefont {Kanasugi}\ and\ \citenamefont
			{Yanase}(2019)}]{yanase2019}%
		\BibitemOpen
		\bibfield  {author} {\bibinfo {author} {\bibfnamefont {Shota}\ \bibnamefont
				{Kanasugi}}\ and\ \bibinfo {author} {\bibfnamefont {Youichi}\ \bibnamefont
				{Yanase}},\ }\bibfield  {title} {\enquote {\bibinfo {title} {{Multiorbital
						ferroelectric superconductivity in doped ${\mathrm{SrTiO}}_{3}$}},}\ }\href
		{\doibase 10.1103/PhysRevB.100.094504} {\bibfield  {journal} {\bibinfo
				{journal} {Phys. Rev. B}\ }\textbf {\bibinfo {volume} {100}},\ \bibinfo
			{pages} {094504} (\bibinfo {year} {2019})}\BibitemShut {NoStop}%
		\bibitem [{\citenamefont {Su}\ \emph {et~al.}(1979)\citenamefont {Su},
			\citenamefont {Schrieffer},\ and\ \citenamefont {Heeger}}]{su1979}%
		\BibitemOpen
		\bibfield  {author} {\bibinfo {author} {\bibfnamefont {W.~P.}\ \bibnamefont
				{Su}}, \bibinfo {author} {\bibfnamefont {J.~R.}\ \bibnamefont {Schrieffer}},
			\ and\ \bibinfo {author} {\bibfnamefont {A.~J.}\ \bibnamefont {Heeger}},\
		}\bibfield  {title} {\enquote {\bibinfo {title} {Solitons in
					polyacetylene},}\ }\href {\doibase 10.1103/PhysRevLett.42.1698} {\bibfield
			{journal} {\bibinfo  {journal} {Phys. Rev. Lett.}\ }\textbf {\bibinfo
				{volume} {42}},\ \bibinfo {pages} {1698--1701} (\bibinfo {year}
			{1979})}\BibitemShut {NoStop}%
		\bibitem [{\citenamefont {Joshua}\ \emph {et~al.}(2012)\citenamefont {Joshua},
			\citenamefont {Pecker}, \citenamefont {Ruhman}, \citenamefont {Altman},\ and\
			\citenamefont {Ilani}}]{joshua2012}%
		\BibitemOpen
		\bibfield  {author} {\bibinfo {author} {\bibfnamefont {Arjun}\ \bibnamefont
				{Joshua}}, \bibinfo {author} {\bibfnamefont {S.}~\bibnamefont {Pecker}},
			\bibinfo {author} {\bibfnamefont {J.}~\bibnamefont {Ruhman}}, \bibinfo
			{author} {\bibfnamefont {E.}~\bibnamefont {Altman}}, \ and\ \bibinfo {author}
			{\bibfnamefont {S.}~\bibnamefont {Ilani}},\ }\bibfield  {title} {\enquote
			{\bibinfo {title} {{A universal critical density underlying the physics of
						electrons at the LaAlO$_3$/SrTiO$_3$ interface}},}\ }\href {\doibase
			10.1038/ncomms2116} {\bibfield  {journal} {\bibinfo  {journal} {Nature
					Communications}\ }\textbf {\bibinfo {volume} {3}},\ \bibinfo {pages} {1129}
			(\bibinfo {year} {2012})}\BibitemShut {NoStop}%
		\bibitem [{\citenamefont {Diez}\ \emph {et~al.}(2015)\citenamefont {Diez},
			\citenamefont {Monteiro}, \citenamefont {Mattoni}, \citenamefont {Cobanera},
			\citenamefont {Hyart}, \citenamefont {Mulazimoglu}, \citenamefont {Bovenzi},
			\citenamefont {Beenakker},\ and\ \citenamefont {Caviglia}}]{diez2015}%
		\BibitemOpen
		\bibfield  {author} {\bibinfo {author} {\bibfnamefont {M.}~\bibnamefont
				{Diez}}, \bibinfo {author} {\bibfnamefont {A.~M. R. V.~L.}\ \bibnamefont
				{Monteiro}}, \bibinfo {author} {\bibfnamefont {G.}~\bibnamefont {Mattoni}},
			\bibinfo {author} {\bibfnamefont {E.}~\bibnamefont {Cobanera}}, \bibinfo
			{author} {\bibfnamefont {T.}~\bibnamefont {Hyart}}, \bibinfo {author}
			{\bibfnamefont {E.}~\bibnamefont {Mulazimoglu}}, \bibinfo {author}
			{\bibfnamefont {N.}~\bibnamefont {Bovenzi}}, \bibinfo {author} {\bibfnamefont
				{C.~W.~J.}\ \bibnamefont {Beenakker}}, \ and\ \bibinfo {author}
			{\bibfnamefont {A.~D.}\ \bibnamefont {Caviglia}},\ }\bibfield  {title}
		{\enquote {\bibinfo {title} {{Giant Negative Magnetoresistance Driven by
						Spin-Orbit Coupling at the ${\mathrm{LaAlO}}_{3}/{\mathrm{SrTiO}}_{3}$
						Interface}},}\ }\href {\doibase 10.1103/PhysRevLett.115.016803} {\bibfield
			{journal} {\bibinfo  {journal} {Phys. Rev. Lett.}\ }\textbf {\bibinfo
				{volume} {115}},\ \bibinfo {pages} {016803} (\bibinfo {year}
			{2015})}\BibitemShut {NoStop}%
		\bibitem [{\citenamefont {Kim}\ \emph {et~al.}(2015)\citenamefont {Kim},
			\citenamefont {Wieder}, \citenamefont {Kane},\ and\ \citenamefont
			{Rappe}}]{kimrappe.2015}%
		\BibitemOpen
		\bibfield  {author} {\bibinfo {author} {\bibfnamefont {Youngkuk}\
				\bibnamefont {Kim}}, \bibinfo {author} {\bibfnamefont {Benjamin~J.}\
				\bibnamefont {Wieder}}, \bibinfo {author} {\bibfnamefont {C.~L.}\
				\bibnamefont {Kane}}, \ and\ \bibinfo {author} {\bibfnamefont {Andrew~M.}\
				\bibnamefont {Rappe}},\ }\bibfield  {title} {\enquote {\bibinfo {title}
				{Dirac line nodes in inversion-symmetric crystals},}\ }\href {\doibase
			10.1103/PhysRevLett.115.036806} {\bibfield  {journal} {\bibinfo  {journal}
				{Phys. Rev. Lett.}\ }\textbf {\bibinfo {volume} {115}},\ \bibinfo {pages}
			{036806} (\bibinfo {year} {2015})}\BibitemShut {NoStop}%
		\bibitem [{\citenamefont {Fang}\ \emph {et~al.}(2015)\citenamefont {Fang},
			\citenamefont {Chen}, \citenamefont {Kee},\ and\ \citenamefont
			{Fu}}]{fang.2015}%
		\BibitemOpen
		\bibfield  {author} {\bibinfo {author} {\bibfnamefont {Chen}\ \bibnamefont
				{Fang}}, \bibinfo {author} {\bibfnamefont {Yige}\ \bibnamefont {Chen}},
			\bibinfo {author} {\bibfnamefont {Hae-Young}\ \bibnamefont {Kee}}, \ and\
			\bibinfo {author} {\bibfnamefont {Liang}\ \bibnamefont {Fu}},\ }\bibfield
		{title} {\enquote {\bibinfo {title} {Topological nodal line semimetals with
					and without spin-orbital coupling},}\ }\href {\doibase
			10.1103/PhysRevB.92.081201} {\bibfield  {journal} {\bibinfo  {journal} {Phys.
					Rev. B}\ }\textbf {\bibinfo {volume} {92}},\ \bibinfo {pages} {081201(R)}
			(\bibinfo {year} {2015})}\BibitemShut {NoStop}%
		\bibitem [{\citenamefont {Bzdu\ifmmode~\check{s}\else \v{s}\fi{}ek}\ and\
			\citenamefont {Sigrist}(2017)}]{bzdusek.2017}%
		\BibitemOpen
		\bibfield  {author} {\bibinfo {author} {\bibfnamefont {Tom\'a\ifmmode
					\check{s}\else~\v{s}\fi{}}\ \bibnamefont {Bzdu\ifmmode~\check{s}\else
					\v{s}\fi{}ek}}\ and\ \bibinfo {author} {\bibfnamefont {Manfred}\ \bibnamefont
				{Sigrist}},\ }\bibfield  {title} {\enquote {\bibinfo {title} {Robust doubly
					charged nodal lines and nodal surfaces in centrosymmetric systems},}\ }\href
		{\doibase 10.1103/PhysRevB.96.155105} {\bibfield  {journal} {\bibinfo
				{journal} {Phys. Rev. B}\ }\textbf {\bibinfo {volume} {96}},\ \bibinfo
			{pages} {155105} (\bibinfo {year} {2017})}\BibitemShut {NoStop}%
		\bibitem [{\citenamefont {Burkov}\ and\ \citenamefont
			{Balents}(2011)}]{burkov.2011}%
		\BibitemOpen
		\bibfield  {author} {\bibinfo {author} {\bibfnamefont {A.~A.}\ \bibnamefont
				{Burkov}}\ and\ \bibinfo {author} {\bibfnamefont {Leon}\ \bibnamefont
				{Balents}},\ }\bibfield  {title} {\enquote {\bibinfo {title} {Weyl semimetal
					in a topological insulator multilayer},}\ }\href {\doibase
			10.1103/PhysRevLett.107.127205} {\bibfield  {journal} {\bibinfo  {journal}
				{Phys. Rev. Lett.}\ }\textbf {\bibinfo {volume} {107}},\ \bibinfo {pages}
			{127205} (\bibinfo {year} {2011})}\BibitemShut {NoStop}%
		\bibitem [{\citenamefont {Hal\'asz}\ and\ \citenamefont
			{Balents}(2012)}]{halasz.2012}%
		\BibitemOpen
		\bibfield  {author} {\bibinfo {author} {\bibfnamefont {G\'abor~B.}\
				\bibnamefont {Hal\'asz}}\ and\ \bibinfo {author} {\bibfnamefont {Leon}\
				\bibnamefont {Balents}},\ }\bibfield  {title} {\enquote {\bibinfo {title}
				{Time-reversal invariant realization of the weyl semimetal phase},}\ }\href
		{\doibase 10.1103/PhysRevB.85.035103} {\bibfield  {journal} {\bibinfo
				{journal} {Phys. Rev. B}\ }\textbf {\bibinfo {volume} {85}},\ \bibinfo
			{pages} {035103} (\bibinfo {year} {2012})}\BibitemShut {NoStop}%
		\bibitem [{\citenamefont {Armitage}\ \emph {et~al.}(2018)\citenamefont
			{Armitage}, \citenamefont {Mele},\ and\ \citenamefont
			{Vishwanath}}]{armitage.2018}%
		\BibitemOpen
		\bibfield  {author} {\bibinfo {author} {\bibfnamefont {N.~P.}\ \bibnamefont
				{Armitage}}, \bibinfo {author} {\bibfnamefont {E.~J.}\ \bibnamefont {Mele}},
			\ and\ \bibinfo {author} {\bibfnamefont {Ashvin}\ \bibnamefont
				{Vishwanath}},\ }\bibfield  {title} {\enquote {\bibinfo {title} {Weyl and
					dirac semimetals in three-dimensional solids},}\ }\href {\doibase
			10.1103/RevModPhys.90.015001} {\bibfield  {journal} {\bibinfo  {journal}
				{Rev. Mod. Phys.}\ }\textbf {\bibinfo {volume} {90}},\ \bibinfo {pages}
			{015001} (\bibinfo {year} {2018})}\BibitemShut {NoStop}%
		\bibitem [{\citenamefont {Klapper}\ and\ \citenamefont {Hahn}(2006)}]{cryst}%
		\BibitemOpen
		\bibfield  {author} {\bibinfo {author} {\bibfnamefont {H.}~\bibnamefont
				{Klapper}}\ and\ \bibinfo {author} {\bibfnamefont {Th.}\ \bibnamefont
				{Hahn}},\ }\enquote {\bibinfo {title} {Point-group symmetry and physical
				properties of crystals},}\ in\ \href {\doibase 10.1107/97809553602060000521}
		{\emph {\bibinfo {booktitle} {International Tables for Crystallography}}}\
		(\bibinfo  {publisher} {Wiley},\ \bibinfo {year} {2006})\ Chap.\ \bibinfo
		{chapter} {10.2}, pp.\ \bibinfo {pages} {804--808}\BibitemShut {NoStop}%
		\bibitem [{\citenamefont {Hertz}(1976)}]{hertz1976}%
		\BibitemOpen
		\bibfield  {author} {\bibinfo {author} {\bibfnamefont {John~A.}\ \bibnamefont
				{Hertz}},\ }\bibfield  {title} {\enquote {\bibinfo {title} {Quantum critical
					phenomena},}\ }\href {\doibase 10.1103/PhysRevB.14.1165} {\bibfield
			{journal} {\bibinfo  {journal} {Phys. Rev. B}\ }\textbf {\bibinfo {volume}
				{14}},\ \bibinfo {pages} {1165--1184} (\bibinfo {year} {1976})}\BibitemShut
		{NoStop}%
		\bibitem [{Note1()}]{Note1}%
		\BibitemOpen
		\bibinfo {note} {See Supplemental Material at [URL will be inserted by
			publisher] for the details of calculations, which includes Refs. \cite
			{fitzpatrick.2013,metlitski.2010,goldefeld1992,Thakur2018,Zhou2018}}\BibitemShut
		{NoStop}%
		\bibitem [{\citenamefont {Huh}\ \emph {et~al.}(2016)\citenamefont {Huh},
			\citenamefont {Moon},\ and\ \citenamefont {Kim}}]{huh.2016}%
		\BibitemOpen
		\bibfield  {author} {\bibinfo {author} {\bibfnamefont {Yejin}\ \bibnamefont
				{Huh}}, \bibinfo {author} {\bibfnamefont {Eun-Gook}\ \bibnamefont {Moon}}, \
			and\ \bibinfo {author} {\bibfnamefont {Yong~Baek}\ \bibnamefont {Kim}},\
		}\bibfield  {title} {\enquote {\bibinfo {title} {Long-range coulomb
					interaction in nodal-ring semimetals},}\ }\href {\doibase
			10.1103/PhysRevB.93.035138} {\bibfield  {journal} {\bibinfo  {journal} {Phys.
					Rev. B}\ }\textbf {\bibinfo {volume} {93}},\ \bibinfo {pages} {035138}
			(\bibinfo {year} {2016})}\BibitemShut {NoStop}%
		\bibitem [{\citenamefont {Jose}\ and\ \citenamefont {Uchoa}(2020)}]{uchoa2019}%
		\BibitemOpen
		\bibfield  {author} {\bibinfo {author} {\bibfnamefont {Geo}\ \bibnamefont
				{Jose}}\ and\ \bibinfo {author} {\bibfnamefont {Bruno}\ \bibnamefont
				{Uchoa}},\ }\bibfield  {title} {\enquote {\bibinfo {title} {Quantum critical
					scaling of gapped phases in nodal-line semimetals},}\ }\href {\doibase
			10.1103/PhysRevB.101.115123} {\bibfield  {journal} {\bibinfo  {journal}
				{Phys. Rev. B}\ }\textbf {\bibinfo {volume} {101}},\ \bibinfo {pages}
			{115123} (\bibinfo {year} {2020})}\BibitemShut {NoStop}%
		\bibitem [{\citenamefont {Rhim}\ and\ \citenamefont {Kim}(2016)}]{rhim2016}%
		\BibitemOpen
		\bibfield  {author} {\bibinfo {author} {\bibfnamefont {Jun-Won}\ \bibnamefont
				{Rhim}}\ and\ \bibinfo {author} {\bibfnamefont {Yong~Baek}\ \bibnamefont
				{Kim}},\ }\bibfield  {title} {\enquote {\bibinfo {title} {Anisotropic density
					fluctuations, plasmons, and friedel oscillations in nodal line semimetal},}\
		}\href {\doibase 10.1088/1367-2630/18/4/043010} {\bibfield  {journal}
			{\bibinfo  {journal} {New Journal of Physics}\ }\textbf {\bibinfo {volume}
				{18}},\ \bibinfo {pages} {043010} (\bibinfo {year} {2016})}\BibitemShut
		{NoStop}%
		\bibitem [{\citenamefont {Alicea}\ and\ \citenamefont
			{Fisher}(2006)}]{alicea2006}%
		\BibitemOpen
		\bibfield  {author} {\bibinfo {author} {\bibfnamefont {Jason}\ \bibnamefont
				{Alicea}}\ and\ \bibinfo {author} {\bibfnamefont {Matthew P.~A.}\
				\bibnamefont {Fisher}},\ }\bibfield  {title} {\enquote {\bibinfo {title}
				{Graphene integer quantum hall effect in the ferromagnetic and paramagnetic
					regimes},}\ }\href {\doibase 10.1103/PhysRevB.74.075422} {\bibfield
			{journal} {\bibinfo  {journal} {Phys. Rev. B}\ }\textbf {\bibinfo {volume}
				{74}},\ \bibinfo {pages} {075422} (\bibinfo {year} {2006})}\BibitemShut
		{NoStop}%
		\bibitem [{\citenamefont {Fuchs}\ and\ \citenamefont
			{Lederer}(2007)}]{Fuchs2007}%
		\BibitemOpen
		\bibfield  {author} {\bibinfo {author} {\bibfnamefont {Jean-No\"el}\
				\bibnamefont {Fuchs}}\ and\ \bibinfo {author} {\bibfnamefont {Pascal}\
				\bibnamefont {Lederer}},\ }\bibfield  {title} {\enquote {\bibinfo {title}
				{Spontaneous parity breaking of graphene in the quantum hall regime},}\
		}\href {\doibase 10.1103/PhysRevLett.98.016803} {\bibfield  {journal}
			{\bibinfo  {journal} {Phys. Rev. Lett.}\ }\textbf {\bibinfo {volume} {98}},\
			\bibinfo {pages} {016803} (\bibinfo {year} {2007})}\BibitemShut {NoStop}%
		\bibitem [{\citenamefont {Gross}\ and\ \citenamefont
			{Neveu}(1974)}]{gross.1974}%
		\BibitemOpen
		\bibfield  {author} {\bibinfo {author} {\bibfnamefont {David~J.}\
				\bibnamefont {Gross}}\ and\ \bibinfo {author} {\bibfnamefont {Andr\'e}\
				\bibnamefont {Neveu}},\ }\bibfield  {title} {\enquote {\bibinfo {title}
				{Dynamical symmetry breaking in asymptotically free field theories},}\ }\href
		{\doibase 10.1103/PhysRevD.10.3235} {\bibfield  {journal} {\bibinfo
				{journal} {Phys. Rev. D}\ }\textbf {\bibinfo {volume} {10}},\ \bibinfo
			{pages} {3235--3253} (\bibinfo {year} {1974})}\BibitemShut {NoStop}%
		\bibitem [{\citenamefont {Zinn-Justin}(1991)}]{zinn.1991}%
		\BibitemOpen
		\bibfield  {author} {\bibinfo {author} {\bibfnamefont {J.}~\bibnamefont
				{Zinn-Justin}},\ }\bibfield  {title} {\enquote {\bibinfo {title}
				{Four-fermion interaction near four dimensions},}\ }\href {\doibase
			https://doi.org/10.1016/0550-3213(91)90043-W} {\bibfield  {journal} {\bibinfo
				{journal} {Nuclear Physics B}\ }\textbf {\bibinfo {volume} {367}},\ \bibinfo
			{pages} {105 -- 122} (\bibinfo {year} {1991})}\BibitemShut {NoStop}%
		\bibitem [{\citenamefont {Mihaila}\ \emph {et~al.}(2017)\citenamefont
			{Mihaila}, \citenamefont {Zerf}, \citenamefont {Ihrig}, \citenamefont
			{Herbut},\ and\ \citenamefont {Scherer}}]{mihaila.2017}%
		\BibitemOpen
		\bibfield  {author} {\bibinfo {author} {\bibfnamefont {Luminita~N.}\
				\bibnamefont {Mihaila}}, \bibinfo {author} {\bibfnamefont {Nikolai}\
				\bibnamefont {Zerf}}, \bibinfo {author} {\bibfnamefont {Bernhard}\
				\bibnamefont {Ihrig}}, \bibinfo {author} {\bibfnamefont {Igor~F.}\
				\bibnamefont {Herbut}}, \ and\ \bibinfo {author} {\bibfnamefont {Michael~M.}\
				\bibnamefont {Scherer}},\ }\bibfield  {title} {\enquote {\bibinfo {title}
				{Gross-neveu-yukawa model at three loops and ising critical behavior of dirac
					systems},}\ }\href {\doibase 10.1103/PhysRevB.96.165133} {\bibfield
			{journal} {\bibinfo  {journal} {Phys. Rev. B}\ }\textbf {\bibinfo {volume}
				{96}},\ \bibinfo {pages} {165133} (\bibinfo {year} {2017})}\BibitemShut
		{NoStop}%
		\bibitem [{\citenamefont {Lang}\ and\ \citenamefont
			{L\"auchli}(2019)}]{lang.2018}%
		\BibitemOpen
		\bibfield  {author} {\bibinfo {author} {\bibfnamefont {Thomas~C.}\
				\bibnamefont {Lang}}\ and\ \bibinfo {author} {\bibfnamefont {Andreas~M.}\
				\bibnamefont {L\"auchli}},\ }\bibfield  {title} {\enquote {\bibinfo {title}
				{Quantum monte carlo simulation of the chiral heisenberg gross-neveu-yukawa
					phase transition with a single dirac cone},}\ }\href {\doibase
			10.1103/PhysRevLett.123.137602} {\bibfield  {journal} {\bibinfo  {journal}
				{Phys. Rev. Lett.}\ }\textbf {\bibinfo {volume} {123}},\ \bibinfo {pages}
			{137602} (\bibinfo {year} {2019})}\BibitemShut {NoStop}%
		\bibitem [{\citenamefont {Kozii}\ \emph {et~al.}(2019)\citenamefont {Kozii},
			\citenamefont {Bi},\ and\ \citenamefont {Ruhman}}]{kozii2019}%
		\BibitemOpen
		\bibfield  {author} {\bibinfo {author} {\bibfnamefont {Vladyslav}\
				\bibnamefont {Kozii}}, \bibinfo {author} {\bibfnamefont {Zhen}\ \bibnamefont
				{Bi}}, \ and\ \bibinfo {author} {\bibfnamefont {Jonathan}\ \bibnamefont
				{Ruhman}},\ }\bibfield  {title} {\enquote {\bibinfo {title}
				{Superconductivity near a ferroelectric quantum critical point in
					ultralow-density dirac materials},}\ }\href {\doibase
			10.1103/PhysRevX.9.031046} {\bibfield  {journal} {\bibinfo  {journal} {Phys.
					Rev. X}\ }\textbf {\bibinfo {volume} {9}},\ \bibinfo {pages} {031046}
			(\bibinfo {year} {2019})}\BibitemShut {NoStop}%
		\bibitem [{\citenamefont {Coleman}\ \emph {et~al.}(2001)\citenamefont
			{Coleman}, \citenamefont {P{\'{e}}pin}, \citenamefont {Si},\ and\
			\citenamefont {Ramazashvili}}]{coleman2001}%
		\BibitemOpen
		\bibfield  {author} {\bibinfo {author} {\bibfnamefont {P}~\bibnamefont
				{Coleman}}, \bibinfo {author} {\bibfnamefont {C}~\bibnamefont {P{\'{e}}pin}},
			\bibinfo {author} {\bibfnamefont {Qimiao}\ \bibnamefont {Si}}, \ and\
			\bibinfo {author} {\bibfnamefont {R}~\bibnamefont {Ramazashvili}},\
		}\bibfield  {title} {\enquote {\bibinfo {title} {How do fermi liquids get
					heavy and die?}}\ }\href {\doibase 10.1088/0953-8984/13/35/202} {\bibfield
			{journal} {\bibinfo  {journal} {Journal of Physics: Condensed Matter}\
			}\textbf {\bibinfo {volume} {13}},\ \bibinfo {pages} {R723--R738} (\bibinfo
			{year} {2001})}\BibitemShut {NoStop}%
		\bibitem [{\citenamefont {Schofield}(1999)}]{schofield1999}%
		\BibitemOpen
		\bibfield  {author} {\bibinfo {author} {\bibfnamefont {A.~J.}\ \bibnamefont
				{Schofield}},\ }\bibfield  {title} {\enquote {\bibinfo {title} {Non-fermi
					liquids},}\ }\href {\doibase 10.1080/001075199181602} {\bibfield  {journal}
			{\bibinfo  {journal} {Contemporary Physics}\ }\textbf {\bibinfo {volume}
				{40}},\ \bibinfo {pages} {95--115} (\bibinfo {year} {1999})},\ \Eprint
		{http://arxiv.org/abs/https://doi.org/10.1080/001075199181602}
		{https://doi.org/10.1080/001075199181602} \BibitemShut {NoStop}%
		\bibitem [{\citenamefont {Vig}\ \emph {et~al.}(2017)\citenamefont {Vig},
			\citenamefont {Kogar}, \citenamefont {Mitrano}, \citenamefont {Husain},
			\citenamefont {Mishra}, \citenamefont {Rak}, \citenamefont {Venema},
			\citenamefont {Johnson}, \citenamefont {Gu}, \citenamefont {Fradkin},
			\citenamefont {Norman},\ and\ \citenamefont {Abbamonte}}]{eels2017}%
		\BibitemOpen
		\bibfield  {author} {\bibinfo {author} {\bibfnamefont {Sean}\ \bibnamefont
				{Vig}}, \bibinfo {author} {\bibfnamefont {Anshul}\ \bibnamefont {Kogar}},
			\bibinfo {author} {\bibfnamefont {Matteo}\ \bibnamefont {Mitrano}}, \bibinfo
			{author} {\bibfnamefont {Ali~A.}\ \bibnamefont {Husain}}, \bibinfo {author}
			{\bibfnamefont {Vivek}\ \bibnamefont {Mishra}}, \bibinfo {author}
			{\bibfnamefont {Melinda~S.}\ \bibnamefont {Rak}}, \bibinfo {author}
			{\bibfnamefont {Luc}\ \bibnamefont {Venema}}, \bibinfo {author}
			{\bibfnamefont {Peter~D.}\ \bibnamefont {Johnson}}, \bibinfo {author}
			{\bibfnamefont {Genda~D.}\ \bibnamefont {Gu}}, \bibinfo {author}
			{\bibfnamefont {Eduardo}\ \bibnamefont {Fradkin}}, \bibinfo {author}
			{\bibfnamefont {Michael~R.}\ \bibnamefont {Norman}}, \ and\ \bibinfo {author}
			{\bibfnamefont {Peter}\ \bibnamefont {Abbamonte}},\ }\bibfield  {title}
		{\enquote {\bibinfo {title} {{Measurement of the dynamic charge response of
						materials using low-energy, momentum-resolved electron energy-loss
						spectroscopy (M-EELS)}},}\ }\href {\doibase 10.21468/SciPostPhys.3.4.026}
		{\bibfield  {journal} {\bibinfo  {journal} {SciPost Phys.}\ }\textbf
			{\bibinfo {volume} {3}},\ \bibinfo {pages} {026} (\bibinfo {year}
			{2017})}\BibitemShut {NoStop}%
		\bibitem [{\citenamefont {Volkov}\ and\ \citenamefont
			{Moroz}(2018)}]{volkov2018}%
		\BibitemOpen
		\bibfield  {author} {\bibinfo {author} {\bibfnamefont {Pavel~A.}\
				\bibnamefont {Volkov}}\ and\ \bibinfo {author} {\bibfnamefont {Sergej}\
				\bibnamefont {Moroz}},\ }\bibfield  {title} {\enquote {\bibinfo {title}
				{Coulomb-induced instabilities of nodal surfaces},}\ }\href {\doibase
			10.1103/PhysRevB.98.241107} {\bibfield  {journal} {\bibinfo  {journal} {Phys.
					Rev. B}\ }\textbf {\bibinfo {volume} {98}},\ \bibinfo {pages} {241107(R)}
			(\bibinfo {year} {2018})}\BibitemShut {NoStop}%
		\bibitem [{\citenamefont {Bersuker}(1966)}]{bersuker1966}%
		\BibitemOpen
		\bibfield  {author} {\bibinfo {author} {\bibfnamefont {I.B.}\ \bibnamefont
				{Bersuker}},\ }\bibfield  {title} {\enquote {\bibinfo {title} {On the origin
					of ferroelectricity in perovskite-type crystals},}\ }\href {\doibase
			https://doi.org/10.1016/0031-9163(66)91127-9} {\bibfield  {journal} {\bibinfo
				{journal} {Physics Letters}\ }\textbf {\bibinfo {volume} {20}},\ \bibinfo
			{pages} {589 -- 590} (\bibinfo {year} {1966})}\BibitemShut {NoStop}%
		\bibitem [{\citenamefont {Fitzpatrick}\ \emph {et~al.}(2013)\citenamefont
			{Fitzpatrick}, \citenamefont {Kachru}, \citenamefont {Kaplan},\ and\
			\citenamefont {Raghu}}]{fitzpatrick.2013}%
		\BibitemOpen
		\bibfield  {author} {\bibinfo {author} {\bibfnamefont {A.~Liam}\ \bibnamefont
				{Fitzpatrick}}, \bibinfo {author} {\bibfnamefont {Shamit}\ \bibnamefont
				{Kachru}}, \bibinfo {author} {\bibfnamefont {Jared}\ \bibnamefont {Kaplan}},
			\ and\ \bibinfo {author} {\bibfnamefont {S.}~\bibnamefont {Raghu}},\
		}\bibfield  {title} {\enquote {\bibinfo {title} {Non-fermi-liquid fixed point
					in a wilsonian theory of quantum critical metals},}\ }\href {\doibase
			10.1103/PhysRevB.88.125116} {\bibfield  {journal} {\bibinfo  {journal} {Phys.
					Rev. B}\ }\textbf {\bibinfo {volume} {88}},\ \bibinfo {pages} {125116}
			(\bibinfo {year} {2013})}\BibitemShut {NoStop}%
		\bibitem [{\citenamefont {Metlitski}\ and\ \citenamefont
			{Sachdev}(2010)}]{metlitski.2010}%
		\BibitemOpen
		\bibfield  {author} {\bibinfo {author} {\bibfnamefont {Max~A.}\ \bibnamefont
				{Metlitski}}\ and\ \bibinfo {author} {\bibfnamefont {Subir}\ \bibnamefont
				{Sachdev}},\ }\bibfield  {title} {\enquote {\bibinfo {title} {Quantum phase
					transitions of metals in two spatial dimensions. i. ising-nematic order},}\
		}\href {\doibase 10.1103/PhysRevB.82.075127} {\bibfield  {journal} {\bibinfo
				{journal} {Phys. Rev. B}\ }\textbf {\bibinfo {volume} {82}},\ \bibinfo
			{pages} {075127} (\bibinfo {year} {2010})}\BibitemShut {NoStop}%
		\bibitem [{\citenamefont {Goldenfeld}(1992)}]{goldefeld1992}%
		\BibitemOpen
		\bibfield  {author} {\bibinfo {author} {\bibfnamefont {N.}~\bibnamefont
				{Goldenfeld}},\ }\href@noop {} {\emph {\bibinfo {title} {{Lectures on phase
						transitions and the renormalization group}}}}\ (\bibinfo {year}
		{1992})\BibitemShut {NoStop}%
		%%CITATION = INSPIRE-345315;%%
		\bibitem [{\citenamefont {Thakur}\ \emph {et~al.}(2018)\citenamefont {Thakur},
			\citenamefont {Sadhukhan},\ and\ \citenamefont {Agarwal}}]{Thakur2018}%
		\BibitemOpen
		\bibfield  {author} {\bibinfo {author} {\bibfnamefont {Anmol}\ \bibnamefont
				{Thakur}}, \bibinfo {author} {\bibfnamefont {Krishanu}\ \bibnamefont
				{Sadhukhan}}, \ and\ \bibinfo {author} {\bibfnamefont {Amit}\ \bibnamefont
				{Agarwal}},\ }\bibfield  {title} {\enquote {\bibinfo {title} {Dynamic
					current-current susceptibility in three-dimensional dirac and weyl
					semimetals},}\ }\href {\doibase 10.1103/PhysRevB.97.035403} {\bibfield
			{journal} {\bibinfo  {journal} {Phys. Rev. B}\ }\textbf {\bibinfo {volume}
				{97}},\ \bibinfo {pages} {035403} (\bibinfo {year} {2018})}\BibitemShut
		{NoStop}%
		\bibitem [{\citenamefont {Zhou}\ and\ \citenamefont {Chang}(2018)}]{Zhou2018}%
		\BibitemOpen
		\bibfield  {author} {\bibinfo {author} {\bibfnamefont {Jianhui}\ \bibnamefont
				{Zhou}}\ and\ \bibinfo {author} {\bibfnamefont {Hao-Ran}\ \bibnamefont
				{Chang}},\ }\bibfield  {title} {\enquote {\bibinfo {title} {Dynamical
					correlation functions and the related physical effects in three-dimensional
					weyl/dirac semimetals},}\ }\href {\doibase 10.1103/PhysRevB.97.075202}
		{\bibfield  {journal} {\bibinfo  {journal} {Phys. Rev. B}\ }\textbf {\bibinfo
				{volume} {97}},\ \bibinfo {pages} {075202} (\bibinfo {year}
			{2018})}\BibitemShut {NoStop}%
	\end{thebibliography}
\end{document}